\documentclass[useAMS,usenatbib]{mnras}

\usepackage{newtxmath}
\RequirePackage{amsfonts}
\usepackage{graphicx}
\usepackage{natbib}
\usepackage{xcolor}

\usepackage[T1]{fontenc}
\usepackage{ae,aecompl}

\usepackage{graphicx}	% Including figure files
\usepackage{amsmath}	% Advanced maths commands
\usepackage{amssymb}	% Extra maths symbols
\usepackage{bm}      	%for bold italic symbol, vector
\usepackage{gensymb}  	%for degree
\usepackage{makecell}
\usepackage{multirow}
\usepackage{url}

\newcommand\rxj{RXJ\,1131$-$1231}
\newcommand\he{HE\,0435$-$1223}
\newcommand\pg{PG\,1115$+$080}

\newcommand{\sref}[1]{Section~\ref{#1}}
\newcommand{\fref}[1]{Figure~\ref{#1}}
\newcommand{\tref}[1]{Table~\ref{#1}}
\newcommand{\eref}[1]{Equation~(\ref{#1})}
\newcommand{\aref}[1]{Appendix~(\ref{#1})}

\def\hst{\textit{HST}}
\def\planck{\textit{Planck}}

\def\zl{z_{\ell}}
\def\zs{z_{s}}

\def\kms {\rm km\,s^{-1}}
\def\kmsmpc{\rm km\,s^{-1}\,Mpc^{-1}}
\def\Msun{M_{\sun}}
\newcommand{\Ddt}{D_{\Delta t}}

\title[$H_{0}~from~three~lenses~with~AO~imaging$]{A SHARP view of H0LiCOW: $H_{0}$ from three time-delay gravitational lens systems with adaptive optics imaging}
\author[Geoff~C.-F.~Chen et al.]{
Geoff~C.-F.~Chen,$^{1}$\thanks{E-mail: chfchen@ucdavis.edu}
Christopher~D.~Fassnacht,$^{1}$
Sherry.~H.~Suyu,$^{2,3,4}$
\newauthor{
Cristian E. Rusu,$^{5,6}$\thanks{Subaru Fellow}
James~H.~H.~Chan,$^{7}$
Kenneth~C.~Wong,$^{8,5}$
}
\newauthor{
Matthew W. Auger,$^{9,10}$
Stefan Hilbert,$^{11,12}$
Vivien Bonvin,$^{7}$
Simon Birrer,$^{13}$
}
\newauthor{
Martin~Millon,$^{7}$
L{\'e}on~V.~E.~Koopmans,$^{14}$
David~J.~Lagattuta,$^{15}$
}
\newauthor{
John~P.~McKean,$^{14,16}$
Simona~Vegetti,$^{2}$
Frederic Courbin,$^{7}$
Xuheng Ding$,^{13}$
}
\newauthor{
Aleksi Halkola,$^{17}$
Inh Jee,$^{2}$
Anowar J. Shajib,$^{13}$
Dominique Sluse,$^{18}$
}
\newauthor{
Alessandro Sonnenfeld,$^{8,19}$
and Tommaso Treu$^{13}$
}
\\
$^{1}$Department of Physics, University of California, Davis, CA 95616, USA\\
$^{2}$Max $\planck$ Institute for Astrophysics, Karl-Schwarzschild-Strasse 1, D-85740 Garching, Germany\\
$^{3}$Physik-Department, Technische Universität München, James-Franck-Str. 1, 85748 Garching, Germany\\
$^{4}$Academia Sinica Institute of Astronomy and Astrophysics (ASIAA), 11F of ASMAB, No.1, Section 4, Roosevelt Road, Taipei 10617,\\
~~Taiwan \\
$^{5}$National Astronomical Observatory of Japan, 2-21-1 Osawa, Mitaka, Tokyo 181-8588, Japan\\
$^{6}$Subaru Telescope, National Astronomical Observatory of Japan, 650 N Aohoku Pl, Hilo, HI 96720\\ 
$^{7}$Institute of Physics, Laboratory of Astrophysics, Ecole Polytechnique F{\'e}d{\'e}rale de Lausanne (EPFL), Observatoire de Sauverny,\\
~~CH-1290 Versoix, Switzerland\\
$^{8}$Kavli Institute for the Physics and Mathematics of the Universe (Kavli IPMU, WPI), University of Tokyo, Chiba 277-8583, Japan\\
$^{9}$Institute of Astronomy, University of Cambridge, Madingley Rd, Cambridge, CB3 0HA, UK\\
$^{10}$Kavli Institute for Cosmology, University of Cambridge, Madingley Road, Cambridge CB3 0HA, UK\\
$^{11}$Exzellenzcluster Universe, Boltzmannstr. 2, D-85748 Garching, Germany\\
$^{12}$Ludwig-Maximilians-Universit¨at, Universit¨ats-Sternwarte, Scheinerstr. 1, D-81679 Munchen, Germany\\
$^{13}$Department of Physics and Astronomy, University of California, Los Angeles, CA 90095, USA\\
$^{14}$Kapteyn Astronomical Institute, University of Groningen, P.O.Box 800, 9700 AV Groningen, The Netherlands\\
$^{15}$Univ Lyon, Univ Lyon1, Ens de Lyon, CNRS, Centre de Recherche Astrophysique de Lyon UMR5574, F-69230, Saint-Genis-Laval,\\ ~~~France\\
$^{16}$ASTRON, Netherlands Institute for Radio Astronomy, P.O. Box 2, 7990 AA Dwingeloo, the Netherlands\\
$^{17}$Pyörrekuja 5 A, 04300 Tuusula, Finland\\
$^{18}$STAR Institute, Quartier Agora~--~All$\acute{\text{e}}$e du six A$\hat{\text{o}}$ut, 19c, B-4000 Li$\grave{\text{e}}$ge, Belgium\\
$^{19}$Leiden Observatory, Leiden University, Niels Bohrweg 2, 2333 CA Leiden, the Netherlands
}
\date{Accepted XXX. Received YYY; in original form ZZZ}
\pubyear{2019}
\begin{document}
\label{firstpage}
\pagerange{\pageref{firstpage}--\pageref{lastpage}}
\maketitle

\begin{abstract}
We present the measurement of the Hubble Constant, $H_0$, with three strong gravitational lens systems.
We describe a blind analysis of both \pg~and \he~as well as an extension of our previous analysis of \rxj. 
For each lens, we combine new adaptive optics (AO) imaging from the Keck Telescope, obtained as part of the SHARP AO effort, with Hubble Space Telescope (\hst) imaging, velocity dispersion measurements, and a description of the line-of-sight mass distribution to build an accurate and precise lens mass model. 
This mass model is then combined with the COSMOGRAIL measured time delays in these systems to determine $H_{0}$.  
We do both an AO-only and an AO+\hst\ analysis of the systems and find that AO and HST results are consistent.
After unblinding, the AO-only analysis gives $H_{0}=82.8\substack{+9.4\\-8.3}~\kmsmpc$ for \pg,  $H_{0}=70.1\substack{+5.3\\-4.5}~\kmsmpc$ for \he,
and $H_{0}=77.0\substack{+4.0\\-4.6}~\kmsmpc$ for \rxj. 
The joint AO-only result for the three lenses is $H_{0}=75.6\substack{+3.2\\-3.3}~\kmsmpc$. 
The joint result of the AO+\hst\ analysis for the three lenses is $H_{0}=76.8\substack{+2.6\\-2.6}~\kmsmpc$. 
All of the above results assume a flat $\Lambda$ cold dark matter cosmology with a uniform prior on $\Omega_{\textrm{m}}$ in [0.05, 0.5] and $H_{0}$ in [0, 150] $\kmsmpc$. This work is a collaboration of the SHARP and H0LiCOW teams, and shows that AO data can be used as the high-resolution imaging component in lens-based measurements of $H_0$. The full time-delay cosmography results from a total of six strongly lensed systems are presented in a companion paper.
\end{abstract}


%%%%%%%%%%%%%%%%%%%%%%%%%%%%%%%%%%%%%%%%%%%%%%%%%%

%%%%%%%%%%%%%%%%% BODY OF PAPER %%%%%%%%%%%%%%%%%%

\section{Introduction}

\subsection{Distance Measurement Discrepancy: A 4.4$\sigma$ Tension on the Value of $H_{0}$}
The temperature anisotropies of the Cosmic Microwave Background (CMB) and density correlations of Baryon Acoustic Oscillations (BAO) obtained with the Wilkinson Microwave Anisotropy Probe, $\planck$ satellite, and BAO surveys provide strong support to the standard flat $\Lambda$CDM cosmological model \citep[e.g.,][]{KomatsuEtal11,HinshawEtal13,planck18parameter}. Under a few strong assumptions, such as flatness and constant dark energy density, these data give sub-percent precision on the parameters of the standard cosmological model \citep[e.g.,][]{AndersonEtal14,KazinEtal14,RossEtal15}.

Intriguingly, distance measurements from Type-Ia supernova (SN) that have been calibrated by the local distance ladder are smaller than the predictions from the CMB data given the flat $\Lambda$CDM model \citep[see the illustration in Fig. 4 in][]{CuestaEtal15}, 
leading to a $\sim 4.4\sigma$ tension in $H_0$ between the value predicted by the CMB and the local value \citep[$=74.03\pm1.42~\kmsmpc$,][]{RiessEtal19}.
The SN data can also be calibrated by the inverse distance ladder method to yield a model-dependent value of $H_{0}$.
Under the assumption of the standard pre-recombination physics,
combining BAO and SN with the CMB-calibrated physical scale of the sound horizon gives $H_{0} =67.3\pm1.1~\kmsmpc$ \citep{AuborugEtal15}. A recent blind analysis with  additional SN data from Dark Energy Survey, gives $H_{0} =67.77\pm1.30~\kmsmpc$ \citep{MacaulayEtal19}. Both results are in excellent agreement with the $\planck$ value ($H_{0} = 67.27\pm0.6~\kmsmpc$) under the assumption of flat $\Lambda$CDM model \citep{planck18parameter}. Furthermore, even without using the CMB anisotropy, the combination of BAO data with light element abundances  
produces $\planck$-like $H_{0}$ values \citep{AddisonEtal18}. This indicates that systematic errors, especially in the $\planck$ data analysis, most likely are not the main driver of the $H_{0}$ discrepancies. 
Similarly, the local distance ladder analyses also have passed a range of systematic checks 
\citep[e.g.,][]{Efstathiou14,CardonaEtal17,ZhangEtal17,FollinKonx18,FeeneyEtal18,DhawanEtal18,RiessEtal19}, 
and rule out the local void scenario
\citep{KeenanEtal13,FleuryEtal17,ShanksEtal19,KenworthyEtal19}.

There have been several attempts to address this $\sim 4.4\sigma$ tension by extending the standard cosmological model, either by changing the size of the sound horizon in the early Universe \citep[e.g.,][]{HeavensEtal14,WymanEtal14,CuestaEtal15,AlamEtal17,KreischEtal19,PoulinEtal18,AgrawalEtal19} or by altering the expansion history \citep{EfstathiouEtal03,Linder04,MorescoEtal16,AlamEtal17}. 

Recent studies \citep[e.g.,][]{BernalEtal16,JoudakiEtal18,LemosEtal19,AylorEtal19} also have tried to directly reconstruct $H(z)$ in order to investigate the $H_{0}$ tension in the context of a possibly poor understanding of the evolution of dark energy density. From an empirical point of view, the current SN and BAO data sets only support $w(z)=-1$ within 
the redshift range where data are available \citep{CuestaEtal16}. A very recent and dramatic decrease in $w$ or the presence of strong dark energy at $3 <z < 1000$ may escape detection and still generate a high value of $H_{0}$ \citep{RiessEtal16}. 
Standard sirens could possibly explore the $z > 3$ range in the future %\citep{ChenHEtal17} 
and provide a high-precision $H_{0}$ measurement \citep{ChenHEtal18natural}. 
Nevertheless, it is also important to note that some $H_{0}$-value tension remains even if we do not consider the distance ladder constraints. 
For example, the high-$\ell$ CMB power spectrum prefers an even lower $H_{0}$ value than that from the low-$\ell$ CMB power spectrum \citep{AddisonEtal16,planck18parameter}.

Given the various tensions across different datasets, any convincing resolution to the $H_0$ tensions, either due to unknown systematics or new physics, needs to simultaneously resolve multiple disagreements. 
Therefore, comparing the distance measurements among independent and robust methodologies to cross-examine the $H_{0}$ tension is probably the only way to shed light on the true answer.

\subsection{Distance Measurement from Time-Delay Cosmography}
Time-delay cosmography is not only a completely independent technique of the distance ladder methods, but it also has the advantage of being a one-step measurement of combined cosmological distances.
In addition, time-delay cosmography is a complementary and cost-effective alternative compared to Type-Ia SN or BAO \citep{SuyuEtal13,TewesEtal13a}.
In a time-delay gravitational lens, the combined cosmological distance we can measure is called the time-delay distance \citep{SuyuEtal10}, which is a ratio of the angular diameter distances in the system: 
\begin{equation}
\label{eq:TDdistance}
\Ddt\equiv\left(1+
\zl\right)\frac{D_{\ell}D_{s}}{{D_{\ell s}}}\propto H_{0}^{-1},
\end{equation}
where $D_{\ell}$ is the distance to the lensing galaxy, $D_s$ is the distance to the background source, and $D_{\ell s}$ is the distance between the lens and the source.  Furthermore, we can make a separate determination of $D_{\ell}$ by measuring the velocity dispersion of the lensing galaxy \citep{JeeEtal15,JeeEtal16,BirrerEtal16,BirrerEtal19}. 
First proposed by \citet{Refsdal64}, the $H_0$ measurement requires modeling the mass in the lensing galaxy and along the line of sight, and measuring the time delays between multiple images via a monitoring program. 
The advantage of this method is that $\Ddt$ is primarily sensitive to $H_{0}$ and insensitive to the neutrino physics and spatial curvature, but still sensitive to the properties of dark energy \citep{BonvinEtal17,BirrerEtal19}.

The H0LiCOW collaboration\footnote{$H_0$ Lenses in COSMOGRAIL's Wellspring, \url{www.h0licow.org}} 
is using strong gravitational lens systems to measure cosmological parameters 
\citep[][]{suyuEtal17}.  
The most recent measurement of $H_0$ from the collaboration used the doubly-lensed quasar system, SDSS J1206+4332, to derive $H_{0}=68.8\substack{+5.4\\-5.1}~\kmsmpc$ for that lens system alone, as well as combining the new system with previous H0LiCOW lenses \citep{SuyuEtal09,SuyuEtal10,SuyuEtal13,SuyuEtal14,SluseEtal17,RusuEtal17,WongEtal17,BonvinEtal17} to obtain a joint inference on $H_0$ with 3\% precision: $H_{0}=72.5\substack{+2.1\\-2.3}~\kmsmpc$ \citep{BirrerEtal19}.
This result agrees with the \citet{RiessEtal19} value within the $1\sigma$ uncertainties.
More recently, the collaboration completed its analysis of WFI2033$-$4723 using the time delays from COSMOGRAIL\footnote{COSmological MOnitoring of GRAvItational Lenses} \citep{BonvinEtal19_WFI2033,SluseEtal19,RusuEtal19_H0LiCOW}.

Achieving the goal of obtaining a 1\%  or better measurement of $H_{0}$ with time-delay cosmography requires a significantly larger sample of lensed quasars with high-quality data than has been analyzed to date.   
Many new lensed quasars have been discovered \citep[e.g.,][]{LinEtal17,SchechterEtal17,AgnelloEtal18,OstrovskiEtal18,WilliamEtal18,RusuEtal19,LemonEtal19} using state-of-the-art lens-finding techniques applied to current large sky surveys \citep[e.g.][]{JosephEtal14,Agnello17,PetrilloEtal17,OstrovskiEtal17,LanusseEtal18,SpinielloEtal18,TreuEtal18,AvestruzEtal19}, and more are expected to be found with the Large Synoptic Survey Telescope \citep{OguriMarshall10}.
Hence, a 1\% $H_{0}$ measurement from time-delay cosmography is a realistic expectation in the near future \citep[e.g.,][Jee et al. 2019 submitted]{JeeEtal15,JeeEtal16,deGrijsEtal17,SuyuEtal18,ShajibEtal18} if we can control the systematic effects to a sub-percent level \citep[][]{DoblerEtal13,LiaoEtal15,DingEtal18}.

\subsection{Lens Modeling with Adaptive Optics Data}

A critical input to achieving robust modeling of the mass distribution in the lensing galaxy is sensitive high-resolution imaging in which extended emission from the background source is detected.  For this reason, many models of the lensing potential are based on imaging from HST
 \citep[e.g.,][]{SuyuEtal09,SuyuEtal10,BirrerEtal15,WongEtal17,BirrerEtal16,BirrerEtal19}. 
Adaptive optics (AO) observations with ground-based telescopes can provide imaging with comparable or better angular resolution to the HST data, especially for systems that are faint in the optical and bright at near infrared wavelengths, thus providing an attractive alternative for modeling the lens mass distribution \citep[e.g.][]{Lagattuta12,GChenEtal16}. 

However, the challenge of using AO data is the unstable point spread function (PSF).
\citet{GChenEtal16} showed that with a new iterative PSF-reconstruction method applied to \rxj, the reconstructed PSF allows one to model the AO imaging down to the noise level, as well as providing tighter constraints on the lens model than were obtained from HST imaging of the system. 
In this new analysis, we apply the PSF-reconstruction method to three lens systems, \he, \pg, and \rxj, which have not only high-resolution AO imaging and HST imaging, but also measured time delays, stellar velocity dispersions, and studies of their environments.  We then infer their time-delay distances via detailed lens mass modelling.  The AO data for \rxj\ have already been analysed by \citet{GChenEtal16}, but the analyses based on the AO imaging of \he\ and \pg\ are presented here for the first time.

The Keck AO imaging data are part of the Strong-lensing High Angular Resolution Programme (SHARP; Fassnacht et al. in preparation), which aims to study the nature of dark matter using high-resolution AO imaging \citep[e.g.,][]{Lagattuta10,Lagattuta12,Vegetti12,Hsueh16,HsuehEtal17_edgeon,HsuehEtal18,SpingolaEtal18}.
The time delay measurements are provided by the COSMOGRAIL group \citep[e.g.,][]{CourbinEtal05,VuissozEtal07,VuissozEtal08,CourbinEtal11,TewesEtal13b,TewesEtal13a,RathnaEtal13,BonvinEtal17,BonvinEtal18_PGTD}, which aims to provide the highest-precision measurements of time delays. The lens environment of \rxj~and \he~studies are provided by the H0LiCOW team \citep{SuyuEtal14,SluseEtal17,RusuEtal17}.

The outline of the paper is as follows. In \sref{sec:basic}, we briefly recap the basics of obtaining inferences on cosmography with time-delay lenses and the statistical tools we use for this process. We describe the observations of \he, \pg, and \rxj\ with the AO imaging system at the Keck Observatory in \sref{sec:data}. In \sref{sec:modeling_tool}, we describe the models that we use to analyze the data. In \sref{sec:lens_modeling}, we elaborate the detailed lens modeling and the properties of the reconstructed PSF for each system. In \sref{sec:cosmoinfer}, we present the joint cosmological inference from the AO imaging only as well as from combined AO plus HST. We summarize in \sref{sec:conclusion}.

\section{Basic theory}
\label{sec:basic}
In this section, we briefly introduce the relation between cosmology and gravitational lensing in \sref{sec:TDcosmo} and the joint inference of all information in \sref{sec:Jointinfer}.

\subsection{Time-Delay Cosmography}
\label{sec:TDcosmo} 
When a compact variable background source, such as an active galactic nucleus (AGN) or SN, sitting inside its host galaxy is strongly lensed by a foreground object, the distorted host galaxy shape, combined with the time delay between the multiple images allows one to precisely determine a particular size of the system. One can express the excess time delays as
\begin{equation}
\label{eq:theory}
\Delta t=(1-\lambda)\frac{\Ddt}{c}\left[\frac{1}{2}
\left(\boldsymbol{\theta}-\boldsymbol{\beta}\right)^{2}-\psi\left(\boldsymbol{\theta}\right)\right],
\end{equation}
where $\bm{\theta}$, $\bm{\beta}$, and
$\psi\left(\bm{\theta}\right)$ are the image location, the source
location, and the projected two-dimension lensing gravitational potential, respectively \citep{Shapiro64, Refsdal64}.  The $\lambda$ parameter represents the lack of perfect knowledge of the full mass distribution, as discussed below.

The advantage of this formulation is the separability of the cosmographic information, contained in the $\Ddt$ parameter, and the lens modeling. This allows one to infer cosmographic information without the need for cosmological priors in the lens modeling.

However, because of the Mass-Sheet Transformation \citep[MST,][]{FalcoEtal85,SchneiderSluse13,SchneiderSluse14}, the determination of $\Ddt$ is subject to an understanding of $\lambda$ \citep[see also the discussion in][]{BirrerEtal19}. Thus, additional priors and information from simulations, environmental data, and the stellar velocity dispersion of the lensing galaxy are required to constrain the degeneracy between different mass profiles and the degeneracy between the mass profile and the mass sheet contributed by the environment \citep[][]{SuyuEtal10,FassnachtEtal11,RusuEtal17,TihhonovaEtal18}. In addition, a prior on the source size \citep{BirrerEtal16} or having a background source of known brightness  \citep[e.g., a lensed SN,][]{GrilloEtal18} can also put constraints on $\lambda$. We refer interested readers to \citet{TreuMarshall16} and \citet{SuyuEtal18} for more details.

\begin{figure*}
\includegraphics*[width=\linewidth]{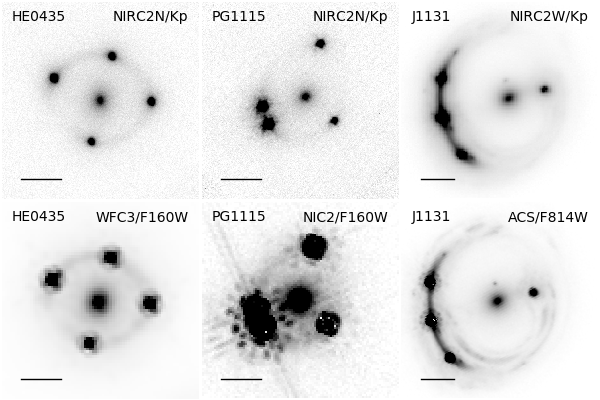}
\caption{Adaptive optics (top row) and HST  (bottom row) images of the three gravitational lens systems. The solid horizontal line represents 1\arcsec scale. The foreground main lenses are located in the center of the lens systems. The multiple lensed images and the extended arc around the lensing galaxy are from the background AGN and its host galaxy. 
}
\label{fig:aoimages}
\end{figure*}

\subsection{Joint Inference}
\label{sec:Jointinfer}
%\subsubsection{Individual lens}
In this work we will present our joint inference on $\Ddt$.
We use $\bm{d}_{i}$ to denote
the imaging data, where $i= \textrm{HE}$, $\textrm{PG}$, and $\textrm{RXJ}$, represent \he, \pg, and \rxj, respectively; we use $\bm{\Delta t}_{i}$ for time delays, $\bm{d}_{\textrm{ENV}_{i}}$ to characterize the lens environments, and $\sigma_{i}$ for the stellar velocity dispersions of the lensing galaxies. Here $\bm{\eta}_{i}$ are the parameters we want to infer from the data, and $\bm{A}$ denotes the discrete assumptions that we made in the models (e.g. whether the lensing galaxy is modeled using a power-law or NFW$+$stellar mass distribution).
The posterior of $\bm{\eta}_{i}$ can be expressed as
\begin{equation}
\begin{split}
&P(\bm{\eta}_{i}|\bm{d}_{i},\bm{\Delta t}_{i},\sigma_{i}, \bm{d}_{\textrm{ENV}_{i}}, \bm{A}) \\
&\propto P(\bm{d}_{i},\bm{\Delta t}_{i},\sigma_{i}, \bm{d}_{\textrm{ENV}_{i}}|\bm{\eta}_{i},\bm{A})P(\bm{\eta}_{i}|\bm{A}), 
\end{split}
\end{equation}
where $P(\bm{d}_{i},\bm{\Delta t}_{i},\sigma_{i}, \bm{d}_{\textrm{ENV}_{i}}|\bm{\eta}_{i},\bm{A}_{i})$ is the joint likelihood for each lens. Since we assume that the environment can be decoupled from the lens, and that the data sets are independent, 
\begin{equation}
\begin{split}
&P(\bm{d}_{i},\bm{\Delta t}_{i},\sigma_{i}, \bm{d}_{\textrm{ENV}_{i}}|\bm{\eta}_{i},\bm{A}_{i}) \\
&=P(\bm{d}_{i}|\bm{\eta}_{i},\bm{A}_{i})P(\bm{\Delta t}_{i}|\bm{\eta}_{i},\bm{A}_{i})P(\sigma_{i}|\bm{\eta}_{i},\bm{A})P(\bm{d}_{\textrm{ENV}_{i}}|\bm{\eta}_{i},\bm{A}_{i}).
\end{split}
\end{equation}
In order to explore the unmodeled systematic uncertainties that may arise from modeling choices, we vary the content of $\bm{A}$ for each lens. The marginalized integral can be expressed as
\begin{equation}
\begin{split}
P(\bm{\eta}_{i}|\bm{d}_{i,\textrm{tot}})&=\int P(\bm{\eta}_{i}|\bm{d}_{i,\textrm{tot}},\bm{A}_{i})P(\bm{A}_{i})d\bm{A}_{i} \\
&\approx \sum_{k}P(\bm{\eta}_{i}|\bm{d}_{i,\textrm{tot}},\bm{A}_{i,k})P(\bm{A}_{i,k}),
\end{split}
\end{equation}
where $\bm{A}_{i,k}$ and $\bm{d}_{i,\textrm{tot}}$ represent the different model choices, and all data sets for the lens system $i$, respectively.

For ranking the models, we follow \citet{BirrerEtal19} to estimate the evidence, $P(\bm{A}_{i,k})$, by using the Bayesian Information Criterion (BIC), which is defined as 
\begin{equation}
\label{eq:BIC_weight}
    \textrm{BIC} = \textrm{ln}(n) k - 2 \textrm{ln}(\hat{L}),
\end{equation}
where $n$ is the number of data points including the lens imaging, 8 AGN positions, three time delays, and one velocity dispersion, $k$ is the number of free parameters in the lens model that are given uniform priors, plus two source position parameters, plus one anisotropy radius to predict the velocity dispersion, and $\hat{L}$ is the maximum likelihood of the model, which is the product of the AGN position likelihood, the time-delay likelihood, the pixelated image plane likelihood, and the kinematic likelihood. The image plane likelihood is the Bayesian evidence of the pixelated source intensity reconstruction using the arcmask imaging data \citep[see][]{SuyuHalkola10} times the likelihood of the lens model parameters within the image plane region that excludes the arcmask. We follow \citet{BirrerEtal19} and calculate the relative BIC and weighting for the SPEMD and composite models separately to avoid biases due to our choice of lens model parameterization.

\section{Data}
\label{sec:data}

The analysis in this paper is based on new Keck AO and archival \hst\ observations of three gravitational lens systems.  In this section we describe the lens sample and the data acquisition and analysis.

\subsection{The Sample}

The sample consists of three well-known lensed quasar systems.  Images of these lens systems are shown in Figure~\ref{fig:aoimages}.
\begin{enumerate}
    \item {\bf HE0435$-$1223:} The \he\ system (J2000: 04
    $^{\textrm{h}}$38$^{\textrm{m}}$14\fs9, −12\degr17\arcmin14\farcs4) is a quadruply-lensed quasar discovered by \citet{WisotzkiEtal02}. The main lensing galaxy is at a redshift of $\zl = 0.4546$ \citep{MorganEtal05}, and the source redshift is $\zs = 1.693$ \citep{SluseEtal12}. The lens resides inside a galaxy group that contains at least 12 galaxies, with a velocity dispersion $\sigma=471\pm100\kms$. \citep[e.g.,][]{MomchevaEtal06,WongEtal11,WilsonEtal16,SluseEtal17}. \citet{WongEtal17} measured the stellar velocity dispersion of the lensing galaxy to be $\sigma=222\pm15~\kms$. The time delays of this system were measured by \citet{BonvinEtal17} with $\sim6.5\%$ uncertainties.
    \item {\bf PG1115+080:} This four-image system was the second strong gravitational lens system to be discovered \citep{WeymannEtal80}.  The system is located at 11$^{\rm h}$18$^{\rm m}$16\fs899 +07\degr45\arcmin58\farcs502 (J2000).  The background quasar with a redshift of $z_{s}$ =1.722 is lensed by a galaxy with $z_{\ell}$ = 0.3098 \citep{HenryEtal86,ChristianEtal87,Tonry98}. The lensed images are in a classic ``fold'' configuration, with an image pair A1 and A2 near the critical curve.  The lens resides inside a galaxy group with 13 known members that has a velocity dispersion of $\sigma=390\pm60\kms$ \citep{WilsonEtal16}. \citet{Tonry98} measured the stellar velocity dispersion of the lensing galaxy to be $\sigma=281\pm25~\kms$. The time delays of this system were measured by \citet{BonvinEtal18_PGTD} with $\sim8.5\%$ uncertainties.
    \item {\bf RXJ1131$-$1231:} The \rxj\ system (J2000: 11$^{\textrm{h}}$31$^{\textrm{m}}$52$^{\textrm{s}}$, $−$12\degr31\arcmin59\arcsec) is a quadruply-lensed quasar discovered by \citet{SluseEtal03}. The spectroscopic redshifts of the lensing galaxy and the backgound source are at $\zl=0.295$ \citep{SuyuEtal13} and $\zs=0.657$ \citep{SluseEtal07}, respectively. \citet{SuyuEtal13} measured the stellar velocity dispersion of the lensing galaxy to be $\sigma=323\pm20\kms$. The time delays were measured by \citet{TewesEtal13a} with $\sim1.5\%$ uncertainties.
\end{enumerate}

\subsection{Keck Adaptive Optics Imaging}

All three lens systems were observed at K$^\prime$-band with the Near-infrared Camera 2 (NIRC2), sitting behind the AO bench on the Keck II Telescope, as part of the SHARP AO effort (Fassnacht et al., in prep).  The targets were observed either with the narrow camera setup, which provides a roughly 10$\times$10\arcsec\ field of view and a pixel scale of 9.942 milliarcsec (mas), or the wide camera that gives a roughly 40$\times$40\arcsec\ field of view and 39.686~mas pixels.  Details of the observations are provided in Table~\ref{tab:obsdata}.
\begin{table}
    \centering
    \begin{tabular}{lllr}
        \hline
         Lens & Instrument &  Date & $t_{\rm exp} $\\
         \hline
         HE0435$-$1223 & Keck/NIRC2-N & 2010-12-02 & 8100 \\
         ... & Keck/NIRC2-N & 2012-01-01 & 2400 \\
         PG1115+080 & Keck/NIRC2-N & 2012-05-15 & 900 \\
         ... & Keck/NIRC2-N & 2017-04-11 & 900 \\
         RXJ1131$-$1231 & Keck/NIRC2-W & 2012-05-16 & 1800 \\
         ... & Keck/NIRC2-W & 2012-05-18 & 1800 \\
    \end{tabular}
    \caption{Details of the AO observations.}
    \label{tab:obsdata}
\end{table}

The NIRC2 data were reduced using the SHARP python-based pipeline, which performs a flat-field correction, sky subtraction, correction of the optical distortion in the images, and a coadditon of the exposures.  During the distortion correction step, the images are resampled to produce final pixel scales of 10~mas pix$^{-1}$ for the narrow camera and 40~mas pix$^{-1}$ for the wide camera.  The narrow camera pixels oversample the PSF, which has typical FWHM values of 60--90~mas.  Therefore, to improve the modeling efficiency for the narrow camera data, we perform a 2$\times$2 binning of the images produced by the pipeline to obtain images that have a 20~mas pix$^{-1}$ scale. Further details on \rxj, which was observed with the NIRC2 wide camera, can be found in \citet{GChenEtal16}.

\subsection{\textit{Hubble Space Telescope} Imaging}

All three lens systems have been observed by \hst\ (GO-9375, PI: Kochanek; GO-9744, PI: Kochanek; GO-12889, PI: Suyu).  The \hst\ imaging of both \rxj\ \citep{SuyuEtal13} and \he\ \citep{WongEtal17} were analyzed in previous work.  Therefore, the inferences from these previous models are combined with those from the new AO models in \sref{sec:lens_modeling}.  In contrast, the \hst\ data for \pg\ have not been modeled using the latest pixelated techniques, although \citet{TreuKoopmans02} combined lensing geometry and velocity dispersion to study the content of the luminous matter and dark matter profiles.  Therefore, we perform a joint modeling procedure on the AO and \hst\ data for \pg.

\subsection{MPIA 2.2~m Imaging \label{mpiadata}}

The contribution of the line-of-sight mass distribution to the lensing requires deep wide-field imaging of the region surrounding the lens system.  For \he\ and \rxj\ we have used {\em HST}/ACS or Subaru SuprimeCam imaging, which have been analyzed as part of our previous work on these systems \citep{SuyuEtal14,RusuEtal17}.  To achieve the requisite combination of depth and area for \pg, we coadded 95 images of the field taken with the Wide Field Imager \citep{BaadeEtal99} mounted at the Cassegrain focus of the MPIA 2.2m telescope.  The camera provides a pixel scale of $0\farcs238$.  The data were obtained through ESO $BB\#R_c/162$ filter.
These are the same data used by COSMOGRAIL to measure the time delay for this system \citep[see Sec. 2.1 of][for details]{BonvinEtal18_PGTD}; each image has been exposed for 330 seconds and covers a field of view 
of $\sim8'\times16'$. 
The data reduction process follows the standard procedure including master bias subtraction, master flat fielding, sky subtraction, fringe pattern removal and finally exposure-to-exposure normalization using \texttt{Sextractor} on field stars prior to coadding the exposures.  The final coadded image has an effective seeing of $0\farcs86$.

\section{Lens models}
\label{sec:modeling_tool}
In this section, we describe the models that we use for fitting the hight resolution imaging data, including the lens mass models in \sref{sec:mass_model}, lens light models in \sref{sec:light_model}, the models for constraining the MST in \sref{subsec:lambda}, and the time-delay prediction models in \sref{sec:Td_model}.
We use {\sc glee}, a strong lens modeling code developed by S.~H.~Suyu and A.~Halkola to model the lens arc, lens light, and lens AGNs simultaneously \citep{SuyuHalkola10,SuyuEtal12a}, and reconstructed the AO PSF by using the PSF correction method developed in \citet{GChenEtal16}.

\subsection{Mass models}
\label{sec:mass_model}
The following three analytical functions are used for modeling the main lens, nearby groups, and nearby galaxies:

\begin{itemize}
  \item \textsf{SPEMD:} many studies have shown that a power-law model provides a good first-order description of the lensing galaxies for galaxy-galaxy lensing \citep[e.g.,][]{KoopmansEtal06,KoopmansEtal09,SuyuEtal09,AugerEtal10,BarnabeEtal11,SonnenfeldEtal13}. Thus, for every lens, we model the mass distribution of the lensing galaxy with a singular power-law elliptical mass distribution \citep[][]{Barkana98}. The main parameters include radial slope ($\gamma'$), Einstein radius ($\theta_{\rm E}$), and the axis ratio of the elliptical isodensity contour ($q$).
  \item \textsf{Composite:} we follow \citet{SuyuEtal14} and test a composite (baryonic + dark matter) model.  The baryonic component is modeled by multiplying  the lens surface brightness distribution by a constant M/L ratio parameter (see \sref{sec:light_model}).  For the dark matter component  we adopt the standard NFW profile \citep{NavarroEtal96,GolseKneib02} with the following parameters:
   halo normalization ($\kappa_{s}$),  halo scale radius ($r_{s}$), and  halo minor-to-major axis ratio ($q$) as well as associated position angle ($\theta_{q}$).
  Note that the ellipticity is implemented in the potential for the dark matter.
  \item \textsf{SIS:} singular isothermal sphere (SIS) models
  are used to describe the nearby group and the individual galaxies inside the group of \pg, the nearby galaxies of \he, and the satellite of \rxj.
\end{itemize}

\subsection{Lens light models}
\label{sec:light_model}
The following two analytical functions are used to model the lens light distribution.
\begin{itemize}
    \item \textsf{2S$\acute{\text{e}}$rsic:} we model the light distribution of the lens galaxy with two concentric elliptical S$\acute{\text{e}}$rsic profiles. For all three lens systems we found that a single S\'{e}rsic profile was insufficient for modeling the light distributions.
    \item \textsf{2Chameleon:} the chameleon profile is the difference of two isothermal profiles. It mimics a S$\acute{\text{e}}$rsic profile and enables computationally efficient lens modeling \citep{DuttonEtal11}. The parametrized Chameleon profile can be found in \citet{SuyuEtal14}.
    We convert the Chameleon light profile to mass with an additional constant M/L ratio parameter when modeling the composite model described in \sref{sec:mass_model}.
\end{itemize}

\subsection{Strategies for mitigating the Mass-sheet Transformation}
\label{subsec:lambda}
The mass-sheet transformation is a known degeneracy in lens modeling, in which one can transform a projected mass distribution, $\kappa(\theta)$, into infinite sets of $\kappa_{\lambda}(\theta)$ via 
\begin{equation}
\label{eq:MST}
    \kappa_{\lambda}(\theta)=(1-\lambda)\kappa(\theta)+\lambda,
\end{equation}
without degrading the fit to the imaging. The corresponding time-delay distance changes via 
\begin{equation}
\label{eq:MST_2}
    D_{\Delta t}=D^{\rm true}_{\Delta t,\lambda}=\frac{\Ddt^{\rm model}}{1-\lambda}.
\end{equation}
This degeneracy can be produced by both the lens environment and an incorrect description of the mass distribution in the lensing galaxy. We discuss the approach for estimating the contribution from the environment in \sref{sec:ENV}, and for ranking the mass models with kinematic information in \sref{sec:kinematics}.

\subsubsection{Mass along the line of sight}
\label{sec:ENV}

If we perfectly know  the true $\kappa(\theta)$, then the role of $\lambda$ in \eref{eq:MST} can be understood by looking at the behaviour far from the lensing galaxy, i.e., as $\theta \rightarrow \infty$.  In this regime, the mass distribution of the lensing galaxy, $\kappa(\theta)$, approaches 0 and, thus,  $\lambda$ can be interpreted as a constant-density mass sheet contributed by the lens environment. It is exactly the first-order form produced by line-of-sight (LOS) structure when its effect is small. Therefore, in our models we identify $\lambda$ with $\kappa_{\textrm{ext}}$, the physical convergence associated with LOS structures that do not affect the kinematics of the strong lens galaxy \citep[e.g.,][]{SuyuEtal10,WongEtal17,BirrerEtal19}: 
\begin{equation}
    \Ddt=\frac{\Ddt^{\textrm{model}}}{1-\kappa_{\textrm{ext}}}.
\end{equation}
In addition, the second-order distortion from the LOS also produces a tidal stretching on the lens images. 

We use the following models to capture these two effects.

\begin{itemize}
    \item \textsf{Shear:} Shear distorts the lensed image shapes and thus it can be detected in the modeling process. We express the lens potential in polar coordinates ($\theta$, $\varphi$) to model the external shear on the imaging plane: 
    \begin{equation}
    \label{eq:arclight0}
    \psi_{\text{ext}}(\theta, \varphi)=\frac{1}{2}\gamma_{\text{ext}}\theta^{2}\cos2(\varphi-\varphi_{\text{ext}}),
    \end{equation}
    where $\gamma_{\text{ext}}$ is the shear strength and $\varphi_{\text{ext}}$ is the shear angle. The shear position angle of $\varphi_{\text{ext}}=0^{\circ}$ corresponds to a shearing along $\theta_{1}$ whereas $\varphi_{\text{ext}}=90^{\circ}$ corresponds to shearing along $\theta_{2}$.\footnote{Our (right-handed) coordinate system $(\theta_{1},\theta_2)$ has $\theta_1$ along the East-West direction and $\theta_2$ along the North-South direction.}
    
    \item \textsf{Millennium Simulation:} 
    Due to the MST, the lens images do not provide direct information on $\kappa_{\textrm{ext}}$. 
    We thus use the results of ray-tracing by \citet{HilbertEtal09} through the Millennium Simulation \citep[][]{SpringelEtal05} to statistically estimate the mass contribution along the line of sight to our lenses. 
    This technique was first employed by \citet{SuyuEtal10}, who took the ratio of observed galaxy number counts in an aperture around a lens to those in a control survey \citep[from][]{FassnachtEtal11}, in order to measure the  local over/under-density of galaxies in the lens fields. They then selected lines of sight of similar over/under-density from the Millennium Simulation, with their corresponding values of the convergence, thus producing a probability distribution for $\kappa_{\rm ext}$: $P(\kappa_{\textrm{ext}}|\bm{d}_\mathrm{ENV})$. \citet{SuyuEtal13} later used, in addition to the number counts, the shear value inferred from lens modeling as an additional constraint on $\kappa_{\textrm{ext}}$. \citet{GreeneEtal13} showed that further constraints,  in the form of weighted number counts, can be derived by incorporating physical quantities relevant to lensing such as the distance of each galaxy to the lens, redshifts, luminosities, and stellar masses. 
    \citet{BirrerEtal19} expressed the technique as an application of Approximate Bayesian Computing and further combined weighted number count constraints from multiple aperture radii. 
    In Sections \ref{subsubsec:kappaPG}, \ref{subsubsec:kappaHE}, and \ref{subsubsec:kappaRXJ} we provide an implementation of this technique for each of our three lenses, customized to the nature of the available environment data. For the main numerical and mathematical details of the implementation, we refer the reader to \citet{RusuEtal17}. 
\end{itemize}

If the mass along the line of sight is large enough such that we cannot ignore the higher order terms (flexion and beyond), we need to model that mass explicitly.
\citet{McCullyEtal14,McCullyEtal17} give a quantitative term, flexion shift ($\Delta_3x$), which estimates the deviations in lensed image positions due to third-order (flexion) terms, and suggest that if $\Delta_3x$ is higher than $10^{-4}\arcsec$ (for the typical galaxy-scale lenses that we are studying here), one should model the perturbers explicitly to avoid biasing $H_{0}$ at the (sub)-percent level. We follow this convention to choose which galaxies are included in our models.
Note that the $\Delta_3x$ threshold is based on mock data that do not include extended arcs.  Therefore, this criterion should already be conservative since real lens imaging with extended arcs provide more constraining power than the point-source imaging that was used to set the threshold. 

\subsubsection{Lens Kinematics}
\label{sec:kinematics}
Conversely, even if we perfectly know the mass along the line of sight, the value of $\lambda$ remains uncertain because we do not know the true $\kappa(\theta)$ distribution. 
As pointed out by \citet{SchneiderSluse13}, when assuming a specific mass profile, one artificially breaks the internal MST \citep[][]{Koopmans04}, a special case of the source-position transformation \citep[SPT;,][]{SchneiderSluse14,WertzEtal18a,WertzEtal18b_pySPT}.
However, recent work based on the Illustris simulation has indicated that this effect may be of less concern for massive galaxies such as those in the H0LiCOW sample \citep{XuEtal16}.
The MST allows many different mass distributions within the Einstein radius of the lens, as long as the integrated $\kappa$ within the Einstein ring is preserved.  However, the stellar velocity dispersion is sensitive to the integrated value of $\kappa$ within the {\em effective} radius of the lensing galaxy, which is often different from the Einstein radius.  Thus, we can use the observed stellar velocity dispersion of the lensing galaxy to rank different mass models \citep{TreuKoopmans02,TreuKoopmans04}. 

In the lens modeling of \rxj, \citet{SuyuEtal14}, 
have shown that by including the velocity dispersion, one can obtain a robust $\Ddt$ when considering both power-law and composite model. 
\citet{Sonnenfeld18} also shows that velocity dispersion is the key to obtaining an unbiased $H_{0}$ measurement. 
Hence, we follow \citet{SuyuEtal14} in adopting the composite model and also incorporating the velocity dispersion into the modeling to mitigate this internal mass-profile degeneracy. 

There are three different components needed to predict the velocity dispersion.
\begin{itemize}
    \item A 3D mass distribution: following  \citet{SuyuEtal10}, one can obtain the 3D lens mass from the lens modeling by assuming spherical symmetry.
    In general, the spherically symmetric 3D mass density of the lens can be expressed as
    \begin{equation}
    \label{eq:3dmass}
    \rho_{\textrm{local}}(r)=\rho_{0}r_{0}^{n}F_{n}(r),
    \end{equation}
    where $\rho_{0}r_{0}^{n}$ and $F_{n}(r)$ are the normalization and the mass density distribution.
    By integrating $\rho_{\textrm{local}}$ within a cylinder with radius given by the Einstein radius, $R_{\textrm{Ein}}$, 
    one obtains
    \begin{equation}
    \label{eq:massinlocal}
    \begin{split}
        M_{\textrm{local}}&=4\pi\int^{\infty}_{0}dz\int^{R_{\textrm{Ein}}}_{0}\rho_{0}r_{0}^{n}F_{n}(\sqrt{s^{2}+z^{2}})sds\\
        &=\rho_{0}r_{0}^{n}M_{\textrm{2D}}(R_{\textrm{Ein}}),
    \end{split}
    \end{equation}
    where $M_{\textrm{2D}}(R_{\textrm{Ein}})$ is the projected mass within $R_{\textrm{Ein}}$.
    The mass contained in $M_{\textrm{local}}$ is
    \begin{equation}
    \label{eq:massinlocaltoMST}
    M_{\textrm{local}}=M_{\textrm{Ein}}-M_{\textrm{ext}}=\pi R_{\textrm{Ein}}^{2}\Sigma_{\textrm{cr}}(1-\kappa_{\textrm{ext}}),
    \end{equation}
    where $M_{\textrm{ext}}$ represents the mass contribution from $\kappa_{\textrm{ext}}$ and 
    \begin{equation}
    \label{eq:crit}
        \Sigma_{\textrm{cr}}=\frac{c^{2}}{4\pi G}\frac{D_{s}}{D_{\ell}D_{\ell s}}
    \end{equation} 
    is the critical surface mass density.
    Combining \eref{eq:massinlocal} with \eref{eq:massinlocaltoMST}, the normalization in \eref{eq:3dmass} can be expressed as
    \begin{equation}
    \label{eq:norm}
    \rho_{0}r_{0}^{n}=\frac{\pi R_{\textrm{Ein}}^{2}\Sigma_{\textrm{cr}}(1-\kappa_{\textrm{ext}})}{M_{\textrm{2D}}(R_{\textrm{Ein}})}.
    \end{equation}
    Substituting this in \eref{eq:3dmass}, we obtain
    \begin{equation}
    \label{eq:localdensity}
    \rho_{\textrm{local}}=\frac{\pi R_{\textrm{Ein}}^{2}\Sigma_{\textrm{cr}}(1-\kappa_{\textrm{ext}})}{M_{\textrm{2D}}(R_{\textrm{Ein}})}F_{n}(r).
    \end{equation}
    Although there is $(1-\kappa_{\textrm{ext}})$ in
    \eref{eq:localdensity}, the normalization of the local mass density distribution remains invariant \citep{YidrimEtal19} as $\Sigma_{\textrm{cr}}$ can be re-expressed as
    \begin{equation}
        \Sigma_{\textrm{cr}}=\frac{c^{2}}{4\pi G}\frac{D_{\Delta t}}{1+z_{\ell}}\frac{1}{D^{2}_{\ell}}=\frac{c^{2}}{4\pi G}\frac{D^{\textrm{model}}_{\Delta t}}{(1+z_{\ell})(1-\kappa_{\textrm{ext}})}\frac{1}{D^{2}_{\ell}},
    \end{equation}
    where $(1-\kappa_{\textrm{ext}})$ term cancels out in \eref{eq:localdensity}.
    \item An anisotropy component:
    we assume the anisotropy component in the form of an anisotropy radius, $r_{\textrm{ani}}$, in the Osipkov-Merritt formulation \citep{Osipkov79,Merritt85}, 
    \begin{equation}
    \beta_{\textrm{ani}}=\frac{r^{2}}{r^{2}_{\textrm{ani}}+r^{2}},
    \end{equation}
    where $r_{\textrm{ani}}=0$ is pure radial orbits and $r_{\textrm{ani}}\rightarrow\infty$ is isotropic with equal radial and tangential velocity dispersions. 
    \item A stellar component: we assume a Hernquist profile \citep{Hernquist90},
    \begin{equation}
    \rho_{\ast}=\frac{I_{0}a}{2\pi r(r+a)^3},
    \end{equation}
    for the power-law model,
    where $I_{0}$ is the normalization term and the scale radius can be related to the effective radius by $a=0.551r_{\textrm{eff}}$. For the composite model, the stellar component is represented by the light profile multiplied by a constant mass-to-light ratio.
\end{itemize}

With the three components mentioned above, we follow \citet{SonnenfeldEtal12} and calculate the three-dimensional radial velocity dispersion %,$\sigma_{\textrm{r}}$, 
by numerically integrating the solutions of the spherical Jeans equation \citep[][]{BinneyTremaine87}
\begin{equation}
    \frac{1}{\rho_{\ast}}\frac{d(\rho_{\ast}\sigma_{\textrm{r}})}{dr}+2\frac{\beta_{\textrm{ani}}\sigma_{\textrm{r}}}{r}=-\frac{GM(r)}{r^{2}},
\end{equation}
given the $\kappa_{\textrm{ext}}$ from \sref{sec:ENV}.
Note that since the LOS velocity dispersion has a degeneracy between its anisotropy and the mass profile \citep{Dejonghe87}, we marginalize the sample of $r_{\textrm{ani}}$ over a uniform distribution $[0.5,5]r_{\textrm{eff}}$.
To compare with the data, we can get the seeing-convolved luminosity-weighted line-of-sight velocity dispersion,
\begin{equation}
    (\sigma^{\textrm{P}})^{2}=\frac{\int_{\mathcal{A}}[I(R)\sigma_{s}^{2}\ast\mathcal{P}]d\mathcal{A}}{\int_{\mathcal{A}}[I(R)\ast\mathcal{P}]d\mathcal{A}},
\end{equation}
where $R$ is the projected radius, $I(R)$ is the light distribution, $\mathcal{P}$ is the PSF 
convolution kernel \citep{MamonEtal05}, and $\mathcal{A}$ is the aperture.
The luminosity-weighted line-of-sight velocity dispersion is given by
\begin{equation}
    I(R)\sigma_{s}^{2}=2\int^{\infty}_{R}(1-\beta_{\textrm{ani}}\frac{R^{2}}{r^{2}})\frac{\rho_{\ast}\sigma^{2}_{\textrm{r}}rdr}{\sqrt{r^{2}-R^{2}}}.
\end{equation}

For a system with significant perturbers at a different redshift from the main lens (e.g., \he), we assume a flat $\Lambda$CDM cosmology with $H_{0}$ uniform in [0, 150]$~\kmsmpc$, $\Omega_{\textrm{m}}=0.3$, and $\Omega_{\textrm{m}}=1-\Omega_{\Lambda}$ to calculate the critical density and rank the models by the predicted velocity dispersion. Note that this assumption does not affect the generality of the conclusion.
For our single lens plane  systems (i.e., \rxj~and \pg), we use the measured velocity dispersion to constrain $D_{s}/D_{\ell s}$ and then combine with the measurement of $\Ddt$ to infer the value of $D_{\ell}$ without assuming any cosmological model \citep{BirrerEtal16,BirrerEtal19}. The further advantage of this method is that $D_{\ell}$ is not affected by $\kappa_{\textrm{ext}}$ \citep{JeeEtal15}.

\subsection{Microlensing time-delay prediction models}
\label{sec:Td_model}
In \sref{sec:basic}, we showed that the time delays between multiple images are due to the geometry and the gravitational potential that the light passes through. \citet{TieKochanek18} introduce a possible new microlensing effect on the time delays that can shift the light curves depending on the structure of the accretion disc in the lensed quasar and the density of the stars in the lensing galaxy. They estimated this effect under the assumption of a lamp-post model for the accretion disc, where a large part of the disc lights up concurrently on light-travel scales that are a significant fraction of the time delays between the lensed images. The observed time delay could thus be affected by differential magnification of the individual images, resulting from microlensing by stars in the lensing galaxy.  However, the lamp-post model is only one choice for how to represent the accretion disc; other accretion disc models for which variability is different from the lamp-post model are possible \citep[e.g.,][]{DexterAgol11}. We follow \citet{GChenEtal18a} and present the $\Ddt$ measurements both with and without the lamp-post assumption. However, we only consider the case without the microlensing effect in our final $H_{0}$ determination since it is not clear at this point which is the proper disc model to use. Note that it also wasn't applied in previous H0LiCOW work to infer the final H0 measurement. 

A more detailed description of this effect and how to estimate the probability distribution of the microlensing time-delay effect (MTDE) can be found in \citet{BonvinEtal18_PGTD}. We briefly summarize the technique here. We generate  magnification maps using GPU-D \citep{Vernardos&Fluke14}, which incorporates a Graphics Processing Unit (GPU) implementation of the inverse ray–shooting technique \citep{KayserEtal86}. 
All magnification maps have dimension of $8192\times8192$ pixels over a scale of $20\langle R_{\textrm{E}}\rangle$, where
\begin{equation}
\langle R_{\textrm{E}} \rangle =\sqrt{\frac{D_{s}D_{\ell s}}{D_{\ell}}\frac{4G\langle M_{\star} \rangle }{c^2}}
\end{equation}
We choose the Salpeter initial mass function with mean mass $\langle M_{\star} \rangle = 0.3 \Msun$ and the ratio between the upper and lower masses $M_{\rm upper}/M_{\rm lower}=100$ \citep{Kochanek04}.  
We consider a standard thin disk model \citep{ShakuraSunyaev73}. 
Given the disk size of lensed quasar, the average microlensing time delay at each position on a magnification map can be derived using Equation 10 of \citet{TieKochanek18}. 
The parameters which are used to estimate the probability distribution of the microlensing time-delay for each system are listed in \tref{table:mltd_param} and \tref{table:mltd_source}.

\begin{table}
\centering
\caption{Lensing parameters for creating the magnification maps of microlensing time-delay effect. The values of $\kappa$,$\gamma$, and $\kappa_*/\kappa$ of \he\ and\pg\ are from our lens models}
\label{table:mltd_param}
 \begin{tabular}{||c c c c c||} 
 \hline
 Name & img & $\kappa$ & $\gamma$ & $\kappa_*/\kappa$ \\
 \hline\hline
 \he & A & 0.473 & 0.358 & 0.347 \\
 (This work) & B & 0.630 & 0.540 & 0.361 \\
     & C & 0.494 & 0.327 & 0.334 \\
     & D & 0.686 & 0.575 & 0.380 \\
 \hline
 \pg & A1 & 0.424 & 0.491 & 0.259 \\
 (This work) & A2 & 0.451 & 0.626 & 0.263 \\
     & B  & 0.502 & 0.811 & 0.331 \\
     & C  & 0.356 & 0.315 & 0.203 \\
 \hline
 \rxj & A & 0.526 & 0.410 & 0.429 \\ 
(This work) & B & 0.459 & 0.412 & 0.434 \\ 
      & C & 0.487 & 0.306 & 0.414 \\ 
      & D & 0.894 & 0.807 & 0.581 \\ 
 \hline
 \end{tabular}
\end{table}

\begin{table*}
\caption{Quasar source parameters: the black hole mass ($M_{\rm bh}$), the Eddington ratio ($L/L_{\rm E}$), the accretion efficiency ($\eta$), the inclination angle (incl.), and the references.}
\label{table:mltd_source}
\begin{tabular}{||c c c c c c||} 
 \hline
 Name & $M_{\rm bh}$ [$\Msun$] & $L/L_{\rm E}$ & $\eta$ & incl.[deg] & reference \\ 
 \hline\hline
 \he &  $5.75\times10^8$ & 0.1 & 0.1 & 0 & \cite{SluseEtal12}\\
 \hline
 \pg & $1.2\times10^9$ & 0.1 & 0.1 & 0 & \cite{MorganEtal10}\\
 \hline
 \rxj & $1.3\times10^8$ & 0.1 & 0.1 & 0 & \cite{DaiEtal10}\\ 
 \hline
 \end{tabular}
\end{table*}

To fold this effect into time-delay modeling, we use \begin{equation}
\label{eq:TDsum}
\Delta t_{ij}=(\Ddt/c)\Delta\tau_{ij}+(t_{i}-t_{j}),
\end{equation}
where the first term on the right hand side is the same as in \eref{eq:theory} and $t_{i}-t_{j}$ is the extra delay caused by the MTDE between images $i$ and $j$ \citep[see details in][]{GChenEtal18a}. 

\section{lens modeling}
\label{sec:lens_modeling}
Both \he\ and \rxj\ have been extensively modeled using the extended lensed emission seen in high-resolution HST imaging of the systems \citep{WongEtal17,SuyuEtal14}, but our modeling techniques have not yet been applied to \pg.  Therefore, in this section we begin with a description of the modeling of \pg\ in \sref{subsec:PGmodeling}, and then describe \he~in \sref{subsec:HEmodeling}, and \rxj~in \sref{subsec:RXJmodeling}.

For \pg\ we model the HST and AO imaging simultaneously.  However, for \he~and \rxj, we only model the AO imaging since the HST imaging has already been modeled \citep{WongEtal17,SuyuEtal14}, and then combine the two modeling outputs to obtain a joint inference on $H_0$ (see \sref{subsec:cosmoAOHST}). 

Our analyses of \pg~and \he\ are blind, as in \citet{SuyuEtal13} and \citet{RusuEtal19_H0LiCOW}, in order to avoid confirmation bias. 
That is, the values of $\Ddt$, $D_\ell$ (if computed), and $H_{0}$ were kept blind until all coauthors came to a consensus to reveal the values during a collaboration telecon on June 5th. 
The analysis was frozen after we unblinded the results and no changes were made to any of the numerical results. The time between unblinding and submission was used to polish the text and figures of the manuscript, and carrying out the detailed comparison of the AO and HST based analysis. 

In contrast, the \rxj\ analysis was not done blindly as the AO data for this system were used to develop the PSF-reconstruction technique. On top of the power-law model we have done in \citet{GChenEtal16}, We further test the composite model and use both models to infer $\Ddt$ and $D_\ell$.

To better control the systematics due to the choice of lens modeling technique, we run Markov chain Monte Carlo (MCMC) sampling with different source resolutions in each model. This approach was used because \citet{SuyuEtal13} have shown that the effects of the pixelated-source grid resolution dominate the uncertainty on the lens modeling when using a modeling code, such as {\tt GLEE}, that implements the pixelated-source reconstruction technique.

\begin{figure*}
\centering
\includegraphics[width=\linewidth]{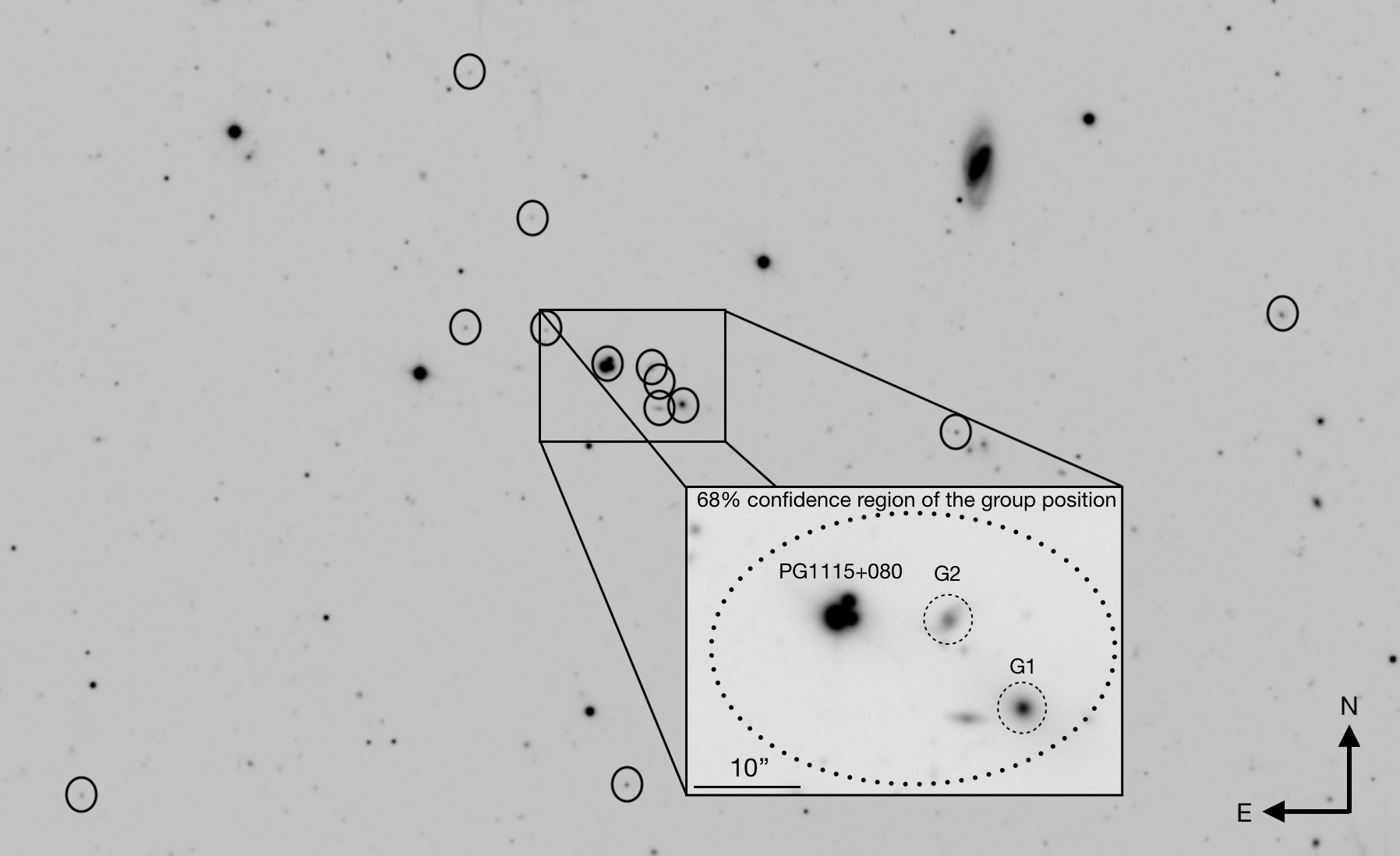}
\caption{
2.2m MPIA telescope image of the FOV around \pg . We show the galaxies/group which we explicitly model here. As \pg~is embedded in a nearby group that consists of 13 galaxies labeled with solid circles, we model not only the main lens but also the group explicitly. The dotted circle represents $1-\sigma$ uncertainty of the priors of the group position ($\Delta \textrm{RA}=23.4\arcsec$ and $\Delta \textrm{DEC}=15.84\arcsec$) measured in \citet{WilsonEtal16}. Since G1 and G2 have the first two largest value of $\Delta_3x$, we model either G1 or both G1 and G2 explicitly in addition to the main lens. We label G1 and G2 with the dashed circles.}
\label{fig:PG1115_envir}
\end{figure*}

\subsection{PG1115+080 Modeling}
\label{subsec:PGmodeling}
\pg~is a single-plane lens system embedded in a nearby group which consists of 13 known galaxies (see the solid circles in \fref{fig:PG1115_envir}). If we model the lens without including the group, the mass profile shows a very steep slope ($\gamma\sim2.35$; note that $\gamma=2$ corresponds to the isothermal profile), which has also been found in previous studies of this system \citep[e.g.,][]{KeetonKochanek97_PG,TreuKoopmans02}, and a strong shear ($\gamma_{
\textrm{ext}}\sim0.15$) which comes from the nearby group. 
\citet{WilsonEtal16} showed that, compared with the other 11 groups along the light of sight, the nearby group contributes the largest convergence at the lens position. 
Furthermore, \cite{McCullyEtal17} indicates that the group produces a significant flexion shift. 
Thus, it is crucial to model not only the main lens but also the group explicitly if we want to obtain an unbiased $H_{0}$ measurement.

\subsubsection{The PSF of PG1115}
\label{subsub:psf}
For the HST imaging, we use {\sc Tinytim} \citep{KristHook97} to generate the PSFs with different spectral index, $\alpha$, of a power-law from -0.4 to -2.5 and different focuses\footnote{The flux per unit frequency interval is F$\nu=C\nu^{\alpha}$, where $\alpha$ is the power-law index and C is a constant; focus is related to the breathing of the second mirror, which is between $0\sim10$.} from 0 to 10. 
We find that the best fit is the PSF with focus equal to 0 and spectral index equal to -1.6. We use this {\sc Tinytim} PSF as the initial guess and then apply the PSF-correction method while modeling the HST imaging.
For the AO imaging, we follow the criteria described in Section 4.4.3 in \citep{GChenEtal16} and perform 8 iterative steps to create the final PSF and make sure the size of the PSF for convolution is large enough so that the results is stable.
The full width half maximum (FWHM) of the AO PSF is 0.07\arcsec, while the FWHM of the HST PSF is 0.15\arcsec. We show the reconstructed AO PSF in \fref{fig:AO_PSF}.

\subsubsection{Main Lens}

We follow two approaches to modeling the mass distribution in the main lensing galaxy.
\begin{itemize}
    \item SPEMD+2S$\acute{\text{e}}$rsic+shear: We first choose the SPEMD density profile to model the extended arc and reconstruct the source structure on a pixelated grid \citep{SuyuEtal06}. 
    We found that a single S$\acute{\text{e}}$rsic profile is not sufficient to describe the light distribution, so we model it with two concentric elliptical S$\acute{\text{e}}$rsic profiles with free relative position angles and ellipticities.
    By comparing the mass and light components, we found a similar result to \citet{YooEtal05}, namely that the position of the center of mass is very close to the center of light, with $|\Delta r|\approx0.015''$. This implies that the offset between the projected center of dark matter and baryonic matter is small. 
    \item Composite+2Chameleon+shear: 
    We also model the main lens with composite model. Because the  SPEMD+2S$\acute{\text{e}}$rsic+shear model indicated that the dark-matter and baryonic centroids were consistent, we link the centroid of NFW profile to the centroid of two concentric chameleon profiles. We also follow \citet{WongEtal17} and \citet{RusuEtal19_H0LiCOW} to iteratively update the relative amplitudes of the associated mass components to match those of the light components, as the relative amplitudes of light components can vary when we run the MCMC chains, whilst the relative amplitudes are fixed in the mass profiles.
\end{itemize}

\begin{figure}
\includegraphics[width=\linewidth]{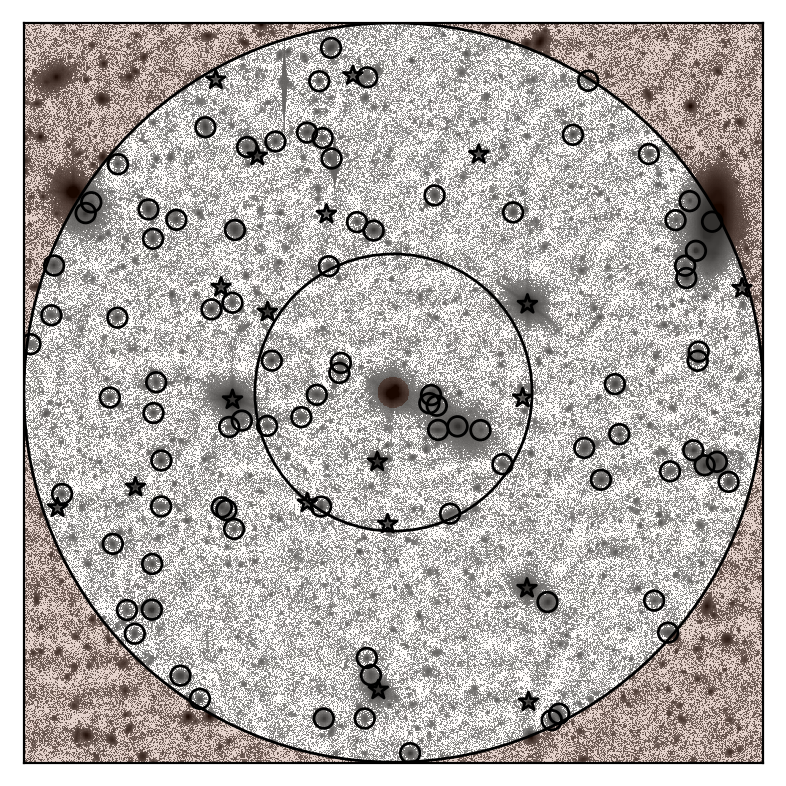}
\caption{$Rc$-band $240\arcsec\times240\arcsec$ image around \pg . This is the same data shown in Figure \ref{fig:PG1115_envir}, but with contrast chosen to better show the detected objects. The large circles mark the $45\arcsec$ and $120\arcsec$ radii apertures centered on the lens. The inner $5\arcsec$ and outer $>120\arcsec$ masked regions are shown in color. For all objects $R\leq23$, small circles mark galaxies and star symbols mark stars.}
\label{fig:fovPG1115}
\end{figure}

\begin{figure}
\includegraphics[width=\linewidth]{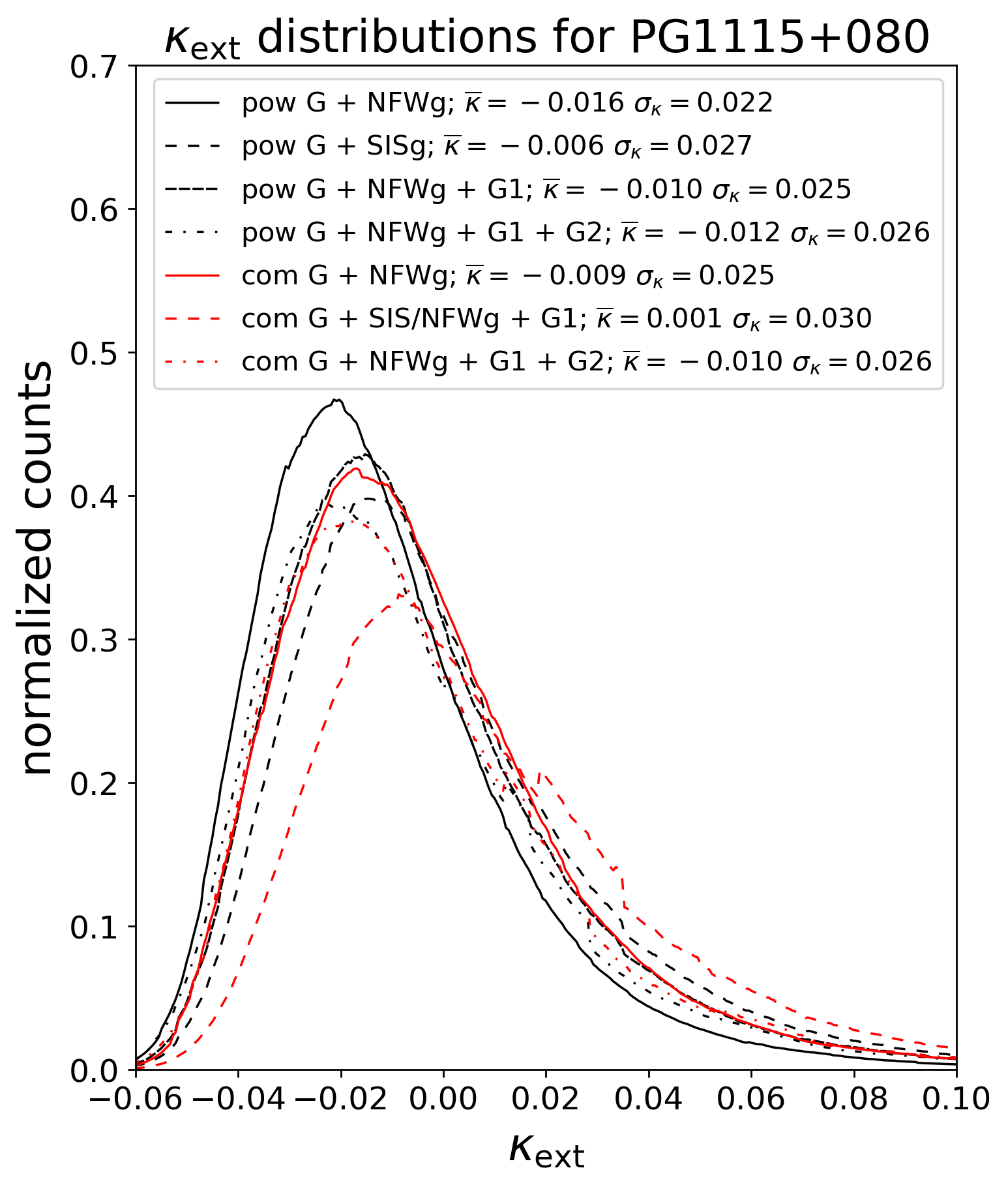}
\caption{Distributions of $\kappa_\mathrm{ext}$ for PG1115+080, for various lens models and their associated shear values. The constraints used to produce the distributions from the Millennium Simulation are the external shear $\gamma$, plus the combination of weighted counts corresponding to galaxy number counts inside the $45\arcsec$ and $120\arcsec$ apertures, as well as number counts weighted by the inverse of the distance of each galaxy to the lens. The numerical constraints are reported in Table~\ref{tab:weights}. The distributions with G1 or G2 marked are calculated by removing those galaxies from the weighted count constraints, since these galaxies are explicitly included in the lens models. The size of the histogram bin is $\Delta\kappa_\mathrm{ext}=0.00055$. As the original distributions are noisy, we plot their convolution with a large smoothing window of size $30\times\Delta\kappa_\mathrm{ext}$. In the legend, ``pow'' refers to the power law model, ``com'' to the composite model, $\overline{\kappa}$ to the median of the distribution, and $\sigma_\kappa$ to the semi-difference of the 84 and 16 percentiles of the distribution.}
\label{fig:kappaPG1115}
\end{figure}

\subsubsection{Nearby Group}
\label{subsubsect:PG1115group}
Based on the velocity dispersion of the nearby group, $\sigma_{\textrm{group}}=390\pm60~\kms$, the inferred group mass is around $10^{13}\sim10^{14}h^{-1}M_{\odot}$ \citep{WilsonEtal16}.  \citet{Oguri06} has shown that in this mass regime the mass profile is too complicated to be described by a either a simple NFW profile or SIS profile, as it is a transition between the two. 
Thus, we use a NFW profile as the fiducial model, but also model the group with a SIS profile as a systematic check. 
In the following, we show how we determine reasonable priors on the NFW and SIS profiles.
\begin{figure*}
\centering
\includegraphics[scale=0.7]{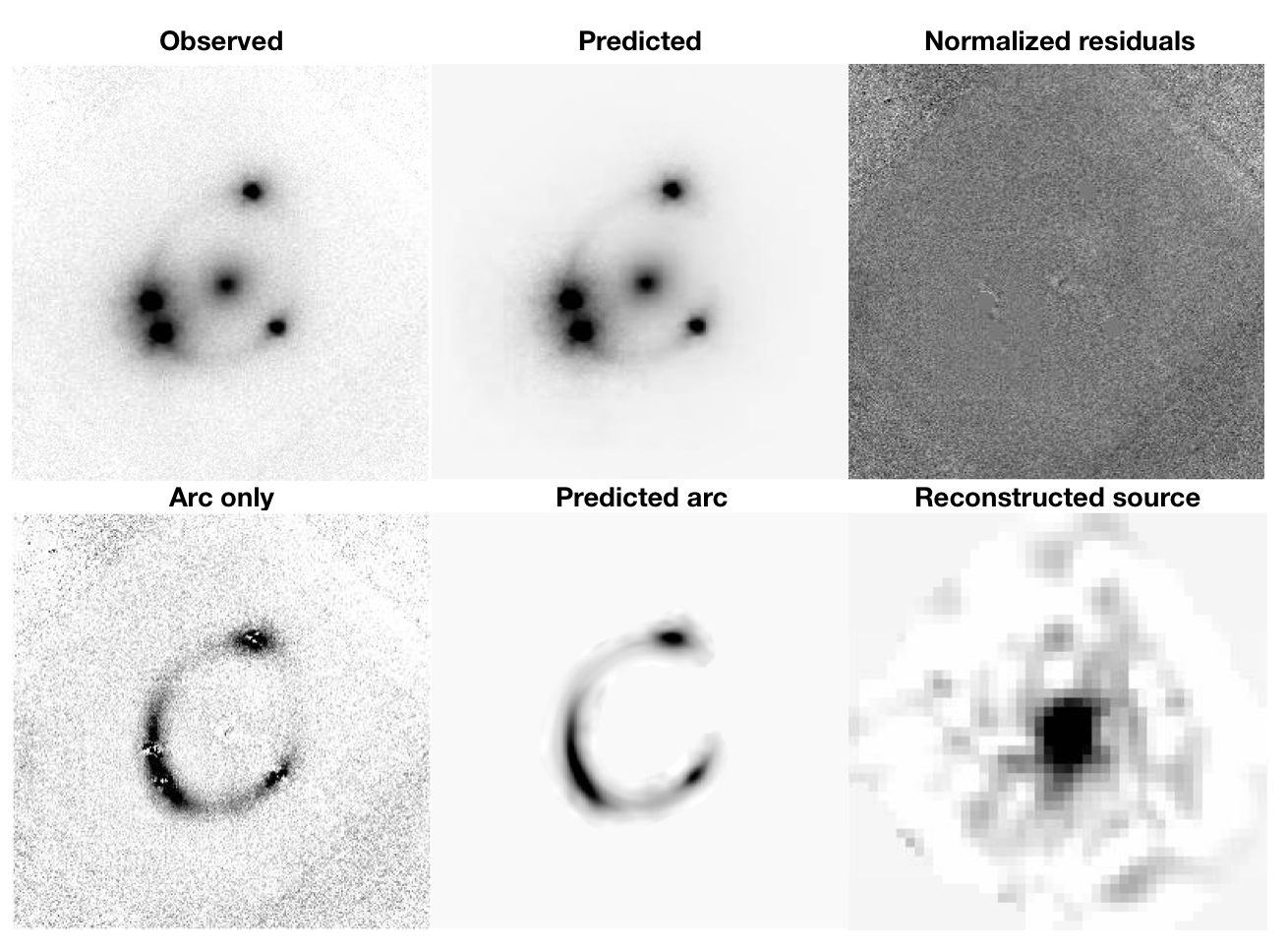}
\caption{\pg~AO image reconstruction of the most probable model with a source grid of $43 \times 43$ pixels and $59 \times 59$ pixels PSF for convolution of spatially extended images. Top left: The \pg~AO image. Top middle: the predicted image of all components including lens light, arc light, and AGN light. Top right: image residuals, normalized by the estimated 1-$\sigma$ uncertainty of each pixel. Bottom left: the arc-only image which removes the lens light and AGN light from the observed image. Bottom middle: predicted lensed image of the background AGN host galaxy. Bottom right: the reconstructed host galaxy of the AGN in the source plane.
}
\label{fig:PG1115_AO_figure}
\end{figure*}

\begin{figure*}
\centering
\includegraphics[width=\linewidth]{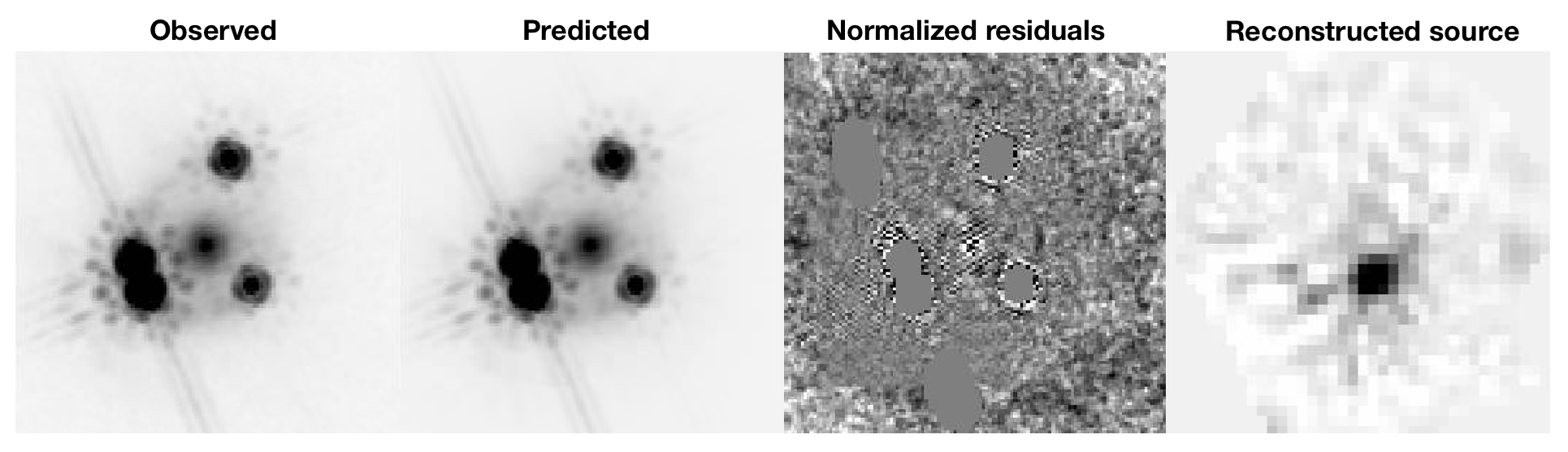}
\caption{\pg~HST image reconstruction of the most probable model with a source grid of $38 \times 38$ pixels. From the left to the right: the \pg~HST image, the predicted image of all components including lens light, arc light, and AGN light, image residuals normalized by the estimated 1-$\sigma$ uncertainty of each pixel, and the reconstructed host galaxy of the AGN in the source plane.}
\label{fig:PG1115_HST_figure}
\end{figure*}

\begin{itemize}
    \item group (NFW): we follow \citet{WongEtal11} and use $M_{\textrm{vir}}-c_{\textrm{vir}}$ relationships based on WMAP5 results in \citet{MacciEtal08} to translate the observed velocity dispersion to scale radius and the normalization of the NFW profile 
    (we compared the priors by assuming WMAP1 and WMAP3 and found that the difference is negligible, so our results are robust to variations in assumed $M_{\textrm{vir}}-c_{\textrm{vir}}$ relation). 
    Here we briefly recap the process. We use the measured velocity dispersion and its uncertainties, assuming it is a Gaussian distribution, to get a probability distribution for the group virial mass, $M_{\textrm{vir}}$. Then we can obtain the concentration, $c_{\textrm{vir}}$, from the $M_{\textrm{vir}}$-$c_{\textrm{vir}}$ relationship assuming a reasonable scatter of 0.14 in $\textrm{log} ~c_{\textrm{vir}}$ \citep{BullockEtal01,WongEtal11}. With the critical density and the characteristic overdensity at the lens redshift \citep{EkeEtal98,EkeEtal01}, we can obtain  $r_{\textrm{vir}}$ and a prior probability on the scale radius, $r_{s}$, via $c_{\textrm{vir}}=r_{\textrm{vir}}/r_{s}$. The prior probability distribution of the normalization can be calculated by combining $r_{s}$, the central density of the halo ($\rho_{0}$), and the critical surface density for lensing ($\Sigma_{\textrm{cr}}$).
    \item group (SIS): we convert the velocity dispersion to an Einstein radius via
    \begin{equation}
    \label{eq:SISVDtoEr}
    \sigma^{2}=\theta_{E}\frac{c^{2}}{4\pi}\frac{D_{s}}{D_{\ell s}},
    \end{equation}
    to get a prior on $\theta_{\rm E}$ of $1.4\arcsec\pm0.2\arcsec$ for G1 and $0.4\arcsec\pm0.4\arcsec$ for G2.
\end{itemize}
For the above two models, we also put a prior on the position of the group (see \fref{fig:PG1115_envir}) based on \citet{WilsonEtal16}.

\subsubsection{Nearby Perturbing Galaxies: G1 and G2}
\label{subsubsec:G1G2}
As some of the galaxies inside the group are close to the main lens, these perturbers could individually affect the main lens beyond the second-order distortion terms. We calculate $\Delta_3x$ of the nearby galaxies using the notation and definition in \citet{McCullyEtal17}. As $\Delta_3x$ is expressed in terms of the Einstein radius of these perturbers, we convert the measured velocity dispersions of G1 and G2 \citep[$250\pm20~\kms$ and $130\pm60~\kms$, respectively, from][]{Tonry98} into corresponding Einstein radii using Equation~\ref{eq:SISVDtoEr}. For the other galaxies lacking a measurement of the velocity dispersion, we assume that they are located at the group redshift (this assumption maximizes the value of $\Delta_3x$), and use their relative luminosities compared to either G1 and G2 (depending on the morphology, since G1 is a spiral), to infer a velocity dispersion from the Faber-Jackson relation \citep{FaberEtal76}. We find $\log\Delta_3x(\textrm{G1})=-3.68^{+0.13}_{-0.14}$ (in units of log(arcseconds)) and $\log\Delta_3x(\textrm{G2})=-4.01^{+0.75}_{-1.07}$, whereas the remaining galaxies have $\log\Delta_3x < -4$, and we therefore neglect them.
Thus, to test for systematic effects, we model the group either as a single group profile or as a group halo plus  either G1, or both G1 and G2, where the galaxies are modeled as SIS mass distributions.

\begin{figure*}
\centering
\includegraphics[width=\linewidth]{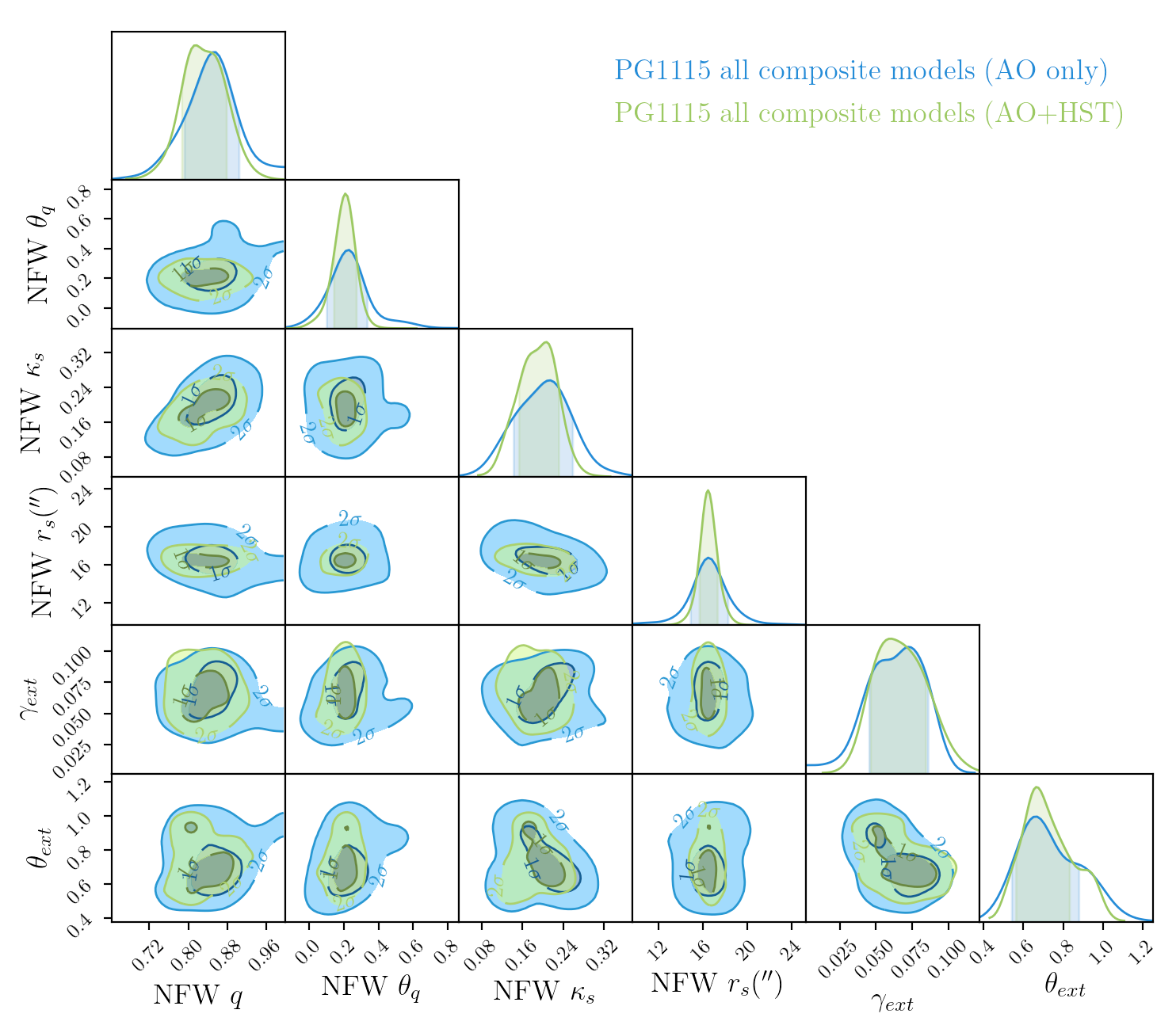}
\caption{Marginalized parameter distributions from the \pg\ composite lens model results. We show the comparison between using only AO imaging data and both AO and HST imaging data. The contours represent the $68.3\%$ and $95.4\%$ quantiles.}
\label{fig:PG_composite}
\end{figure*}

\begin{figure*}
\centering
\includegraphics[width=\linewidth]{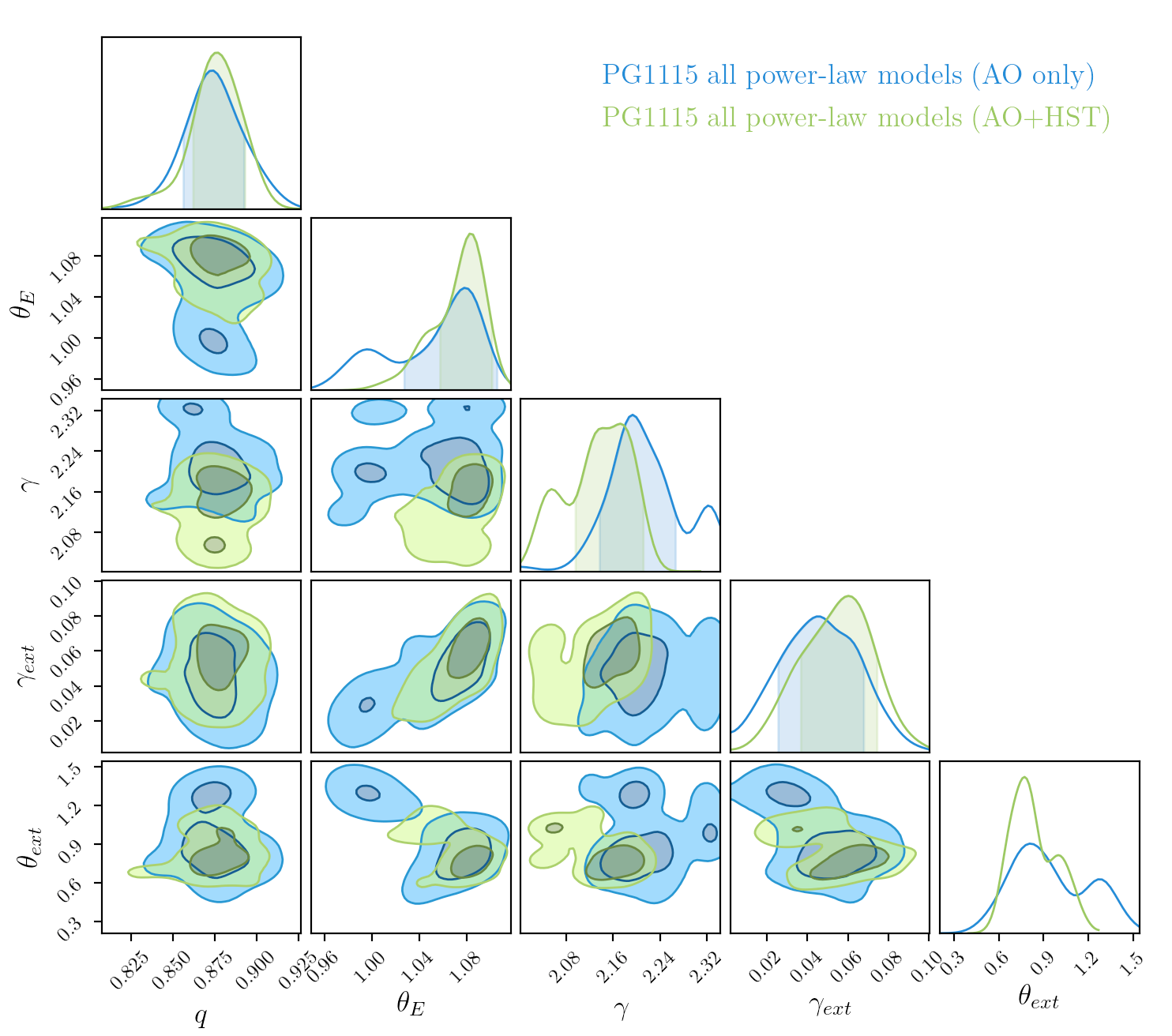}
\caption{Marginalized parameter distributions from the \pg\ power-law lens model results. We show the comparison between using only AO imaging data and both AO and HST imaging data. The contours represent the $68.3\%$ and $95.4\%$ quantiles.}
\label{fig:PG_pl}
\end{figure*}

\subsubsection{LOS Analysis and the External Convergence}
\label{subsubsec:kappaPG}

The technique of inferring $P(\kappa_{\textrm{ext}}|\bm{d}_\mathrm{ENV})$, based on the Millennium Simulation and observed weighted galaxy number counts, was briefly described in \sref{sec:ENV}. 
As implemented by \citet{RusuEtal17}, it requires wide-field, broad-band images to compute photometric redshifts and other physical properties of the galaxies surrounding the lens. 
The deepest multi-band images currently available are provided by the Sloan Digital Sky Survey \citep[SDSS;][]{AdelmanEtal08} and the Panoramic Survey Telescope and Rapid Response System \citep[][]{ChambersEtal16}, but these are still relatively shallow, which may result in a biased $\kappa_{\textrm{ext}}$ \citep{CollettEtal13}.  For our $\kappa_{\rm ext}$ analysis for \pg\ we therefore use the deep coadded data set from the MPIA 2.2~m Telescope described in \S\ref{mpiadata}.

The coadded image has a limiting magnitude of $25.36\pm0.08$, deeper than the control survey, CFHTLenS ($r=24.88\pm0.16$). We perform source detection in this image using \texttt{Sextractor} \citep{BertinArnouts96}. 
For a fair comparison with the control survey, we need to convert our $R_c$ magnitudes to $r$-band magnitudes. 
However, as we only have a single band, we cannot compute color terms. 
Fortunately, a cross-match of the detections in our field with the ones in SDSS\footnote{We ignore the negligible differences between the SDSS and CFHTLenS $r$-band filters.} shows that, after correcting for the zero-point offset, the scatter is small, with an rms of $\sim0.10$ mag, which we add to the photometric error budget. 
However, we choose a brighter magnitude limit $r\leq 23$ mag, in order to be able to use the purely morphological galaxy-star classification of CFHTLenS \citep{HildebrandtEtal07}, where objects down to $i<23$ mag are classified based solely on the \texttt{FLUX\_RADIUS} parameter measured by \texttt{Sextractor}, and because, roughly, $r-i\sim 0.5$ for our cross-matches with SDSS. 
Due to the good seeing of our data, we find a clear stellar locus which allows us to determine a good classification threshold for \texttt{FLUX\_RADIUS}, using the methodology in \citet{CouponEtal09}. 
We show the $240\arcsec\times240\arcsec$ cutout of the field of view (FOV) in Figure~\ref{fig:fovPG1115}, marking the sources detected down to our magnitude limit. 

\begin{figure*}
\includegraphics[width=\linewidth]{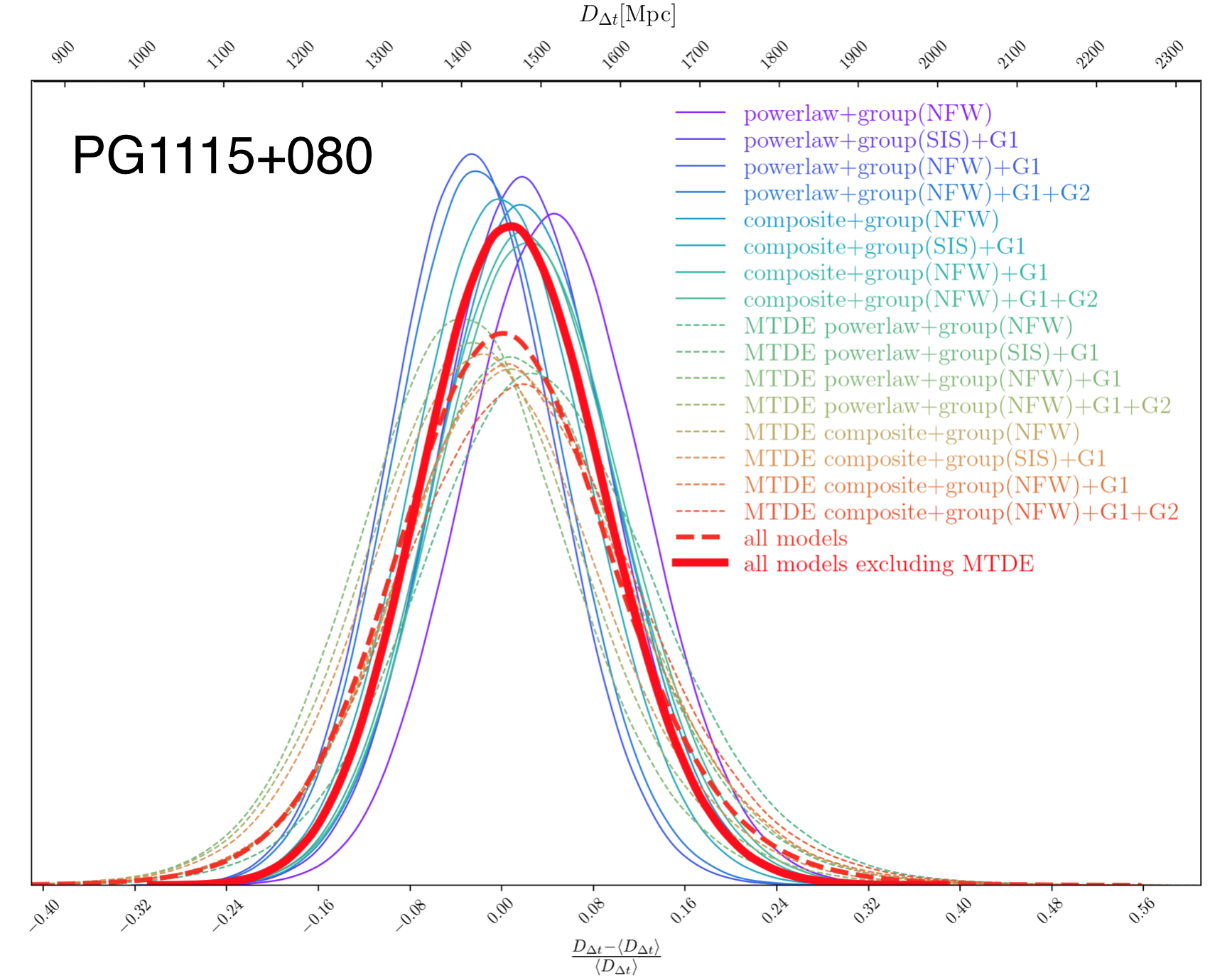}
\caption{The $\Ddt$ of different model choices from \pg. We explore 160 model choices in total with all different combination of choices among two kinds of main lens models, various mass models for the group, five different resolutions of the reconstructed source, and three different priors of the accretion disk sizes (or no MTDE). The solid lines are the cases without including the MTDE, while the dashed lines are the cases with including MTDE. Each dashed line has marginalized three different kind of accretion disk sizes. The last chain, which excludes the MTDE cases, is used to infer the value of $H_{0}$.}
\label{fig:PG1115_Ddt}
\end{figure*}

\begin{figure*}
\includegraphics[width=\linewidth]{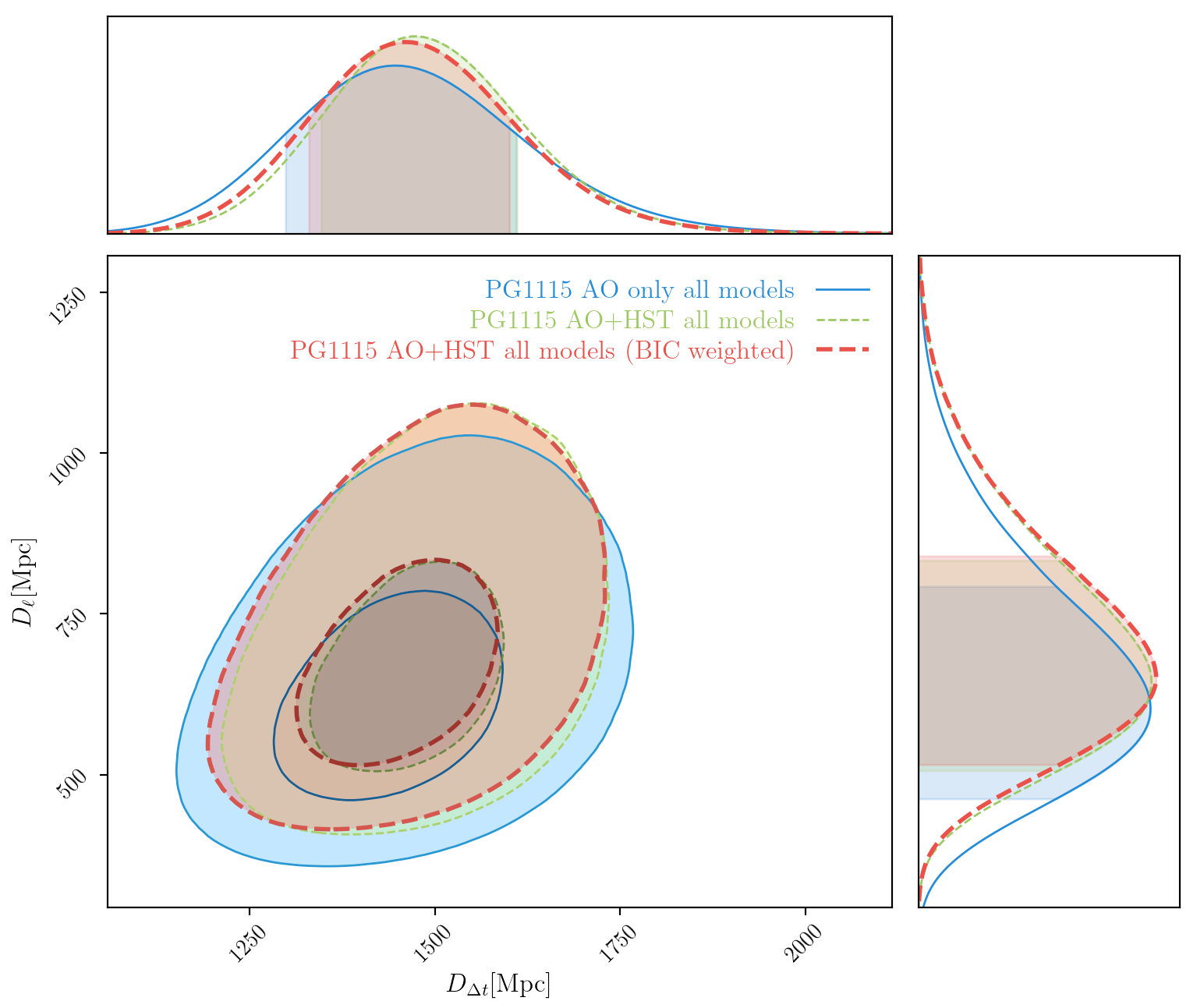}
\caption{The unblinded $\Ddt$ and $D_{\ell}$ of \pg. We plot the results with only using AO imaging, using both AO and HST imaging, and the BIC weighted results.}
\label{fig:PG1115_Ddt_Dd}
\end{figure*}

We compute relative weighted galaxy number counts in terms of simple counts (a weight of unity) as well as using as weight the inverse of the distance between each galaxy in the field and the lens (weighting by $1/r$). 
We do this inside both the $45\arcsec$- and $120\arcsec$-radius apertures, using the technique from \citet{RusuEtal17} and the galaxy catalogue produced above, with the exception that we use $r$-band magnitudes for both the lens field and the CFHTLenS fields, whereas \citet{RusuEtal17} used $i$-band magnitudes. 
When doing this, we ignore the galaxies confirmed as part of the galaxy group, as the group is explicitly incorporated in our lensing models, and we need to compute $\kappa_{\textrm{ext}}$ without its contribution. 
In addition, we account for the galaxies that are expected to be part of the group, but are missed due to the spectroscopic incompleteness, as described in Appendix~\ref{missinggroup_PG1115}. 
We report our results in Table~\ref{tab:weights}. 
These numbers are mostly consistent with the unit value, indicating that, after removing the contribution of the galaxy group, the field around the lens is of average density. 
Finally, we compute $P(\kappa_{\textrm{ext}}|\bm{d}_\mathrm{ENV},\gamma)$ following the technique presented in \citet{BirrerEtal19}, which combines the constraints from both apertures.
The combination of apertures results in a tighter distribution, as shown by \citet{RusuEtal19_H0LiCOW}. 
We show the resulting distributions, corresponding to the various tests of systematics from \sref{subsubsec:pg_system}, in Figure~\ref{fig:kappaPG1115} and the summary table in \aref{appendix:kappaconstraints}.

\begin{figure*}
\centering
\includegraphics[scale=0.6]{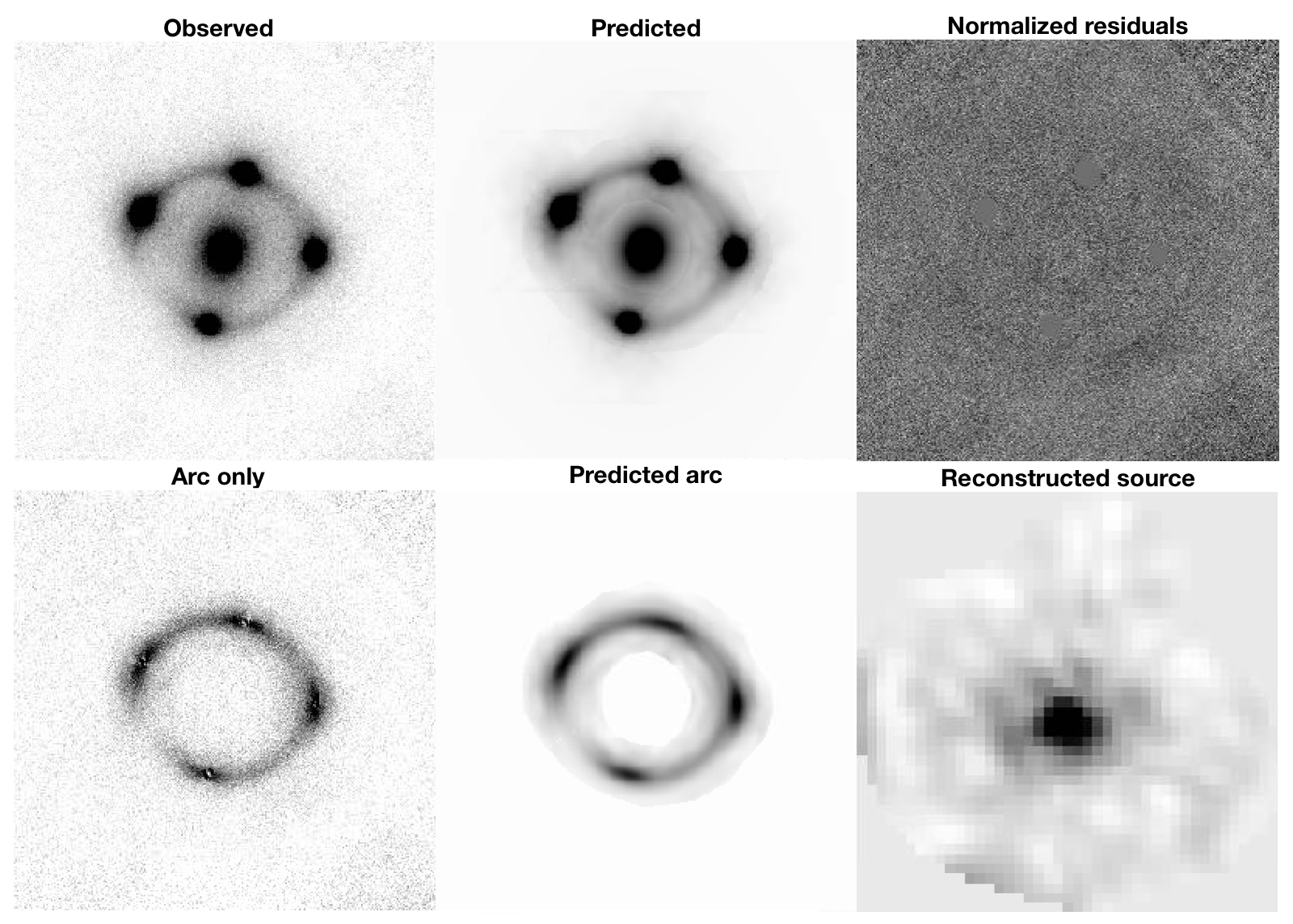}
\caption{\he~AO image reconstruction of the most probable model with a source grid of $50 \times 50$ pixels and $69 \times 69$ pixels PSF for convolution of spatially extended images. Top left: \he~AO image. Top middle: predicted lensed image of the background AGN host galaxy. Top right: predicted light of the lensed AGNs and the lens galaxies. Bottom left: the arc-only image which removes the lens light and AGN light from the observed image. Bottom middle: predicted lensed image of the background AGN host galaxy. Bottom right: the reconstructed host galaxy of the AGN in the source plane.}
\label{fig:HE0435_figure}
\end{figure*}

\begin{figure}
 \includegraphics[width=\linewidth]{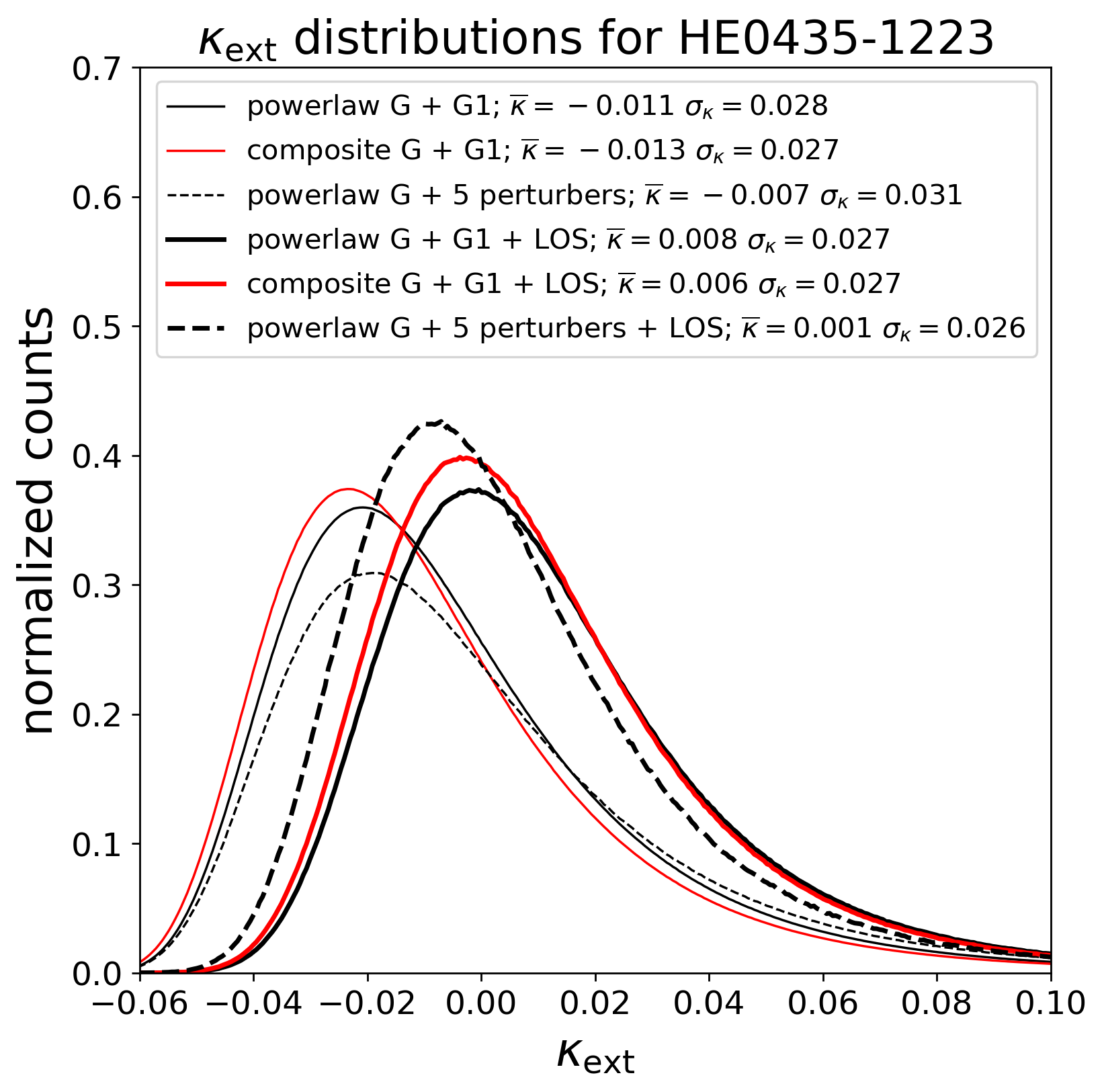}
\caption{Distributions of $\kappa_\mathrm{ext}$ for HE0435-1223, for three different lens models and their associated shear values. The constraints used to produce the distributions from the Millennium Simulation are either only the external shear $\gamma$, or the shear plus the combination of weighted counts corresponding to galaxy number counts inside the $45\arcsec$ aperture, as well as number counts weighted by the inverse of the distance of each galaxy to the lens (the ``+LOS'' models). The numerical constraints are reported in Table~\ref{tab:weights}. For the \texttt{powerlaw + G1} and \texttt{composite + G1} models we used an inner mask of $5\arcsec$ around the lens, and for the \texttt{powerlaw + 5 perturbers} model we used a mask of $12\arcsec$ radius. The five perturbers are indicated in Figure 3 from \citet{WongEtal17}. Three additional galaxies enter the 12\arcsec-radius inner mask, and we slightly boosted their distance from the lens, in order to avoid masking them. See caption of Figure~\ref{fig:kappaPG1115} for additional details.
}
\label{fig:kappaHE0435}
\end{figure}

\begin{figure*}
\centering
\includegraphics[width=\linewidth]{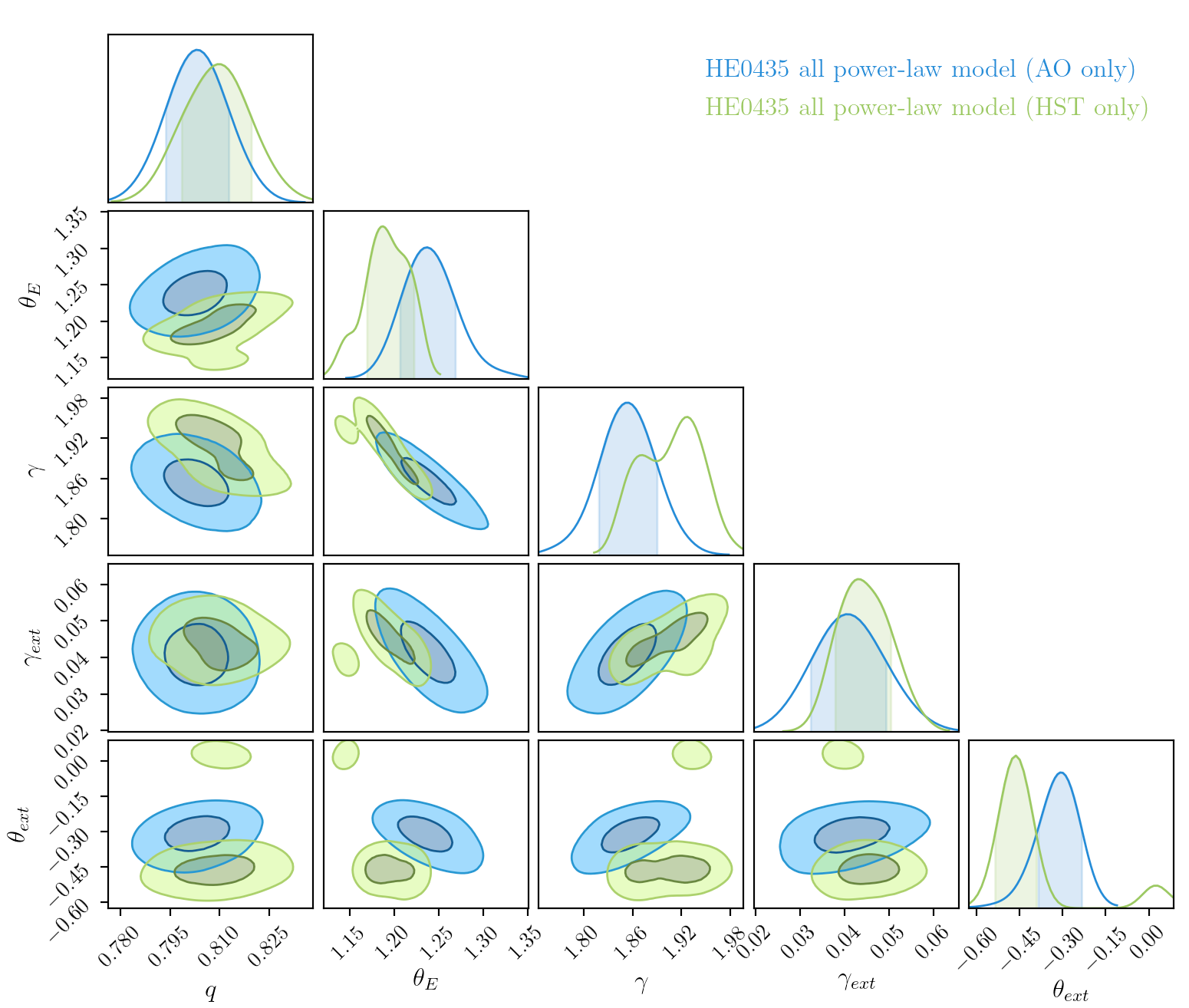}
\caption{Marginalized parameter distributions from the power-law lens model results for \he. We show the comparison between using only AO imaging data and using only HST imaging data. The contours represent the $68.3\%$ and $95.4\%$ quantiles.}
\label{fig:he_pl}
\end{figure*}

\begin{figure*}
\centering
\includegraphics[width=\linewidth]{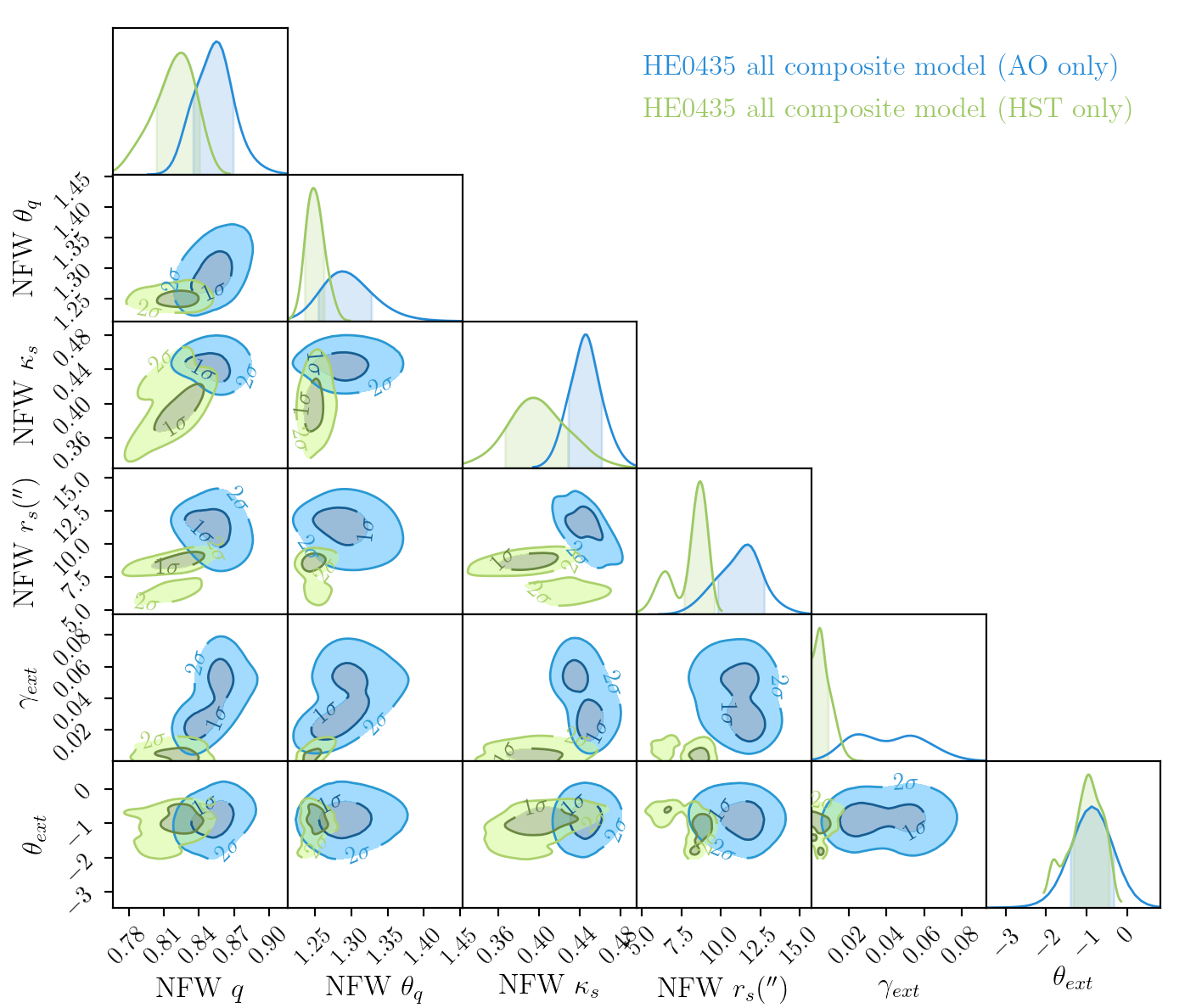}
\caption{Marginalized parameter distributions from the composite lens model results for \he. We show the comparison between using only AO imaging data and using only HST imaging data. The contours represent the $68.3\%$ and $95.4\%$ quantiles.}
\label{fig:he_comp}
\end{figure*}

\begin{figure*}
\centering
\includegraphics[width=\linewidth]{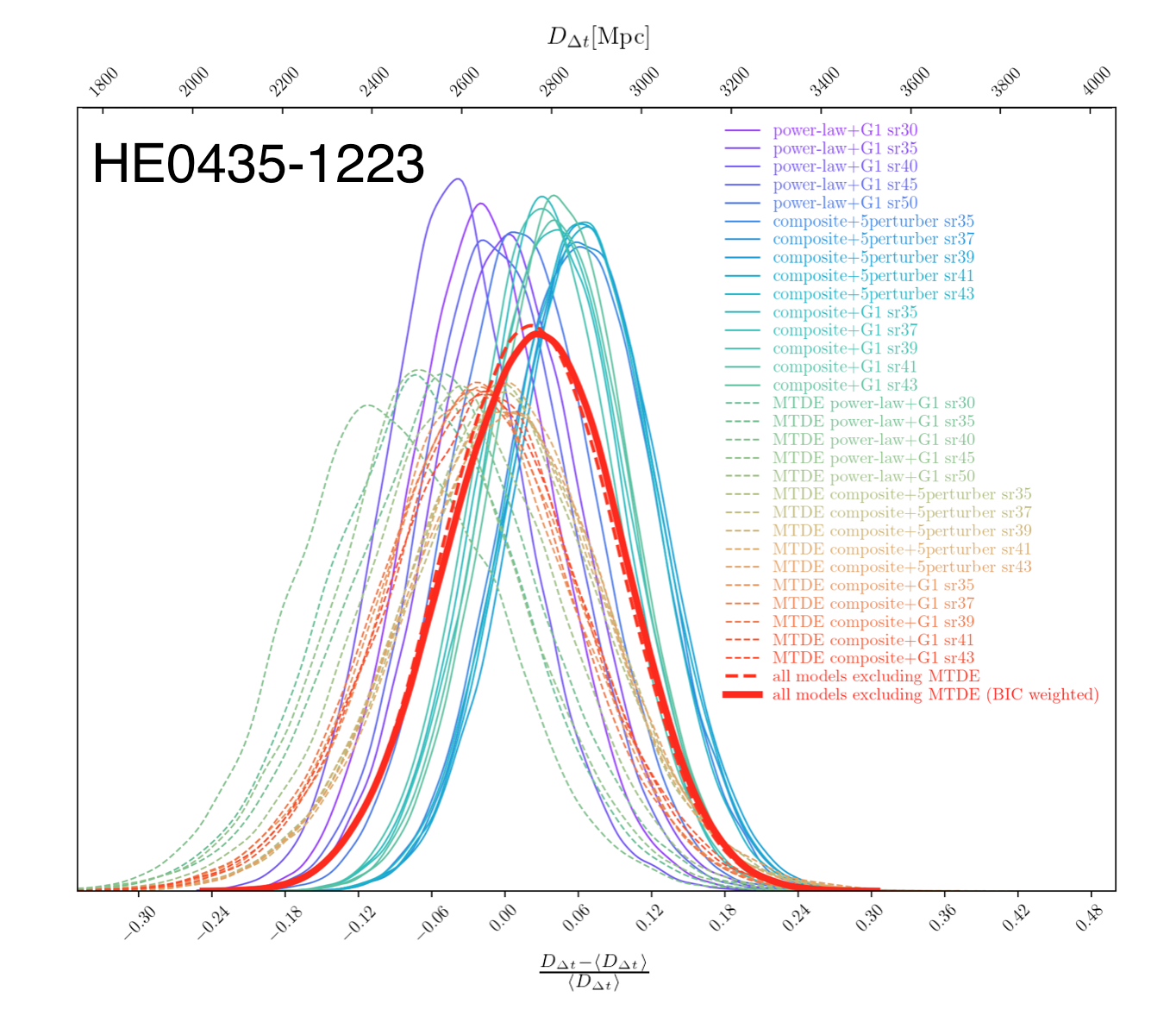}
\caption{The various model choices of \he\ with and without considering MTDE.}
\label{fig:HE0435_Ddt}
\end{figure*}

\begin{figure*}
\centering
\includegraphics[width=\linewidth]{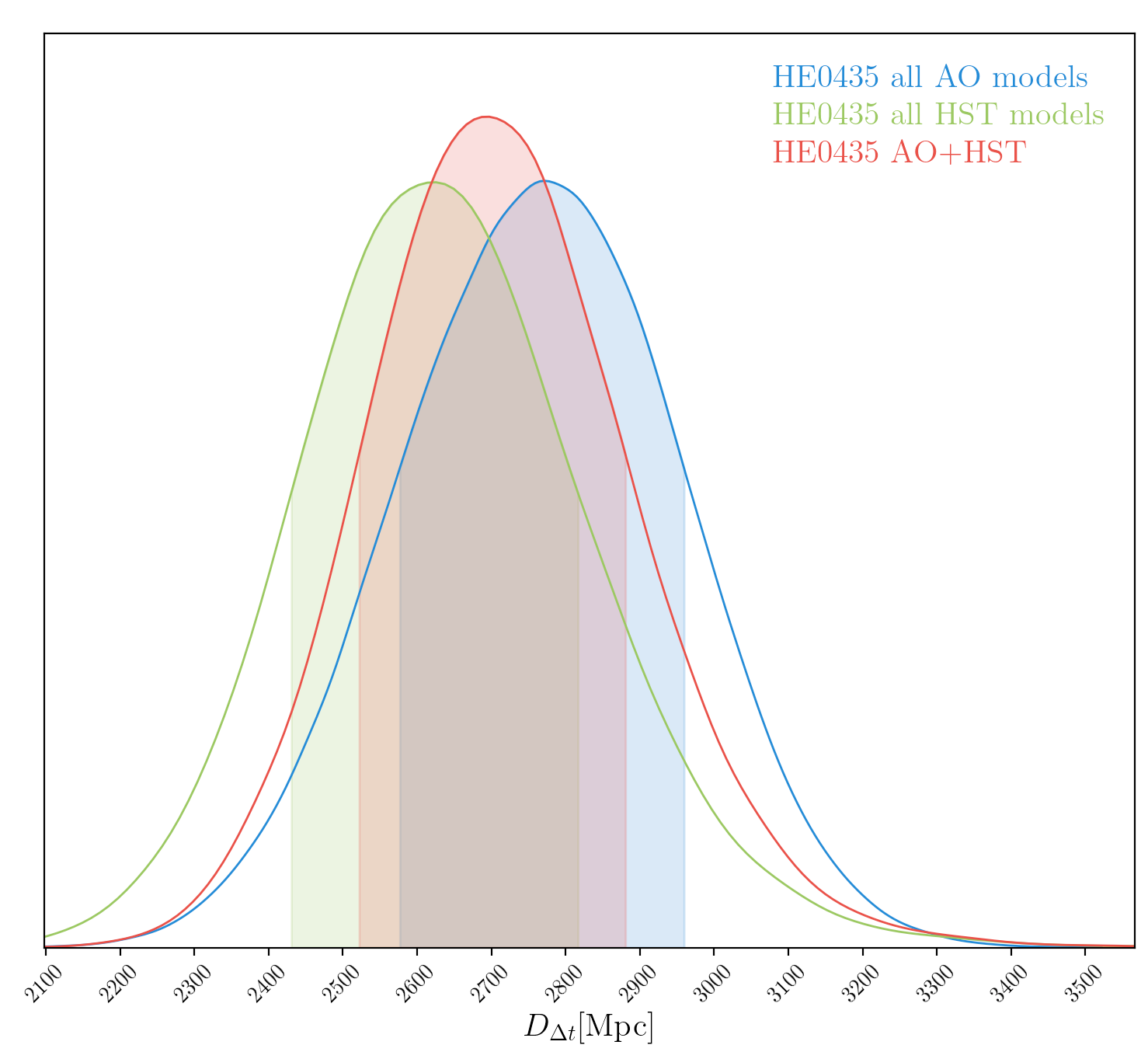}
\caption{Constraints on $\Ddt$ from \he\ when using AO imaging data only, HST imaging data only, or AO+HST.  Because of the significant multiplane lensing needed for this system, we cannot compute a $D_\ell$.}
\label{fig:HE0435_Ddt_compare}
\end{figure*}

\begin{figure}
\centering
\includegraphics[width=\linewidth]{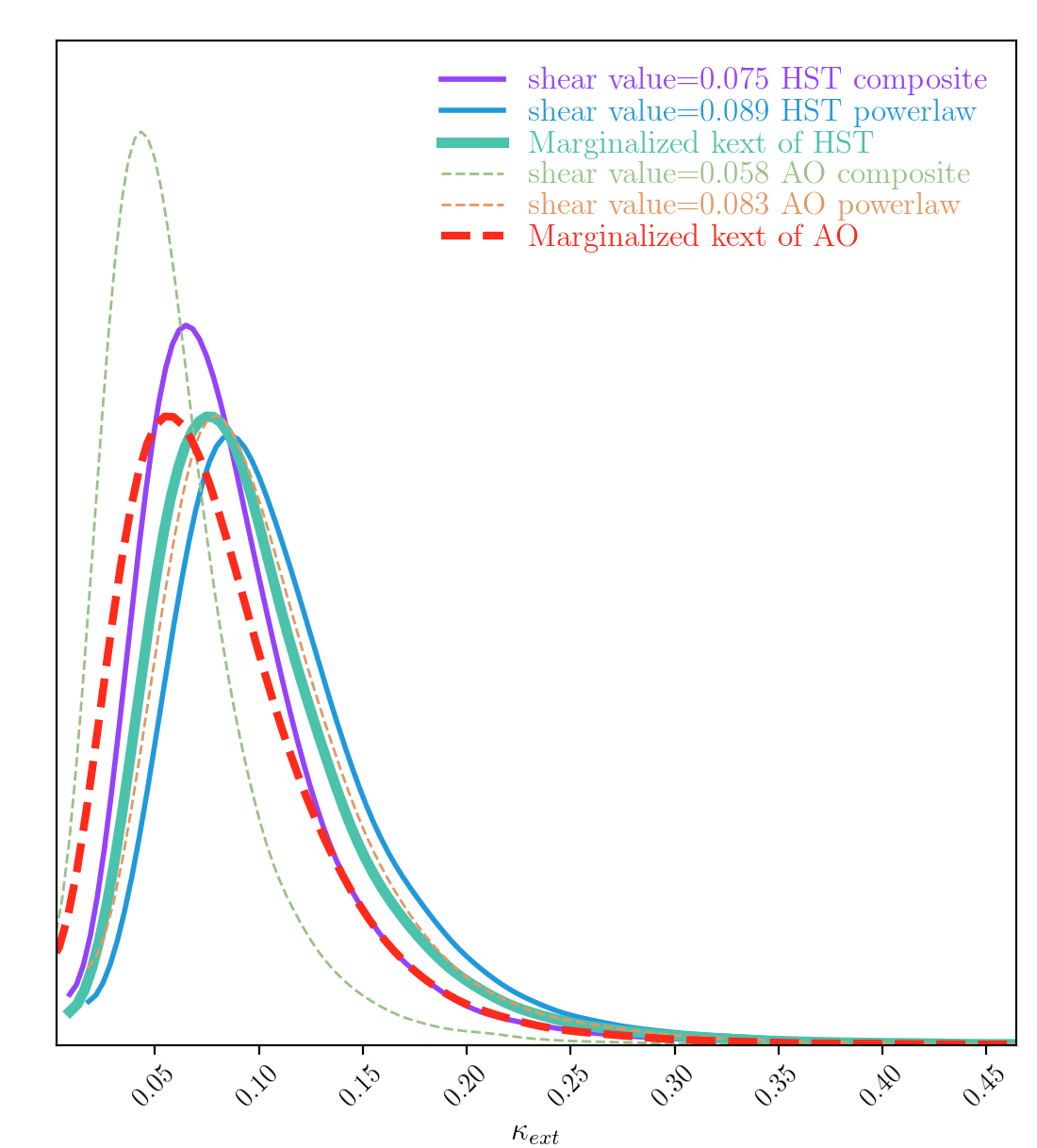}
\caption{The comparison of the  $\kappa_{\textrm{ext}}$ for \rxj, based on the number counts and the shear values inferred from the AO imaging and HST imaging.}
\label{fig:compare_kappa_rxj}
\end{figure}

\begin{figure*}
\centering
\includegraphics[width=\linewidth]{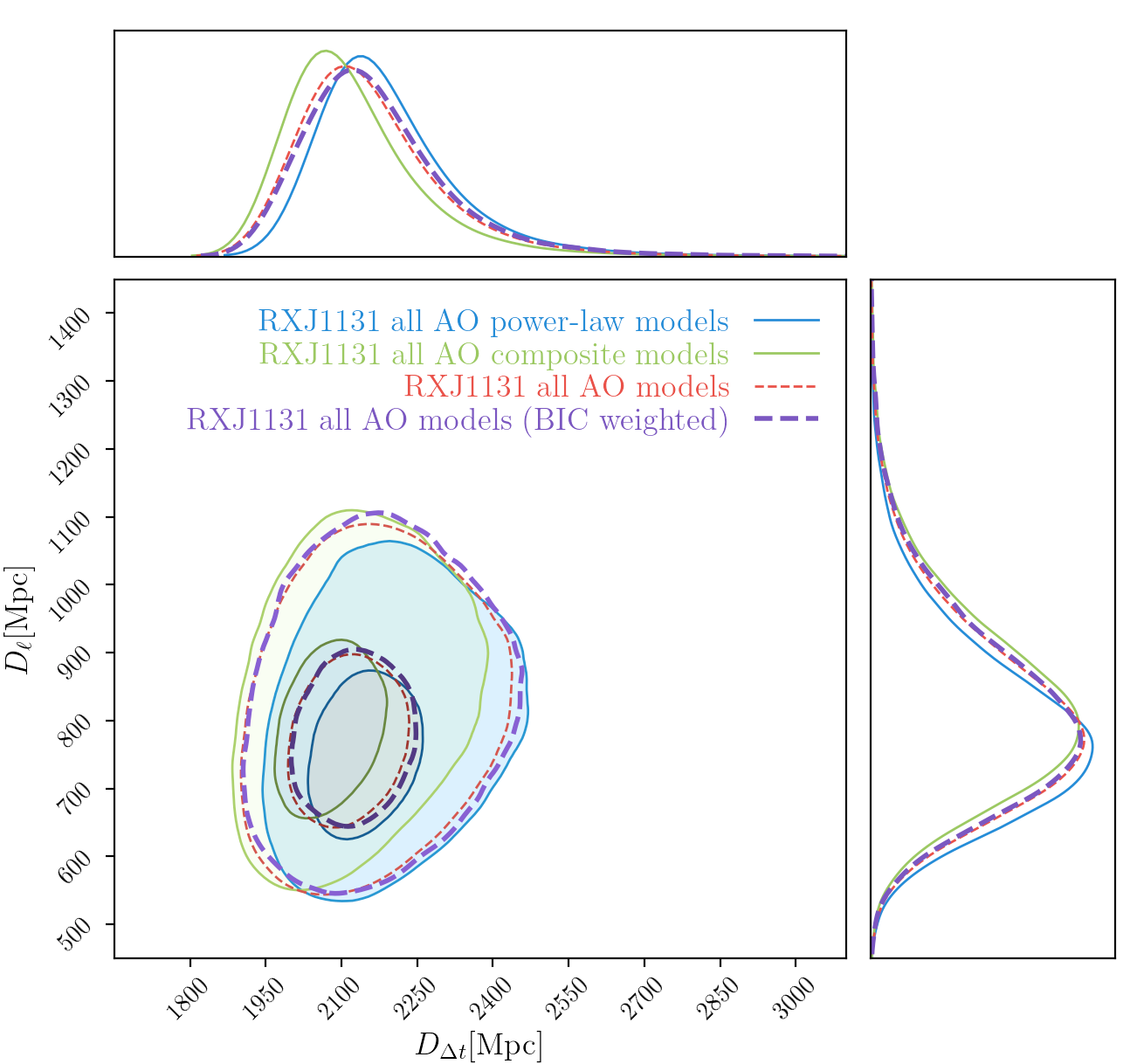}
\caption{The $\Ddt$ and $D_{\ell}$ of \rxj\ based on the analysis of the AO data.}
\label{fig:RXJ1131_Ddt}
\end{figure*}

\begin{figure*}
\centering
\includegraphics[width=\linewidth]{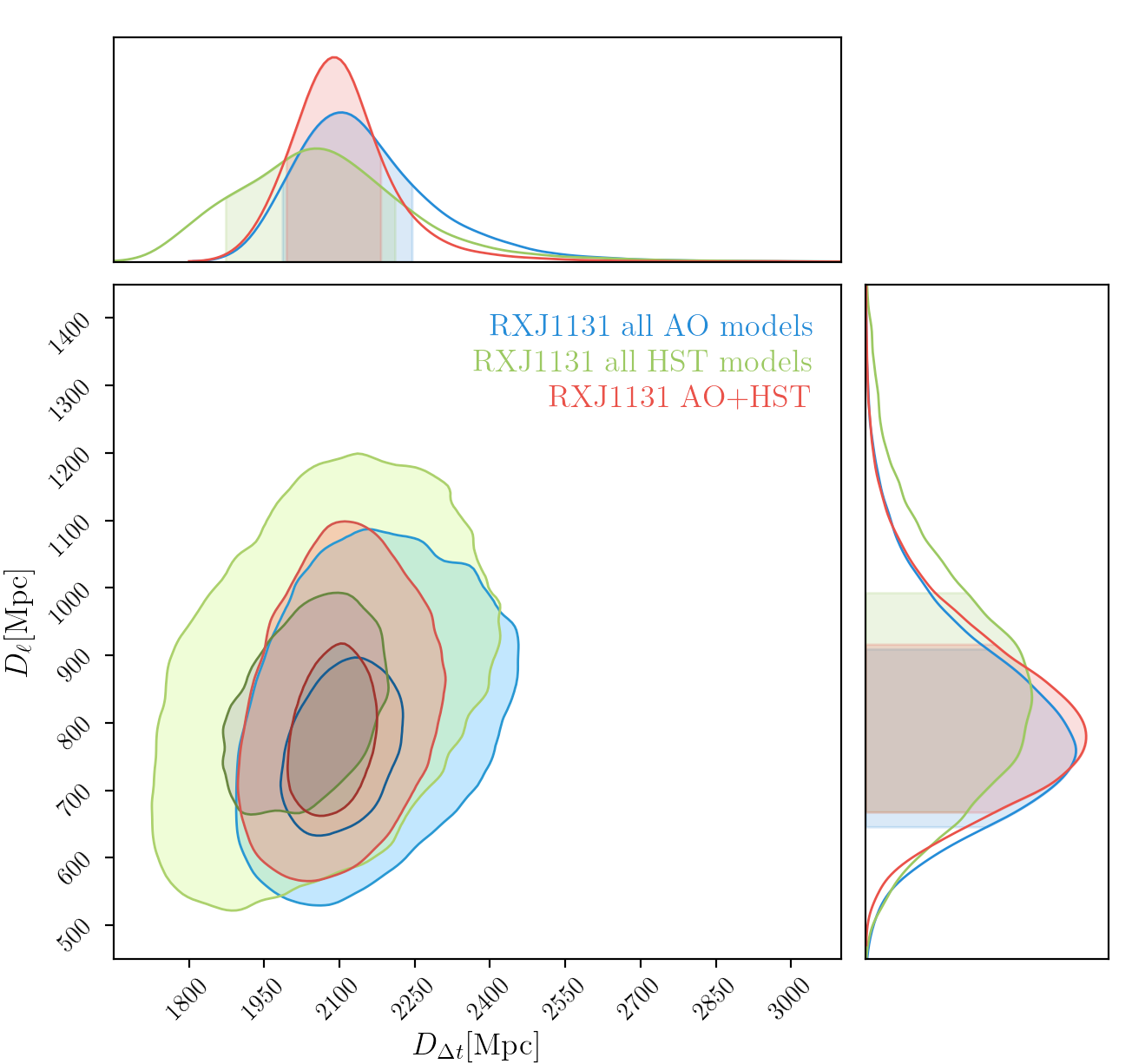}
\caption{The $\Ddt$ and $D_{\ell}$ of \rxj\ based on the combination of AO and HST.}
\label{fig:RXJ1131_Ddt_compare}
\end{figure*}

\subsubsection{Systematics Tests and Unblinding Results}
\label{subsubsec:pg_system}
We summarize below the choices that we explore for the mass modeling, including the nearby group/galaxies. 
For each of the models, we set the weights for the regions containing the AGN images to zero and fix the mass centroid for G1 and G2 at the center of its light distribution.
When modeling the galaxy group as a NFW profile, we use the $M_{\rm vir} - c_{\rm vir}$ from \citet{MacciEtal08}, based on a WMAP5 cosmology (we found that the impact of using different cosmology is negligible.).
\begin{description}
  \item[$\bullet$] SPEMD + 2S$\acute{\text{e}}$rsic + NFW group.
  \item[$\bullet$] SPEMD + 2S$\acute{\text{e}}$rsic + NFW group + G1
  \item[$\bullet$] SPEMD + 2S$\acute{\text{e}}$rsic + SIS group + G1
  \item[$\bullet$] SPEMD + 2S$\acute{\text{e}}$rsic + NFW group + G1 + G2
  \item[$\bullet$] Composite lens + NFW group
  \item[$\bullet$] Composite lens + NFW group + G1
  \item[$\bullet$] Composite lens + SIS group + G1
  \item[$\bullet$] Composite lens + NFW group + G1 + G2
  \item[$\bullet$] For all of the above models, we test five different source resolutions.
\end{description}
To address the MST, we use the
$P(\kappa_{\textrm{ext}}|\bm{d}_\mathrm{ENV},\gamma)$ from the previous section and importance sample the measured velocity dispersion, $\sigma=281\pm25~\kms$ \citep{Tonry98}, which was obtained inside a 1.0-arcsec$^{2}$ aperture with a seeing of $0.8\arcsec$. 
When sampling $\Ddt$ with the time-delay measurements from \citet{BonvinEtal18_PGTD}, we follow \citet{GChenEtal18a} to sample the $\Ddt$ both with and without considering the MTDE. The parameters for estimating the star density and accretion disk model can be found in \citet{GChenEtal18a}. 
In sum, we explore 160 modeling choices in total, with all different combination of choices among two kinds of main lens models, various mass models for the group, five different resolutions of the reconstructed source, and three different priors of the accretion disk sizes (or we ignore the MTDE).  We show the AO imaging reconstruction in \fref{fig:PG1115_AO_figure} and HST imaging reconstruction in \fref{fig:PG1115_HST_figure}. In \fref{fig:PG_composite} and \fref{fig:PG_pl}, we present the posteriors of the important parameters of AO-only results and HST+AO results for the composite model and the power-law model, respectively.  
The $\Ddt$ distributions from the 160 different model choices are shown in \fref{fig:PG1115_Ddt}. The final $\Ddt$-$D_{\ell}$ without considering MTDE is shown in \fref{fig:PG1115_Ddt_Dd}. We show the $\Delta$BIC value of each model without considering MTDE in \aref{appendix:pg_BIC}. We found that the $\Ddt$ inferred from various model choices are statistically consistent. The uncertainties of the final marginalized $\Ddt$ are $\sim9.6\%$ and $D_{\ell}$ are $\sim 29\%$.

\subsection{HE0435-1223 modeling}
\label{subsec:HEmodeling}
For the \he\ system we model only the AO data, and then combine the results with the HST-based modeling of \citet{WongEtal17}. 
This lens system presents a bit of complexity because there are significant contributions to the lensing signal from galaxies at multiple redshifts.
In particular, there are five important perturbers (G1 -- G5) that are close in projection to \he~\citep[see Fig. 3 in][]{WongEtal17}. Based on the $\Delta_3x$ criterion of \citet{McCullyEtal14,McCullyEtal17}, we should include the most massive nearby perturber, G1, explicitly in the model. However, \citet{SluseEtal17} show that although the $\Delta_3x$ values of the other four galaxies are not above the threshold when considered individually, when considered together they do show a significant effect.   Since G1--G5 are located at different redshifts, we follow \citet{WongEtal17} and model this system through the multi-plane lens equation \citep[e.g.,][]{BlandfordNarayan86,SEF92,Collett&Auger14,McCullyEtal14,WongEtal17}. In this case, there is no single time-delay distance, and therefore a particular cosmological model needs to be applied to the analysis. However, if the lens system is dominated by a single primary lens, as is the case for \he, then we can define an effective time delay distance, $\Ddt^{\textrm{eff}}\left(\zl,\zs\right)$, which is fairly robust to changes in the assumed cosmology.

\subsubsection{The AO PSF of \he}
We follow the same criteria described in \sref{subsub:psf} and perform 13  iterative correction steps to obtain the final AO PSF of \he.
The FWHM of the reconstructed \he~AO PSF is 0.07\arcsec (see \fref{fig:AO_PSF}). 

\subsubsection{Lens Model Choices}
As we did for \pg, we model the main lens with either a SPEMD or composite model. For the composite model, we follow \citet{WongEtal17} and set the Gaussian prior for the scale radius to $4.3\arcsec\pm2.0\arcsec$ based on scaling relations derived from the SLACS sample \citep{GavazziEtal07}.
The most massive perturber, G1, is modeled as a SIS profile. 
When modeling G1 -- G5 simultaneously as SIS distributions, we fix the ratios of their Einstein radii by estimating their stellar masses~\citep{RusuEtal17} and then using \citet{BernnardiEtal11} to convert these to velocity dispersions and then to Einstein radii.
We follow \citet{WongEtal17} and fix the ratio of Einstein radii, but the global scaling is allowed to vary.

\subsubsection{LOS Analysis and the External Convergence}
\label{subsubsec:kappaHE}

For this system, we have gathered wide-field imaging in a variety of filters, as well as conducting targeted spectroscopy \citep{RusuEtal17,SluseEtal17}. Our results on $P(\kappa_{\textrm{ext}}|\bm{d}_\mathrm{ENV},\gamma)$ are presented in \citet{RusuEtal17}. In this work, we use the shear values determined from our lens modeling of the AO data to update the weighted number counts for the system.  Otherwise we follow the analysis of \citet{RusuEtal17} in order to conduct a direct comparison of $H_0$ from the HST and AO datasets. We show our updated results in Figure~\ref{fig:kappaHE0435}.

\subsubsection{Systematics Tests and Unblinding results}
We list the systematic tests we have done here. For each of the models, we 
set the weights for the regions containing the AGN images to zero
and fix the mass centroid for G1 at the center of its light distribution.
 
\begin{description}
  \item[$\bullet$] A power-law model plus G1 as a SIS. 
  \item[$\bullet$] A composite model plus G1 as a SIS.
  \item[$\bullet$] A composite model plus  the five perturbers (G1 -- G5).
  \item[$\bullet$] For all of the above models, we test five different source resolutions. 
\end{description}
We show an example of the AO imaging reconstruction in \fref{fig:HE0435_figure} and present the posteriors of $\Ddt$ in \fref{fig:HE0435_Ddt}.

To assess the MST, for each model we perform the importance sampling given the measured velocity dispersion, $\sigma=222\pm15~\kms$ inside a $0.54\arcsec\times0.7\arcsec$ aperture with a seeing of $0.8\arcsec$ \citep{WongEtal17}.

Note that for the baryonic component in the composite model, the mass distribution is based on the light distribution in the HST imaging.
This is because an insufficient knowledge about the structure in the wings of the AO PSF introduces a degeneracy between the reconstructed PSF structure and lens galaxy light (see the discussion in Appendix~\ref{Degeneracy}).

\subsection{RXJ1131-1231 modeling}
\label{subsec:RXJmodeling}
A detailed discussion of the \rxj~lens modeling of the AO imaging can be found in the paper of \citet{GChenEtal16}.
To summarize, we used the power-law mass distribution to model the lens potential and used two concentric S$\acute{\text{e}}$rsic profiles to model the lens light. 
The satellite galaxy of the main deflector was modeled as an SIS profile. 
We modeled only the lensing galaxy plus satellite, and did not consider $\kappa_{\rm ext}$.  The modeling marginalized over five different source resolutions
in order to better control the systematics. 
In this paper, we further explore a different mass model and turn the previous work and the new results in this paper into cosmology.
\subsubsection{Main Lens and Satellite}
To add to the previous power-law model, we test a composite model with different source resolutions in this paper. 
We follow \citet{SuyuEtal14} to set a gaussian prior on the NFW scale radius of $18.6\arcsec\pm2.6\arcsec$,
based on the weak lensing analysis of SLACS lenses \citep{GavazziEtal07} that have similar velocity dispersions to \rxj. For the other parameters, we set uniform priors. We model the satellite light distribution with a circular S$\acute{\text{e}}$rsic profile, and the satellite mass as a SIS distribution whose centroid is linked to the light centroid.  

Note that due to the degeneracy between the reconstructed PSF structure and lens galaxy light, the baryonic mass distribution in the composite model is also based on the light distribution in the HST imaging.

\subsubsection{LOS Analysis and the External Convergence}
\label{subsubsec:kappaRXJ}

As in \citet{SuyuEtal13}, we use a combination of observations and simulations to estimate the contribution of the LOS mass distribution 
for \rxj, i.e.,
$P(\kappa_{\textrm{ext}})$,
$P(\kappa_{\textrm{ext}}|\gamma)$, and
$P(\kappa_{\textrm{ext}}|\bm{d}_\mathrm{ENV},\gamma)$.
Here $\gamma$ is the external shear required by the mass models of the main lensing galaxy, while  $\bm{d}_\mathrm{ENV}$ is the relative overdensity of galaxies within a 45\arcsec\ aperture that is centred on the lens.
This overdensity, $\zeta_1^{45\arcsec} = 1.4\pm0.05$ (following the notation in \citet{BirrerEtal19}), is calculated from galaxies with apparent HST/ACS F814W  magnitudes $18.5 \leq m \leq 24.5$ in both the lens and control samples \citep{FassnachtEtal11}.
The overdensity and shear values are combined with the simulated lensing data based on the Millennium Simulation \citep{SpringelEtal05, HilbertEtal09} together with the semi-analytic galaxy model of \citet{HenriquesEtal15}, to get the probability distributions for $\kappa_{\rm ext}$.

\subsubsection{Systematics Tests}
We list the systematic tests we have done including those done in our previous work. 
In all of the models the regions near the AGN images are given zero weight.
\begin{description}
  \item[$\bullet$] SPEMD+2S\'{e}rsic lens model.
  We rerun the model since we did not link the satellite mass position to its light position in \citet{GChenEtal16}. 
  \item[$\bullet$] A composite model.
   \item[$\bullet$] For the above models, we test five different source resolutions.  
\end{description}

We use the observed velocity dispersion, $323\pm20\kms$ \citep{SuyuEtal13} given the $\kappa_{\textrm{ext}}$ in \sref{subsubsec:kappaRXJ} to sample $\Ddt$ and $D_{\ell}$ without assuming cosmology. We plot the posteriors of $\Ddt$ and $D_{\ell}$ in \fref{fig:RXJ1131_Ddt}.

\section{Cosmological inference}
\label{sec:cosmoinfer}
We present the cosmological inferences based on our sample of three gravitational lenses that have AO imaging data, HST imaging data, velocity dispsersion measurements, line-of-sight studies, and time-delay measurements. In particular, we present the cosmological inferences based on only the AO imaging data in \sref{subsec:cosmoAO}, while we present the cosmological inferences based on a combination of both the AO and HST imaging in \sref{subsec:cosmoAOHST}. 

\subsection{Cosmological inference from AO Strong Lensing}
\label{subsec:cosmoAO}
\begin{figure*}
  \centering
  \includegraphics[width=\textwidth]{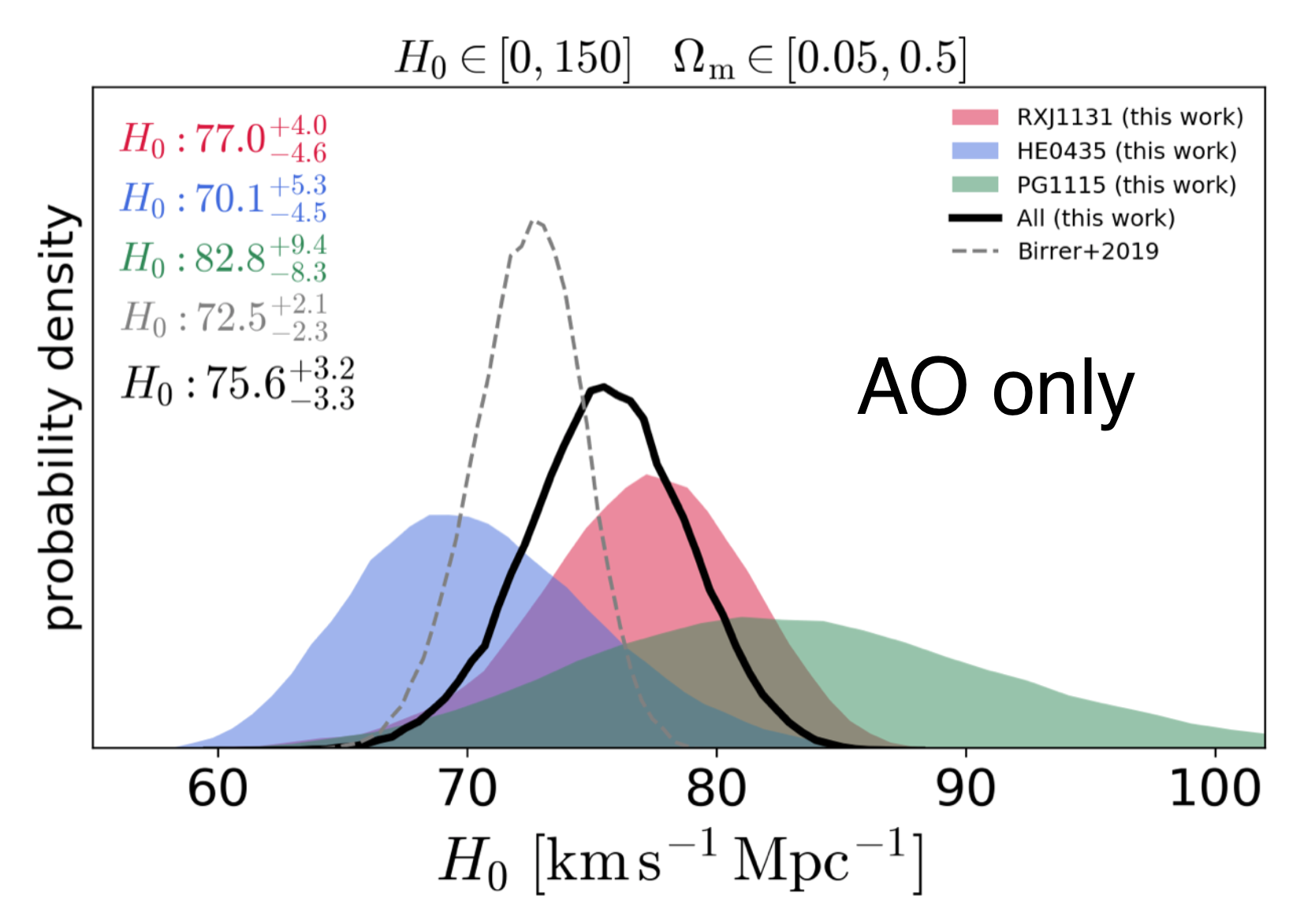}
  \caption{Marginalized posterior probability distributions for $H_{0}$ in the U$\Lambda$CDM cosmology using the constraints from the three AO-only strong lenses (\rxj,~\he,~\pg ). The overlaid histograms present the distributions for each individual strong lens, and the solid black line corresponds to the distribution resulting from the joint inference from all three datasets. The dashed line shows the latest joint $H_{0}$ from the H0LiCOW collaboration \citep{BirrerEtal19}. The quoted values of $H_{0}$ in the top-left corner of each panel are the median, 16th, and 84th percentiles.}
\label{fig:uniformcosmologies_AO}
\end{figure*}

\fref{fig:uniformcosmologies_AO} presents the marginalized posterior PDF for $H_{0}$ assuming a flat $\Lambda$CDM model which has a uniform prior on $H_{0}$ in the range [0, 150] $\kmsmpc$ and a uniform prior on $\Omega_{\textrm{m}}$ in the range of [0.05, 0.5]. After unblinding, we find that \pg\ AO imaging yields $H_{0}=82.8\substack{+9.4\\-8.3}~\kmsmpc$; \he\ AO imaging yields $H_{0}=70.1\substack{+5.3\\-4.5}~\kmsmpc$. \rxj\ AO imaging yields $H_{0}=77.0\substack{+4.0\\-4.6}~\kmsmpc$. The joint analysis of three AO lenses yields $H_{0}=75.6\substack{+3.2\\-3.3}~\kmsmpc$.

\subsection{Cosmological Inference from AO and HST Strong Lensing imaging}
\label{subsec:cosmoAOHST}
As the AO imaging of \he\ and \rxj\ are modeled separately from the HST imaging, we have developed a Bayesian approach to properly combine the HST and AO results for these two lenses (see details in \aref{appendix:bayesAOplusHST}). In short, since the HST and AO images are independent data sets, we can get the joint probability distribution by multiplying their probability distributions, as long as the prior on the joint parameters are used only once. 
We can express the joint posterior as
\begin{equation}
\label{joint}
\begin{split}
P(\bm{\eta}&|\bm{d}_{\textrm{HST}},\bm{\bm{d}_{\textrm{AO}}},\sigma, \bm{d_{\textrm{ENV}}},\textrm{H},\textrm{MS})\\
&\propto \frac{P(\bm{\eta}|\bm{d}_{\textrm{HST}},\sigma, \bm{d_{\textrm{ENV}}},\textrm{H},\textrm{MS})P(\bm{\eta}|\bm{d}_{\textrm{AO}},\textrm{H},\textrm{MS})}{P(\bm{\eta}|\textrm{H},\textrm{MS})},
\end{split}
\end{equation}
where $P(\bm{\eta}|\bm{d}_{\textrm{HST}},\sigma,\bm{d_{\textrm{ENV}}},\textrm{H},\textrm{MS})$ reflects the previous modeling results, $P(\bm{\eta}|\bm{d}_{\textrm{AO}},\textrm{H},\textrm{MS})$ is obtained by this work, $P(\bm{\eta}|\textrm{H},\textrm{MS})$ is the prior used in both analysis, H is the lens model, and MS is the Millennium Simulation. 

\begin{figure*}
  \centering
  \includegraphics[width=\textwidth]{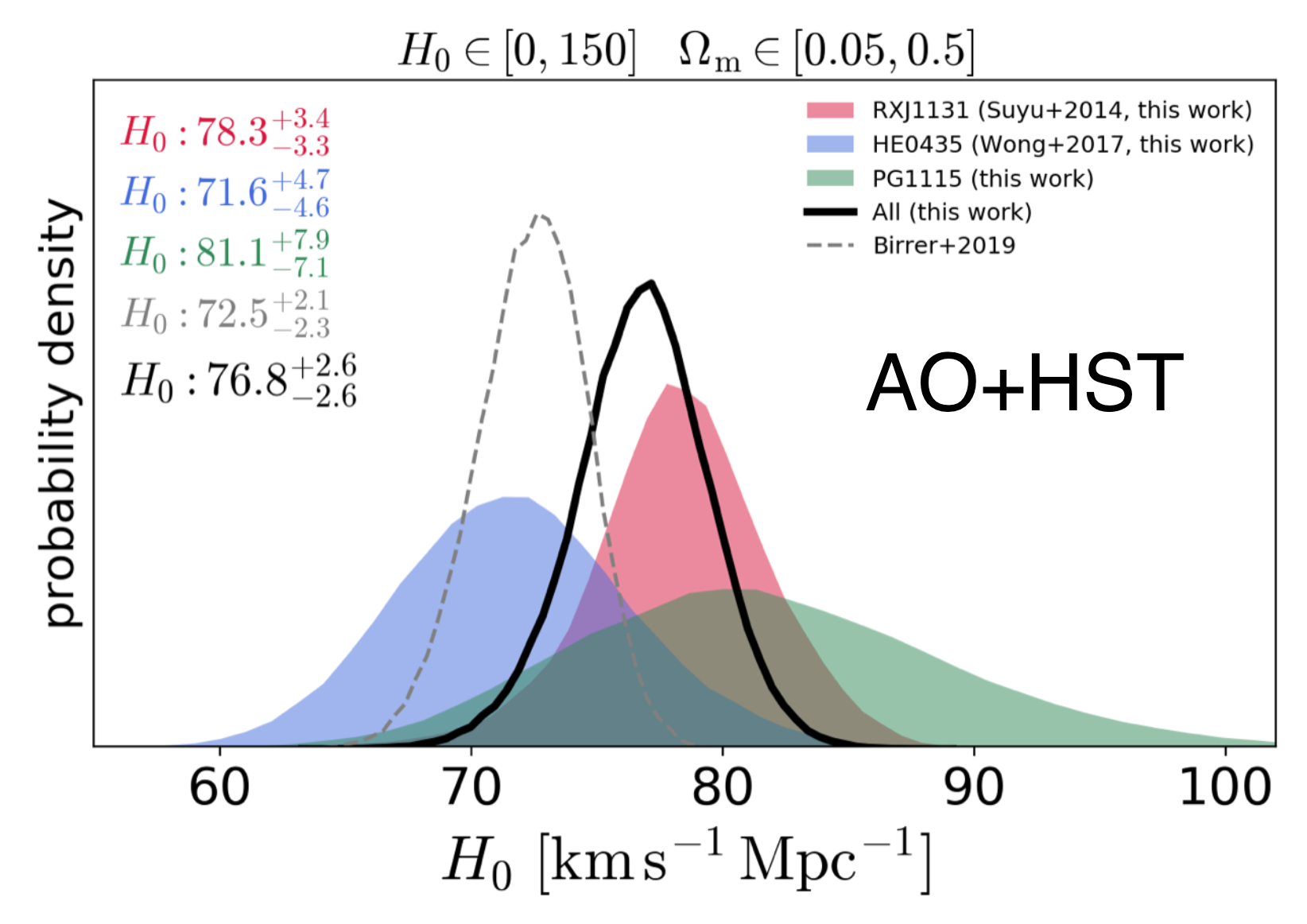}
  \caption{Marginalized posterior probability distributions for $H_{0}$ in the U$\Lambda$CDM cosmology using the constraints from the three AO+HST strong lenses (\rxj,~\he,~\pg ). The overlaid histograms present the distributions for each individual strong lens, and the solid black line corresponds to the distribution resulting from the joint inference from all three datasets. The dashed line shows the latest joint $H_{0}$ from the H0LiCOW collaboration \citep{BirrerEtal19}. The quoted values of $H_{0}$ in the top-left corner of each panel are the median, 16th, and 84th percentiles.}
\label{fig:uniformcosmologies_AOHST}
\end{figure*}

\fref{fig:uniformcosmologies_AOHST} presents the marginalized posterior PDF for $H_{0}$ assuming flat-$\Lambda$CDM model. We found that the joint AO+HST of \he~implies a value of the Hubble constant of $H_{0}=71.6\substack{+4.7\\-4.6}~\kmsmpc$; the joint AO+HST of \pg~implies $H_{0}=81.1\substack{+7.9\\-7.1}~\kmsmpc$. The joint AO+HST of \rxj~implies $H_{0}=78.3\substack{+3.4\\-3.3}~\kmsmpc$. The combination of three AO+HST lenses yields $H_{0}=76.8\substack{+2.6\\-2.6}~\kmsmpc$.

We found that after combining AO and HST imaging, the dominant sources of uncertainty of both \pg\ and \he\ are time-delay measurements, while the dominant sources of uncertainty of \rxj\ is the line-of-sight mass distribution.

\section{Conclusions}
\label{sec:conclusion}
We did the blind analysis on both \pg\ and \he\ as well as an extension of our previous analysis of \rxj. For each system, we combined the AO imaging, HST imaging,
the measurements of the lens galaxy’s velocity dispersion, the line-of-sight studies from deep wide-area spectroscopic as well as photometric data, and the time-delay measurements from state-of-the-art light curve fitting algorithm to infer the value of $H_{0}$. We find that the high S/N AO and HST imaging data yield consistent results, providing and important validation of the AO PSF reconstruction techniques for high precision lensing work. 

This paper demonstrates the ability of using AO imaging to constrain the mass model as well as the value of $H_{0}$. Furthermore, we show that combining AO imaging with HST imaging can further tighten the uncertainties from the lens mass model and thus improve the precision of the determination of $H_{0}$.

In this paper, we infer the value of $H_{0}$ under the assumption of a flat $\Lambda$CDM model which has a uniform prior on $H_{0}$ in the range [0, 150] $\kmsmpc$ and a uniform prior on $\Omega_{\textrm{m}}$ in the range of [0.05, 0.5].
After unblinding, \pg\ AO imaging yields $H_{0}=82.8\substack{+9.4\\-8.3}~\kmsmpc$; \he\ AO imaging yields $H_{0}=70.1\substack{+5.3\\-4.5}~\kmsmpc$. \rxj\ AO imaging yields $H_{0}=77.0\substack{+4.0\\-4.6}~\kmsmpc$. The joint analysis of three AO lenses yields $H_{0}=75.6\substack{+3.2\\-3.3}~\kmsmpc$. 

The joint AO+HST of \pg\ yields $H_{0}=81.1\substack{+7.9\\-7.1}~\kmsmpc$.
The joint AO+HST of \he\ yields  $H_{0}=71.6\substack{+4.7\\-4.6}~\kmsmpc$. The joint AO+HST of \rxj\ yields $H_{0}=78.3\substack{+3.4\\-3.3}~\kmsmpc$. The combination of three AO+HST lenses yields $H_{0}=76.8\substack{+2.6\\-2.6}~\kmsmpc$.

We refer the reader to the paper by Wong et al. where the results presented here will be combined with a self-consistent analysis of three previously published systems to carry out a full cosmological investigation.

\section*{Acknowledgements}
G.~C.-F.~Chen acknowledges support from the Ministry of Education in Taiwan via Government Scholarship to Study Abroad (GSSA).
C.D.F. and G.~C.-F.~Chen  acknowledge support for this work from the National Science Foundation under Grant No. AST-1715611.
S.H. acknowledges support by the DFG cluster of excellence \lq{}Origin and Structure of the Universe\rq{} (\href{http://www.universe-cluster.de}{\texttt{www.universe-cluster.de}}). 
L.~V.~E.~K. acknowledges the support of an NWO-VIDI grant (nr. 639.043.308). 

This work was supported by World Premier International
Research Center Initiative (WPI Initiative), MEXT, Japan. K.C.W. is supported in part by an EACOA Fellowship awarded by the East Asia Core Observatories Association, which consists of the Academia Sinica Institute of Astronomy and Astrophysics, the National Astronomical Observatory of Japan, the National Astronomical Observatories of the Chinese Academy of Sciences, and the Korea Astronomy and Space Science Institute. SHS thanks the Max Planck Society for support through the Max Planck Research Group. J. Chan, V. Bonvin and F. Courbin acknowledge support from the Swiss National Science Foundation (SNSF). This project has received funding from the European Research Council (ERC) under the European Union’s Horizon 2020 research and innovation programme (COSMICLENS: grant agreement No 787866). X.D. and T.T. acknowledges support from the Packard Foundation through a Packard Research Fellowship and from the NSF through grant AST-1714953. 
G.~C.-F.~Chen thank Jen-Wei Hsueh for the useful discussions and feedback. The data presented herein were obtained at the W. M. Keck Observatory, which is operated as a scientific partnership among the California Institute of Technology, the University of California, and the National Aeronautics and Space Administration. The Observatory was made possible by the generous financial support of the W. M. Keck Foundation. The authors wish to recognize and acknowledge the very significant cultural role and reverence that the summit of Maunakea has always had within the indigenous Hawaiian community. We are most fortunate to have the opportunity to conduct observations from this mountain. 

This work is based on observations obtained with MegaPrime/MegaCam, a joint project of CFHT and CEA/IRFU, at the Canada-France-Hawaii Telescope (CFHT) which is operated by the National Research Council (NRC) of Canada, the Institut National des Sciences de l'Univers of the Centre National de la Recherche Scientifique (CNRS) of France, and the University of Hawaii. This research used the facilities of the Canadian Astronomy Data Centre operated by the National Research Council of Canada with the support of the Canadian Space Agency. CFHTLenS data processing was made possible thanks to significant computing support from the NSERC Research Tools and Instruments grant program.

This work used computational and storage services associated with the Hoffman2 Shared Cluster provided by UCLA Institute for Digital Research and Education’s Research Technology Group. Data analysis was in part carried out on common use data analysis computer systems at the Astronomy Data Center, ADC, of the National Astronomical Observatory of Japan.

%%%%%%%%%%%%%%%%%%%%%%%%%%%%%%%%%%%%%%%%%%%%%%%%%%



\bibliographystyle{mnras}
\bibliography{AO_cosmography} % if your bibtex file is called example.bib

%\begin{thebibliography}{99}
%\end{thebibliography}

%%%%%%%%%%%%%%%%%%%%%%%%%%%%%%%%%%%%%%%%%%%%%%%%%%

%%%%%%%%%%%%%%%%% APPENDICES %%%%%%%%%%%%%%%%%%%%%

\appendix
\section{Comparison of the reconstructed AO PSF from three lenses}
\begin{figure*}
  \centering
  \includegraphics[width=\textwidth]{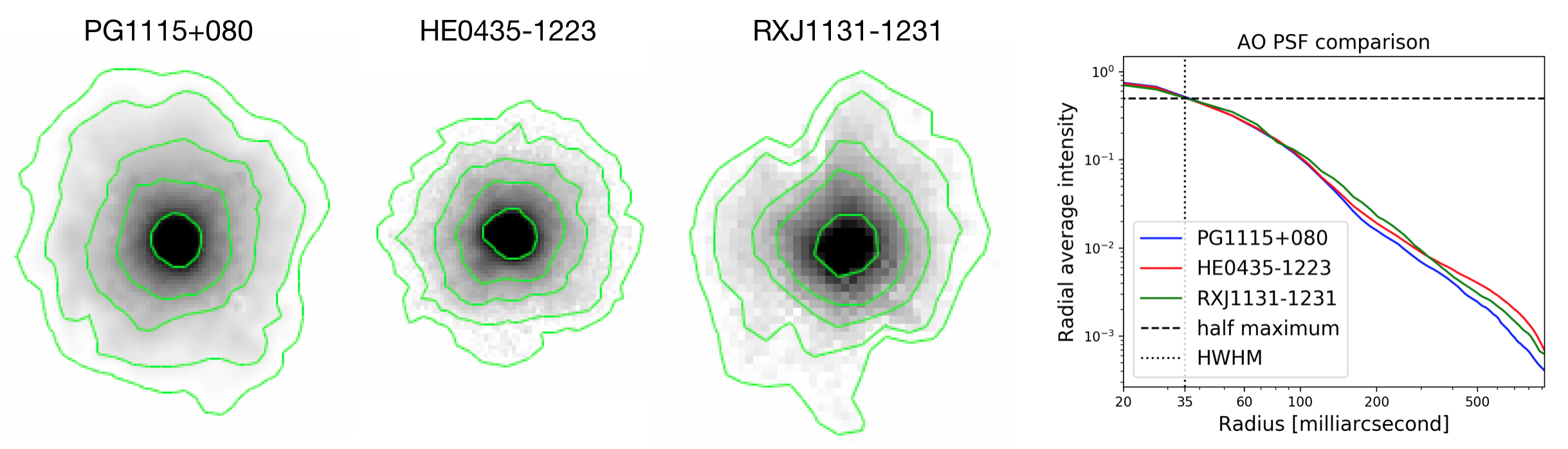}
  \caption{The left three figures are the reconstructed AO PSF of \pg, \he, and \rxj, respectively. The right panel is the comparison of the radial average intensity of the PSF, which shows the core plus its wings. The isointensity contours represent 0.07, 0.01, 0.005, 0.0025, and 0.0015.}
\label{fig:AO_PSF}
\end{figure*}

We show the comparison of the reconstructed AO PSF in \fref{fig:AO_PSF}. The 2D plots of the three PSF clearly show that the atmosphere disturbance produce various structures of PSF with a core plus a wing, while the radial average intensity indicates that the intensity gradient of the cores are very similar inside 0.1\arcsec.

\section{Degeneracy between AO PSF wing and lens light}
\label{Degeneracy}
The PSF-reconstructed method developed in \citet{GChenEtal16} allow us to model the AO imaging down to the noise level and recover $\Ddt$ when we adopt the power-law model. 
However, we found that there exist a degeneracy between AO PSF wing and the lens light if we reconstruct the AO PSF from the AO imaging only. 
While it does not cause a problem for the power-law model since the power-law mass is constrained by the arc, the composite model instead could yield a  baised $H_{0}$ because we assume the baryonic matter distribution follow the lens light. 
Thus, in the case of \he~and \rxj~where we do not reconstruct the PSF simultaneously from HST and AO imaging, we fix the mass distribution of the baryonic matter to that baryonic mass inferred from HST imaging. 

In the case of \pg, we found that the degeneracy can be broken by simultaneously modeling the AO imaging and HST imaging as they share the same mass model. The additional constraint allow us to better characterize the AO PSF wing. Thus, given the same PSF, both power-law model and composite model can be modeled down to the noise level. In the future, 2D kinematic data could also potentially further break the degeneracy between the PSF wing and lens light, as it provides the information to characterize the dark and baryonic matter content \citep{CappellariEtal13}.

\section{Constraints used to estimate $\kappa_{\textrm{ext}}$}
\label{appendix:kappaconstraints}
Here we show the $\kappa_{\textrm{ext}}$ distribution of \he~in \fref{fig:kappaHE0435} and present the summary table of the constraints used to estimate $\kappa_{\textrm{ext}}$ for all three systems in \tref{table:summaryconstraints}.

\addtocounter{footnote}{1} %3=n
\stepcounter{footnote}\footnotetext{The five perturbers are indicated in Figure 3 from \citet{WongEtal17}. Three additional galaxies enter the 12\arcsec-radius inner mask, and we slightly boosted their distance from the lens, in order to avoid masking them.}

\begin{table*}
\label{table:summaryconstraints}
 \centering
 \begin{minipage}{\linewidth}
   \caption{Joined constraints used to estimate $\kappa_\textrm{ext}$}
  \begin{tabular}{@{}lllllll@{}}
  \hline 
  \hline 
Lens & model name & $\gamma_\mathrm{ext}$ & $\zeta_1^{45\arcsec}$ & $\zeta_{1/r}^{45\arcsec}$ & $\zeta_{1}^{120\arcsec}$ & $\zeta_{1/r}^{120\arcsec}$ \\
\hline
\hline 
\rxj & power law G & $0.083\pm0.003$ & $1.4^{+0.05}_{-0.05}$ & & & \\
\rxj & composite G & $0.058\pm0.005$ & $1.4^{+0.05}_{-0.05}$ & & & \\
\hline 
\he & power law G + G1 & $0.041\pm0.018$ &  &  &  &  \\
\he & composite G + G1 & $0.026\pm0.026$ &  &  &  &  \\
\he & composite G + 5 perturbers & $0.056\pm0.026$ &  &  &  &  \\
\he & power law G + G1 + LOS & $0.041\pm0.018$ & $1.27^{+0.05}_{-0.05}$ & $1.31^{+0.05}_{-0.05}$ &  &  \\
\he & composite G + G1 + LOS & $0.026\pm0.026$ & $1.27^{+0.05}_{-0.05}$ & $1.31^{+0.05}_{-0.05}$ &  &  \\
\he & composite G + 5 perturbers + LOS & $0.056\pm0.026$ & $1.21^{+0.05}_{-0.05}$ & $1.17^{+0.05}_{-0.05}$ &  &  \\
\hline 
%\pg & power law G + NFW group (WMAP1) & $0.042\pm0.010$ & $1.11^{+0.01}_{-0.11}$ & $1.27^{+0.01}_{-0.23}$ & $1.10^{+0.05}_{-0.09}$ & $0.98^{+0.03}_{-0.10}$ \\
%\pg & power law G + NFW group (WMAP3) & $0.031\pm0.011$ & $1.11^{+0.01}_{-0.11}$ & $1.27^{+0.01}_{-0.23}$ & $1.10^{+0.05}_{-0.09}$ & $0.98^{+0.03}_{-0.10}$ \\
\pg & power law G + NFW group & $0.027\pm0.009$ & $1.11^{+0.01}_{-0.11}$ & $1.27^{+0.01}_{-0.23}$ & $1.10^{+0.05}_{-0.09}$ & $0.98^{+0.03}_{-0.10}$ \\
\pg & power law G + SIS group & $0.061\pm0.009$ & $1.11^{+0.01}_{-0.11}$ & $1.27^{+0.01}_{-0.23}$ & $1.10^{+0.05}_{-0.09}$ & $0.98^{+0.03}_{-0.10}$ \\
\pg & power law G + NFW group + G1 & $0.054\pm0.009$ & $1.00^{+0.01}_{-0.11}$ & $1.02^{+0.03}_{-0.12}$ & $1.09^{+0.04}_{-0.09}$ & $0.96^{+0.04}_{-0.10}$ \\
\pg & power law G + NFW group + G1 + G2 & $0.058\pm0.012$ & $0.89^{+0.01}_{-0.11}$ & $0.84^{+0.06}_{-0.24}$ & $1.07^{+0.05}_{-0.09}$ & $0.94^{+0.04}_{-0.10}$ \\
%\pg & composite G + NFW group (WMAP1) & $0.043\pm0.008$ & $1.11^{+0.01}_{-0.11}$ & $1.27^{+0.01}_{-0.23}$ & $1.10^{+0.05}_{-0.09}$ & $0.98^{+0.03}_{-0.10}$ \\
%\pg & composite G + NFW group (WMAP3) & $0.050\pm0.012$ & $1.11^{+0.01}_{-0.11}$ & $1.27^{+0.01}_{-0.23}$ & $1.10^{+0.05}_{-0.09}$ & $0.98^{+0.03}_{-0.10}$ \\
\pg & composite G + NFW group & $0.048\pm0.009$ & $1.11^{+0.01}_{-0.11}$ & $1.27^{+0.01}_{-0.23}$ & $1.10^{+0.05}_{-0.09}$ & $0.98^{+0.03}_{-0.10}$ \\
\pg & composite G + SIS group + G1 & $0.072\pm0.008$ & $1.00^{+0.01}_{-0.11}$ & $1.02^{+0.03}_{-0.12}$ & $1.09^{+0.04}_{-0.09}$ & $0.96^{+0.04}_{-0.10}$ \\
\pg & composite G + NFW group + G1 & $0.072\pm0.008$ & $1.00^{+0.01}_{-0.11}$ & $1.02^{+0.03}_{-0.12}$ & $1.09^{+0.04}_{-0.09}$ & $0.96^{+0.04}_{-0.10}$ \\
\pg & composite G + NFW group + G1 + G2 & $0.060\pm0.008$ & $0.89^{+0.01}_{-0.11}$ & $0.84^{+0.06}_{-0.24}$ & $1.07^{+0.05}_{-0.09}$ & $0.94^{+0.04}_{-0.10}$ \\
 
\hline
\hline 
\end{tabular}
\\ 
%\bigskip
\label{tab:weights}
\end{minipage}
\end{table*}
\normalsize

\section{Accounting for the missing galaxy group members in PG1115+080 due to spectroscopic incompleteness}
\label{missinggroup_PG1115}

The spectroscopic coverage of the FOV around \pg~is incomplete. Down to $R_c\leq22.5$ and within $120\arcsec$-radius around the lens there are 63 galaxies, out of which 33 have spectroscopy \citep{WilsonEtal16}, 11 of which are part of the galaxy group at $z=0.31$, including the lensing galaxy. This means that there may be other galaxies within this magnitude range and radius from the lens which are also part of the galaxy group associated with the lensing galaxy, but which are missed due to spectroscopic incompleteness. In \sref{subsubsect:PG1115group} we have specifically computed the lensing properties of this group, based on its physical properties derived by \citet{MomchevaEtal15}. As a result, when we compute $\kappa_\mathrm{ext}$ at the location of the lens, using the weighted number counts approach, we must remove the galaxies which are part of this group, as the convergence from the group has already been included in the lensing models, and must not be double-counted. While the galaxies which are known to be part of the group can easily be removed, we must also account for the galaxies expected to be missed due to our spectroscopic incompleteness. 

\begin{figure}
\includegraphics[width=\linewidth]{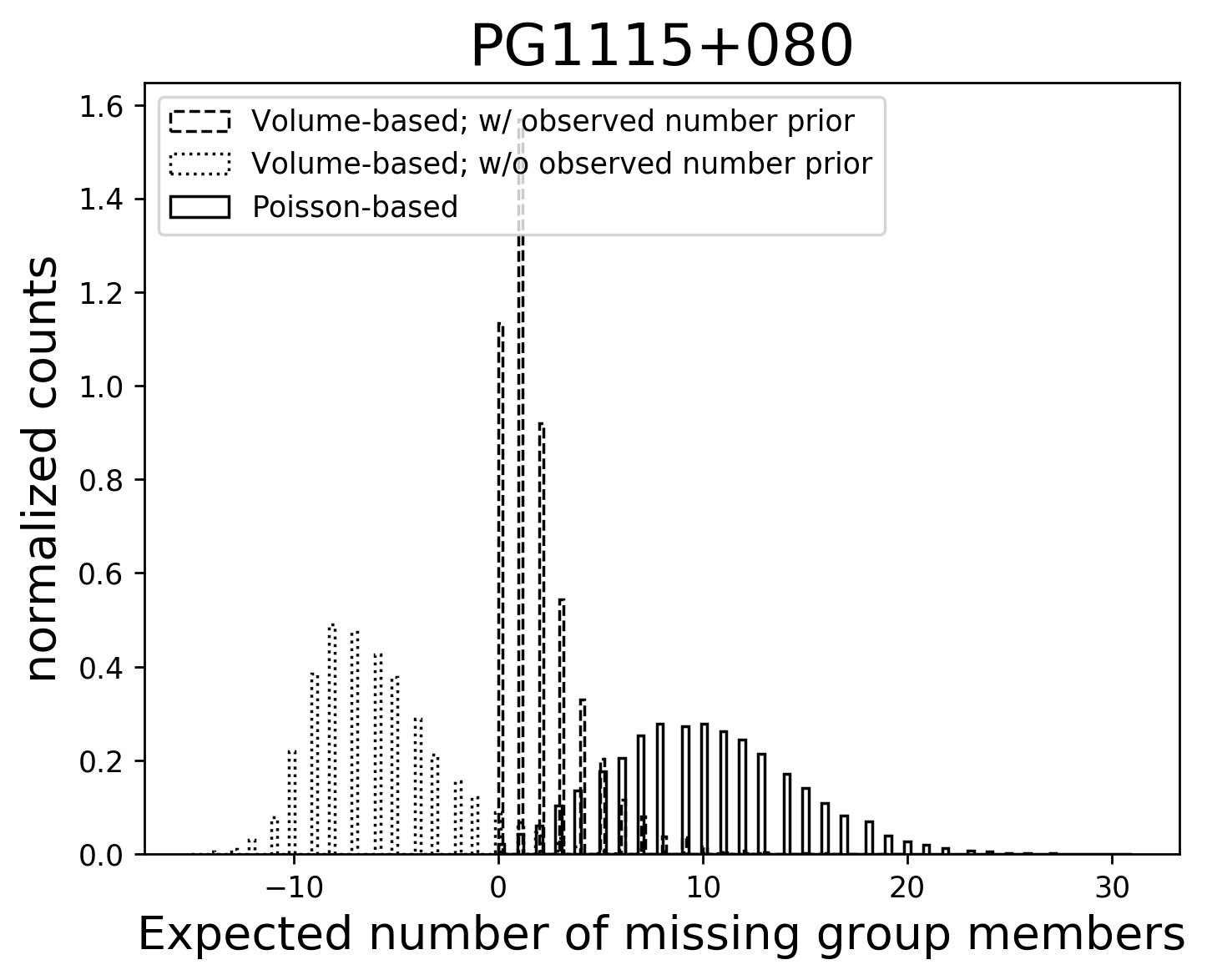}
\caption{Estimated number of missing galaxy group members inside the $\leq120\arcsec$-radius from the lens system for PG1115+080, computed with two methods, with or without imposing the prior that the group consists of at least the number of galaxies spectroscopically confirmed to be members.}
\label{fig:missinggroupPG1115}
\end{figure}

Following the technique presented in \citet{RusuEtal19_H0LiCOW}, we use two different approaches to estimate the number of missing galaxies part of the group. In the first approach we use the knowledge provided by the number of known group members, the number of galaxies with spectroscopy, and the total number of detected galaxies (within the given magnitude and aperture radius), and we apply Poisson statistics to estimate a number of $10\pm5$ missing galaxies (median and 16th, 84th percentiles). In the second approach we use the group velocity dispersion and virial radius from \citet{WilsonEtal16}, and we estimate the expected number of galaxies inside the virial radius using the empirical relation from \citet{AndreonEtal10}. Using the measured offset from the group centroid to the lens, and propagating all uncertainties, we measure the expected number of missing galaxies at the intersection of the sphere of virial radius and the $120\arcsec$-radius cylinder centered on the lens to be $1^{+3}_{-1}$. We plot the distributions of these numbers in Figure~\ref{fig:missinggroupPG1115}. The first approach predicts a significantly larger number of missing galaxies than the second. In fact, due to the small value of the velocity dispersion, the second method would only predict a total number of $8^{+6}_{-4}$ galaxies, therefore less than the number of confirmed group members, unless we enforce this constraint. This discrepancy may be due to the shallow absolute magnitude limit of M$_V=-20$ used by \citet{AndreonEtal10}, corresponding to $r\sim21 $ at $z\sim0.3$, therefore significantly brighter than our limiting magnitude. We note, however, that the two techniques produced results which were in agreement for a different lens, described in Rusu et. al., submitted.

In view of the above, and also due to the fact that it avoids any physical assumption, we consider the first method to be more reliable. 
Finally, when computing weighted galaxy counts, we do this by randomly sampling 10 times from the distribution of missing galaxy numbers, and then randomly excluding that number of galaxies from our catalogue of galaxies inside the $120\arcsec$ apertures. 

\section{Combining the independent measurements}
\label{appendix:bayesAOplusHST}
Consider that we have following sets of information, AO imaging ($\bm{d}_{\textrm{AO}}$), HST imaging ($\bm{d}_{\textrm{HST}}$), velocity dispersion ($\sigma$) and environment data ($\bm{d_{\textrm{ENV}}}$), which can be used to constrain the difference of the fermat potential. We already have two independent measurements, $P(\bm{\eta}|\bm{d}_{\textrm{AO}},\textrm{H},\textrm{MS})$ and $P(\bm{\eta}|\bm{d}_{\textrm{HST}}$, $\sigma$, $\bm{d_{\textrm{ENV}}}$,$\textrm{H},\textrm{MS})$, 
 The goal is to combine them and obtain the joint constraint, $P(\bm{\eta}|\bm{d}_{\textrm{HST}},\bm{d}_{\textrm{AO}},\sigma$, $\bm{d_{\textrm{ENV}}},\textrm{H},\textrm{MS})$, where $\bm{\eta}=(\phi_{ij}^{\textrm{H}},\kappa_{\textrm{ext}},\gamma_{\textrm{ext}},\bm{\zeta})$, $\phi_{ij}^{\textrm{H}}$ are the difference of the model fermat potentials (not the true fermat potentail,
$\phi_{ij}^{\textrm{True}}$) at imaging i and j which we are interested, $\kappa_{\textrm{ext}}$ is the convergence from the line of sight, $\gamma_{\textrm{ext}}$ is the external shear inferred from the imaging data, and $\bm{\zeta}$ are the other parameters which we want to marginalize upon. H is one kind of the lens models (i.g., the power-law model or the composite model) and MS is Millennium Simulation.

We start with the joint constraint and step by step link to the independent measurements. With Bayes’ theorem, the joint constraint can be expressed as
\begin{equation}
\label{bayes}
\begin{split}
P(&\phi_{ij}^{\textrm{H}},\kappa_{\textrm{ext}},\gamma_{\textrm{ext}},\bm{\zeta}|\bm{d}_{\textrm{HST}},\bm{d}_{\textrm{AO}},\sigma, \bm{d_{\textrm{ENV}}},\textrm{H},\textrm{MS})\\
=&P(\bm{d}_{\textrm{HST}},\bm{d}_{\textrm{AO}},\sigma, \bm{d_{\textrm{ENV}}}|\phi_{ij}^{\textrm{H}},\kappa_{\textrm{ext}},\gamma_{\textrm{ext}},\bm{\zeta},\textrm{H},\textrm{MS})\\
&\cdot\frac{P(\phi_{ij}^{\textrm{H}},\kappa_{\textrm{ext}},\gamma_{\textrm{ext}},\bm{\zeta}|\textrm{H},\textrm{MS})}{P(\bm{d}_{\textrm{HST}},\bm{d}_{\textrm{AO}},\sigma, \bm{d_{\textrm{ENV}}}|\textrm{H},\textrm{MS})}, 
\end{split}
\end{equation}
where 
\begin{equation}
\begin{split}
P&(\phi_{ij}^{\textrm{H}},\kappa_{\textrm{ext}},\gamma_{\textrm{ext}},\bm{\zeta}|\textrm{H},\textrm{MS})\\
&=P(\phi_{ij}^{\textrm{H}}|\textrm{H})P(\kappa_{\textrm{ext}}|\textrm{H,MS})P(\gamma_{\textrm{ext}}|\textrm{H})P(\bm{\zeta}|\textrm{H})
\end{split}
\end{equation}
are the priors on the parameters of the mass model and MS. Note that Millennium Simulation naturally provides more $\kappa_{\textrm{ext}}$ which close to mean density and lens model implicitly assumes $\kappa_{\textrm{ext}}<1$, so  $P(\kappa_{\textrm{ext}}|\textrm{H,MS})$ is a non-flat prior with an upper bound ($<1$).

The next step is to separate the datasets into [$\bm{d}_{\textrm{HST}},\sigma, \bm{d_{\textrm{ENV}}}$] and [$\bm{d}_{\textrm{AO}}$]. Since the data are all independent, we can write down
\begin{equation}
\label{eq:separate}
    \begin{split}
        P(&\bm{d}_{\textrm{HST}},\bm{d}_{\textrm{AO}},\sigma, \bm{d_{\textrm{ENV}}}|\phi_{ij}^{\textrm{H}},\kappa_{\textrm{ext}},\gamma_{\textrm{ext}},\bm{\zeta},\textrm{H},\textrm{MS})\\
        =&P(\bm{d}_{\textrm{HST}},\sigma, \bm{d_{\textrm{ENV}}}|\phi_{ij}^{\textrm{H}},\kappa_{\textrm{ext}},\gamma_{\textrm{ext}},\bm{\zeta},\textrm{H},\textrm{MS})\\
        &\cdot P(\bm{d}_{\textrm{AO}}|\phi_{ij}^{\textrm{H}},\kappa_{\textrm{ext}},\gamma_{\textrm{ext}},\bm{\zeta},\textrm{H},\textrm{MS}).
    \end{split}
\end{equation}
Although $\bm{d}_{\textrm{AO}}$ do not have the direct constraint power on $\kappa_{\textrm{ext}}$, the shear value inferred from $\bm{d}_{\textrm{AO}}$ implicitly help constrain $\kappa_{\textrm{ext}}$. Furthermore, we leave $\kappa_{\textrm{ext}}$ in the last term in \eref{eq:separate} because we want to link to $P(\bm{\eta}|\bm{d}_{\textrm{AO}},\textrm{H},\textrm{MS})$ in the future steps. 
\eref{bayes} become
\begin{equation}
\label{inde}
\begin{split}
    P(&\phi_{ij}^{\textrm{H}},\kappa_{\textrm{ext}},\gamma_{\textrm{ext}},\bm{\zeta}|\bm{d}_{\textrm{HST}},\bm{d}_{\textrm{AO}},\sigma, \bm{d_{\textrm{ENV}}},\textrm{H},\textrm{MS})\\
   =&P(\bm{d}_{\textrm{HST}},\sigma, \bm{d_{\textrm{ENV}}}|\phi_{ij}^{\textrm{H}},\kappa_{\textrm{ext}},\gamma_{\textrm{ext}},\bm{\zeta},\textrm{H},\textrm{MS})\\ &\cdot\frac{P(\bm{d}_{\textrm{AO}}|\phi_{ij}^{\textrm{H}},\kappa_{\textrm{ext}},\gamma_{\textrm{ext}},\bm{\zeta},\textrm{H},\textrm{MS})P(\phi_{ij}^{\textrm{H}},\kappa_{\textrm{ext}},\gamma_{\textrm{ext}},\bm{\zeta}|\textrm{H},\textrm{MS})}{P(\bm{d}_{\textrm{HST}},\bm{d}_{\textrm{AO}},\sigma, \bm{d_{\textrm{ENV}}}|\textrm{H},\textrm{MS})},
\end{split}
\end{equation}
where
\begin{equation}
\label{eqe}
\begin{split}
    P(&\bm{d}_{\textrm{HST}},\sigma, \bm{d_{\textrm{ENV}}}|\phi_{ij}^{\textrm{H}},\kappa_{\textrm{ext}},\gamma_{\textrm{ext}},\bm{\zeta},\textrm{H},\textrm{MS})\\
    =&P(\bm{d}_{\textrm{HST}}|\phi_{ij}^{\textrm{H}},\gamma_{\textrm{ext}},\bm{\zeta},\textrm{H})P(\sigma|\phi_{ij}^{\textrm{H}},\kappa_{\textrm{ext}},\gamma_{\textrm{ext}},\bm{\zeta},\textrm{H},\textrm{MS})\\
    &\cdot P(\bm{d_{\textrm{ENV}}}|\gamma_{\textrm{ext}},\kappa_{\textrm{ext}},\textrm{MS}).
\end{split}
\end{equation}
The last term in \eref{eqe} tells us that the shear value inferred from the lens imaging can also help us to further constrain the convergence because of the correlation between the $\gamma_{\textrm{ext}}$ and $\kappa_{\textrm{ext}}$ in MS.

Bayes theorem tells us that 
\begin{equation}
\label{eqqee}
\begin{split}
P(&\bm{d}_{\textrm{HST}},\sigma, \bm{d_{\textrm{ENV}}}|\phi_{ij}^{\textrm{H}},\kappa_{\textrm{ext}},\gamma_{\textrm{ext}},\bm{\zeta},\textrm{H},\textrm{MS})\\
=&P(\phi_{ij}^{\textrm{H}},\kappa_{\textrm{ext}},\gamma_{\textrm{ext}},\bm{\zeta}|\bm{d}_{\textrm{HST}},\sigma,\bm{d_{\textrm{ENV}}},\textrm{H},\textrm{MS})\\
&\cdot\frac{P(\bm{d}_{\textrm{HST}},\sigma,\bm{d_{\textrm{ENV}}}|\textrm{H},\textrm{MS})}{P(\phi_{ij}^{\textrm{H}},\kappa_{\textrm{ext}},\gamma_{\textrm{ext}},\bm{\zeta}|\textrm{H},\textrm{MS})},
\end{split}
\end{equation}
and
\begin{equation}
\label{eqqe}
    \begin{split}
        P&(\bm{d}_{\textrm{AO}}|\phi_{ij}^{\textrm{H}},\kappa_{\textrm{ext}},\gamma_{\textrm{ext}},\bm{\zeta},\textrm{H},\textrm{MS})\\
        &=\frac{P(\phi_{ij}^{\textrm{H}},\kappa_{\textrm{ext}},\gamma_{\textrm{ext}},\bm{\zeta}|\bm{d}_{\textrm{AO}},\textrm{H},\textrm{MS})P(\bm{d}_{\textrm{AO}}|\textrm{H},\textrm{MS})}{P(\phi_{ij}^{\textrm{H}},\kappa_{\textrm{ext}},\gamma_{\textrm{ext}},\bm{\zeta}|\textrm{H},\textrm{MS})}.
    \end{split}
\end{equation}
By substituting \eref{eqqee} and \eref{eqqe} into \eref{inde}, finally we obtain
\begin{equation}
\label{final}
\begin{split}
P(&\phi_{ij}^{\textrm{H}},\kappa_{\textrm{ext}},\gamma_{\textrm{ext}},\bm{\zeta}|\bm{d}_{\textrm{HST}},\bm{d}_{\textrm{AO}},\sigma, \bm{d_{\textrm{ENV}}},\textrm{H},\textrm{MS})\\
\propto &P(\phi_{ij}^{\textrm{H}},\kappa_{\textrm{ext}},\gamma_{\textrm{ext}},\bm{\zeta}|\bm{d}_{\textrm{HST}},\sigma, \bm{d_{\textrm{ENV}}},\textrm{H},\textrm{MS})\\
&\cdot\frac{P(\phi_{ij}^{\textrm{H}},\kappa_{\textrm{ext}},\gamma_{\textrm{ext}},\bm{\zeta}|\bm{d}_{\textrm{AO}},\textrm{H},\textrm{MS})}{P(\phi_{ij}^{\textrm{H}},\kappa_{\textrm{ext}},\gamma_{\textrm{ext}},\bm{\zeta}|\textrm{H},\textrm{MS})}.
\end{split}
\end{equation} 
The first term in the right hand side in \eref{final} is done by \citet{WongEtal17}, while the numerator in the second term is from AO data alone. Thus, based on \eref{final}, in order to get the joint constraint, we need to multiply this two posteriors and divide it by the non-uniform priors (e.g., $P(\kappa_{\textrm{ext}}|\textrm{H,MS})$) used in both datasets. This is because we need to get rid of the extra constraining power from doubly using the same non-uniform priors.
Therefore, the denominator in \eref{final} become $P(\kappa_{\textrm{ext}}|\textrm{H,MS})$.
Note that $\phi_{ij}^{\textrm{H}}$ is the model fermat potential, but what we want to obtain is $P(\phi_{ij}^{\textrm{True}})$, which can be expressed as

\begin{equation}
\begin{split}
P&(\phi_{ij}^{\textrm{True}}|\bm{d}_{\textrm{HST}},\bm{d}_{\textrm{AO}},\sigma, \bm{d_{\textrm{ENV}}},\textrm{H},\textrm{MS})\\
    &=\int d\kappa_{\textrm{ext}}\int d\phi^{\textrm{H}}_{ij}P(\phi^{\textrm{H}}_{ij},\kappa_{\textrm{ext}})\delta(\phi^{\textrm{True}}_{ij}-\phi^{H}_{ij}(1-\kappa_{\textrm{ext}}))\\
    &=\int d\kappa_{\textrm{ext}}\int d\phi^{\textrm{H}}_{ij}P(\phi^{\textrm{H}}_{ij},\kappa_{\textrm{ext}})\frac{\delta(\phi^{\textrm{H}}_{ij}-\phi^{\textrm{True}}_{ij}/(1-\kappa_{\textrm{ext}}))}{|-(1-\kappa_{\textrm{ext}})|}\\
    &=\int d\kappa_{\textrm{ext}}\frac{P(\phi^{\textrm{True}}_{ij}/(1-\kappa_{\textrm{ext}}),\kappa_{\textrm{ext}})}{1-\kappa_{\textrm{ext}}}
\end{split}
\end{equation}
where 
\begin{equation}
\begin{split}
&P(\phi^{\textrm{H}}_{ij},\kappa_{\textrm{ext}})\\
&=\int P(\phi_{ij}^{\textrm{H}},\kappa_{\textrm{ext}},\gamma_{\textrm{ext}},\bm{\zeta}|\bm{d}_{\textrm{HST}},\bm{d}_{\textrm{AO}},\sigma, \bm{d_{\textrm{ENV}}},\textrm{H},\textrm{MS}) d\bm{\zeta} d\gamma_{\textrm{ext}}.
\end{split}
\end{equation}

\section{SUMMARY OF PG1115+080~LENS MODELS WITH RESPECT TO THE BIC VALUE}
\label{appendix:pg_BIC}
We present the BIC of the power-law models in \tref{tab:6TD_1} and composite models in \tref{tab:6TD_2}.

\begin{table*}
    \centering
    \caption{Total 20 power-law models of PG1115+080 ordered in increased $\Delta$ BIC value.}
    \label{tab:6TD_1}
    \begin{tabular}{llccccrr}
        \hline
        Main lens model & perturbers & $S_{r}$ & $\Delta$ BIC & posterior weight\\
        \hline
        SPEMD     & group (NFW)          &$37\times37$ & 0 & 1\\
        SPEMD     & group (NFW)+G1+G2    &$37\times37$ & 11 & 0.9820\\
        SPEMD     & group (NFW)+G1+G2    &$39\times39$ & 18 & 0.9469\\
        SPEMD     & group (NFW)+G1       &$37\times37$ & 31 & 0.8967\\
        SPEMD     & group (NFW)          &$38\times38$ & 35 & 0.8338\\
        SPEMD     & group (NFW)+G1+G2    &$33\times33$ & 41 & 0.7614\\
        SPEMD     & group (NFW)+G1       &$41\times41$ & 46 & 0.6827\\ 
        SPEMD     & group (NFW)+G1+G2    &$35\times35$ & 46 & 0.6827\\
        SPEMD     & group (NFW)+G1       &$39\times39$ & 54 & 0.5198\\
        SPEMD     & group (SIS)~~~+G1    &$36\times36$ & 72 & 0.4415\\
        SPEMD     & group (SIS)~~~+G1    &$34\times34$ & 74 & 0.3684\\
        SPEMD     & group (NFW)+G1       &$35\times35$ & 88 & 0.3019\\
        SPEMD     & group (NFW)+G1+G2    &$41\times41$ & 90  & 0.2433\\ 
        SPEMD     & group (SIS)~~~+G1    &$32\times32$ & 91 & 0.1930\\
        SPEMD     & group (NFW)          &$32\times32$ & 92 & 0.1510\\
        SPEMD     & group (NFW)          &$36\times36$ & 95 & 0.1172\\
        SPEMD     & group (NFW)          &$34\times34$ & 114 & 0.0911\\
        SPEMD     & group (NFW)+G1       &$43\times43$ & 126 & 0.0721\\ 
        SPEMD     & group (SIS)~~~+G1    &$40\times40$ & 127 & 0.0597\\
        SPEMD     & group (SIS)~~~+G1    &$38\times38$ & 148 & 0.0536\\
        \hline
    \end{tabular}
\end{table*}
\begin{table*}
    \centering
    \caption{Total 20 composite models of PG1115+080 ordered in increased $\Delta$ BIC value.}
    \label{tab:6TD_2}
    \begin{tabular}{llccccrr}
        \hline
        Main lens model & perturbers & $S_{r}$ & $\Delta$ BIC & posterior weight\\
        \hline
        COMPOSITE & group (SIS)~~~+G1  &$37\times37$ & 0 & 1\\
        COMPOSITE & group (NFW)        &$30\times30$ & 34 & 0.9799\\
        COMPOSITE & group (NFW)        &$32\times32$ & 55 & 0.9409\\
        COMPOSITE & group (NFW)        &$36\times36$ & 67 & 0.8853\\
        COMPOSITE & group (NFW)+G1+G2  &$39\times39$ & 69 & 0.8162\\
        COMPOSITE & group (NFW)+G1+G2  &$37\times37$ & 80 & 0.7374\\
        COMPOSITE & group (NFW)+G1     &$33\times33$ & 88 & 0.6529\\
        COMPOSITE & group (NFW)+G1     &$35\times35$ & 90 & 0.5664\\
        COMPOSITE & group (NFW)+G1     &$32\times32$ & 96 & 0.4815\\
        COMPOSITE & group (NFW)+G1+G2  &$33\times33$ & 118 & 0.4010\\
        COMPOSITE & group (NFW)+G1+G2  &$35\times35$ & 122 & 0.3273\\
        COMPOSITE & group (NFW)        &$34\times34$ & 132 & 0.2620\\
        COMPOSITE & group (SIS)~~~+G1  &$31\times31$ & 138 & 0.2057\\
        COMPOSITE & group (SIS)~~~+G1  &$35\times35$ & 145 & 0.1584\\
        COMPOSITE & group (NFW)+G1     &$39\times39$ & 146 & 0.1200\\
        COMPOSITE & group (SIS)~~~+G1  &$33\times33$ & 149 & 0.0897\\
        COMPOSITE & group (SIS)~~~+G1  &$39\times39$ & 156 & 0.0669\\
        COMPOSITE & group (NFW)+G1+G2  &$31\times31$ & 163 & 0.0506\\
        COMPOSITE & group (NFW)+G1     &$37\times37$ & 193 & 0.0402\\
        COMPOSITE & group (NFW)        &$38\times38$ & 232 & 0.0352\\
		\hline
    \end{tabular}
\end{table*}

%%%%%%%%%%%%%%%%%%%%%%%%%%%%%%%%%%%%%%%%%%%%%%%%%%
\section{SUMMARY OF \he~LENS MODELS WITH RESPECT TO THE BIC VALUE}
\label{appendix:he_BIC}
We present the BIC of the power-law models in \tref{tab:he_BIC_power} and composite models in \tref{tab:he_BIC_comp}.
\begin{table*}
    \centering
    \caption{Total 5 powerlaw models of \he\ ordered in increased $\Delta$ BIC value.}
    \label{tab:he_BIC_power}
    \begin{tabular}{llccccrr}
        \hline
        Main lens model & perturbers & $S_{r}$ & $\Delta$ BIC & posterior weight\\
        \hline
        SPEMD     & G1          &$50\times50$ & 0 & 1\\
        SPEMD     & G1          &$40\times40$ & 5 & 0.9743\\
        SPEMD     & G1          &$45\times45$ & 7 & 0.9328\\
        SPEMD     & G1          &$35\times35$ & 14 &0.8914\\
        SPEMD     & G1          &$30\times30$ & 16 &0.8658\\
        \hline
    \end{tabular}
\end{table*}

\begin{table*}
    \centering
    \caption{Total 10 composite models of \he\ ordered in increased $\Delta$ BIC value.}
    \label{tab:he_BIC_comp}
    \begin{tabular}{llccccrr}
        \hline
        Main lens model & perturbers & $S_{r}$ & $\Delta$ BIC & posterior weight\\
        \hline
        COMPOSITE   & G1--G5        &$43\times43$ & 0 & 1\\
        COMPOSITE     & G1          &$43\times43$ & 10 & 0.9855\\
        COMPOSITE     & G1          &$41\times41$ & 11 & 0.9580\\
        COMPOSITE   & G1--G5        &$41\times41$ & 12 & 0.9201\\
        COMPOSITE     & G1          &$39\times39$ & 16 & 0.8756\\
        COMPOSITE   & G1--G5        &$39\times39$ & 19 & 0.8288\\
        COMPOSITE   & G1--G5        &$37\times37$ & 20 & 0.7844\\
        COMPOSITE     & G1          &$35\times35$ & 21 & 0.7466\\
        COMPOSITE     & G1          &$37\times37$ & 21 & 0.7466\\
        COMPOSITE   & G1--G5        &$35\times35$ & 23 & 0.7046\\
    \end{tabular}
\end{table*}

\section{SUMMARY OF \rxj~LENS MODELS WITH RESPECT TO THE BIC VALUE}
\label{appendix:rxj_BIC}
We present the BIC of the power-law models in \tref{tab:rxj_BIC_power} and composite models in \tref{tab:rxj_BIC_comp}.
\begin{table*}
    \centering
    \caption{Total 5 power-law model of \rxj\ ordered in increased $\Delta$ BIC value.}
    \label{tab:rxj_BIC_power}
    \begin{tabular}{llccccrr}
        \hline
        Main lens model & perturbers & $S_{r}$ & $\Delta$ BIC & posterior weight\\
        \hline
        SPEMD     & satellite (SIS) &$79\times79$ & 0 & 1\\
        SPEMD     & satellite (SIS) &$77\times77$ & 85 & 0.4618\\
        SPEMD     & satellite (SIS) &$75\times75$ & 176 & 0.0912\\
        SPEMD     & satellite (SIS) &$73\times73$ & 499 & 0.0071\\
        SPEMD     & satellite (SIS) &$71\times71$ & 795 & 0.0002\\
        \hline
    \end{tabular}
\end{table*}

\begin{table*}
    \centering
    \caption{Total 5 composite models of \rxj~ordered in increased $\Delta$ BIC value.}
    \label{tab:rxj_BIC_comp}
    \begin{tabular}{llccccrr}
        \hline
        Main lens model & perturbers & $S_{r}$ & $\Delta$ BIC & posterior weight\\
        \hline
        COMPOSITE     & satellite (SIS) &$71\times71$ & 0 & 1\\
        COMPOSITE     & satellite (SIS) &$70\times70$ & 398 & 0.4050\\
        COMPOSITE     & satellite (SIS) &$73\times73$ & 915 & 0.0575\\
        COMPOSITE     & satellite (SIS) &$74\times74$ & 1641 & 0.0025\\
        COMPOSITE     & satellite (SIS) &$72\times72$ & 1798 & 0.0000\\
        \hline
    \end{tabular}
\end{table*}

% Don't change these lines
\bsp	% typesetting comment
\label{lastpage}
\end{document}